\documentclass{ws-rv961x669}
\usepackage[square]{ws-rv-van} 

\usepackage{atbegshi,picture}

\AtBeginShipout{\AtBeginShipoutFirst{%
  \put(\dimexpr\paperwidth-5cm\relax,-1cm){\makebox[0pt][r]{LPTENS-18/02}} %
}}

\crop[off] 

\usepackage[utf8]{inputenc}
\usepackage{setspace}
\usepackage{verbatim}
\usepackage[usenames,dvipsnames]{xcolor}
\usepackage{amsmath}
\usepackage{latexsym}
\usepackage{tikz}
\usepackage{caption}
\usepackage{subcaption}
\usepackage{mathtools}
\usepackage[colorlinks=true,linkcolor=black, citecolor=black, urlcolor=black]{hyperref}

\usepackage{epstopdf}

\newcommand{\p}{\partial}

\newcommand{\CP}{{\cal P}}
\newcommand{\bP}{{\bf P}}
\newcommand{\bQ}{{\bf Q}}
\newcommand{\bp}{{\bf p}}

\newcommand{\tr}{{\rm tr }}

\newcommand{\Tr}{{\rm Tr }}

\makeatletter
\newcommand{\subalign}[1]{%
  \vcenter{%
    \Let@ \restore@math@cr \default@tag
    \baselineskip\fontdimen10 \scriptfont\tw@
    \advance\baselineskip\fontdimen12 \scriptfont\tw@
    \lineskip\thr@@\fontdimen8 \scriptfont\thr@@
    \lineskiplimit\lineskip
    \ialign{\hfil$\m@th\scriptstyle##$&$\m@th\scriptstyle{}##$\crcr
      #1\crcr
    }%
  }
}
\makeatother

\usepackage{amssymb}
\usepackage{varioref}
\usepackage{graphicx}

\renewcommand{\thefootnote}{\fnsymbol{footnote}}     

\begin{document}



\chapter*{Quantum Spectral Curve of  \(\gamma\)-twisted \({\cal N}=4\) SYM theory \\ and fishnet CFT$^*$}

\footnotetext{$^*$This work will appear in {\it Ludwig Faddeev Memorial Volume: A Life in Mathematical Physics}, edited by Molin Ge, Antti Niemi, Kok Khoo Phua and Leon A Takhtajan (World Scientific, 2018); \url{http://www.worldscientific.com/worldscibooks/10.1142/10811}}

\addtocounter{footnote}{1}

\author[V. Kazakov]{Vladimir Kazakov\,\,\(^{a,b}\)}

\address{\(^a\)Laboratoire de Physique Th\'eorique, Ecole Normale Sup\'erieure,
\\  24 rue Lhomond, F-75231 Paris Cedex 05, France\footnote{kazakov AT lpt.ens.fr}
\\  \(^b\)PSL Universit\'es, Sorbonne Universit\'es, UPMC Universit\'e Paris 6, CNRS}

\begin{abstract}
We review the quantum spectral curve (QSC) formalism for the spectrum of anomalous dimensions of \({\cal\ N}=4\) SYM, including its \(\gamma\)-deformation. Leaving aside its derivation, we concentrate on the formulation of the ``final product" in its most general form: a minimal set of assumptions   about the algebraic structure and the analyticity of the \(Q\)-system -- the full system of Baxter \(Q\)-functions of the underlying integrable model.  The algebraic structure of the  \(Q\)-system is entirely based on (super)symmetry of the model and is efficiently described by Wronskian formulas for \(Q\)-functions organized into the Hasse diagram. When  supplemented with analyticity conditions on \(Q\)-functions, it fixes completely the set of physical solutions for the spectrum of an integrable model. First we demonstrate the spectral equations on the example of \(gl(N)\) and \(gl(K|M     )\) Heisenberg (super)spin chains. Supersymmetry \(gl(K|M)\)  occurs as a simple ``rotation" of the  Hasse diagram for a \(gl(K+M)\) system. Then we apply this method to the spectral problem of AdS\(_5\)/CFT\(_4\)-duality, describing the QSC formalism. The main difference with the spin chains consists in  more complicated analyticity constraints on \(Q\)-functions which involve an infinitely branching Riemann surface and a set of Riemann-Hilbert conditions. As an example of application of QSC, we consider a special double scaling  limit of \(\gamma\)-twisted \({\cal\ N}=4\) SYM, combining  weak coupling and  strong imaginary twist. This leads to a new type of  non-unitary CFT dominated by particular integrable, and often computable,   4D fishnet Feynman graphs. For the simplest of such models -- the bi-scalar  theory -- the QSC degenerates into the \(Q\)-system for integrable non-compact Heisenberg spin chain with conformal, \(SU(2,2)\) symmetry.
We describe the QSC derivation of  Baxter equation and the quantisation condition for particular fishnet graphs -- wheel graphs, and review numerical and analytic results for them.\\[6pt]
\end{abstract}

\markboth{V. Kazakov}{Quantum Spectral Curve of  \(\gamma\)-twisted \({\cal N}=4\) SYM and fishnet CFT}

\body

\section{Introduction}

In the past 40  years, a multitude of super-symmetric conformal quantum field theories (CFT) in four dimensions has been discovered and studied\cite{Leigh1995}.
Typically, they are various deformations of super-Yang-Mills theories, with \({\cal N}=1-4\) supersymmetries (see \cite{Cordova:2016} for modern classification).       On  the other hand, well identified non-supersymmetric and/or non-gauge CFTs in four dimensions are  rare species. Apart from  a rather exotic Banks-Zaks theory \cite{Banks:1981nn} or critical Potts model\cite{Vasseur:2013baa} there are hardly known examples which are explicitly constructed and well understood.\footnote{In 3 dimensions a well defined and studied example of non-supersymmetric  CFT's is the Ising model, but not much beyond that.}

Even more rare are the  integrable four-dimensional CFT's. The \({\cal\ N}=4\) SYM theory is the emblematic example of such a theory: it is conformal for  \(SU(N_c)\) gauge group for any \(N_c\) but integrable  only in the large \(N_c\), t'~Hooft limit~(see the review\cite{Beisert:2010jr} and references therein).\footnote{Its 3-dimensional cousin is the ABJM model, also integrable at large \(N_c\).} It was long believed that its large global \(PSU(2,2|4)\) super-conformal symmetry is responsible for the integrability. However, under specific deformations breaking the supersymmetry partially or entirely, the theory seems to retain its integrability.

An important, and rather general class of such deformations is the  \(\gamma\)-twist~\cite{Leigh1995,Lunin:2005jy,Frolov:2005dj}. It breaks the global symmetry to \(SU(2,2)\times U(1)^3\), i.e. only three Cartan subgroups are left from \(R\)-symmetry and preserves, at least on the tree level, the  4D conformal symmetry. The last one could be endangered by various conformal anomalies\cite{Fokken:2013aea} but, remarkably, it survives when adding to the action a well defined set of double-trace counter-terms~\cite{Witten:2001ua,Dymarsky:2005uh,Tseytlin:1999ii,Sieg:2016vap,Grabner:2017pgm} and tuning the double-trace couplings to certain critical values. The critical double-trace couplings are, in general, complex functions of the 't~Hooft coupling which parameterizes the whole family of these non-unitary CFTs.

All the quantum integrability properties known from the undeformed  \({\cal\ N}=4\) SYM seem to survive as well this \(\gamma\)-deformation\cite{Grabner:2017pgm}. In particular, the quantum spectral curve (QSC)\cite{Gromov2014a,Gromov:2014caa,Kazakov:2015efa} -- the most advanced formalism of AdS/CFT integrability, giving a comprehensive solution of the problem of spectrum of anomalous dimensions of local (and some non-local) operators -- remains valid the \(\gamma\)-deformation, with minor modifications, and describes this non-unitary CFT precisely at the (complex) critical line~\cite{Grabner:2017pgm}. The QSC method  already found numerous applications in the study of planar spectrum of  \({\cal\ N}=4\) SYM theory    \cite{Bombardelli:2017vhk,Cavaglia:2014exa,Cavaglia:2016ide,Gromov:2015dfa,Gromov:2015vua,Gromov:2015wca,Gromov:2016rrp,Gromov:2017cja,Kazakov:2015efa,Klabbers:2017vtw,Marboe:2014gma,Marboe:2014sya,Marboe:2017dmb} (see also  recent review~\cite{Gromov:2017blm} and references therein).

Recently, \"O.G\"urdogan and the author proposed in \cite{Gurdogan:2015csr}  a special double scaling (DS) limit of the \(\gamma\)-deformed  \({\cal\ N}=4\) SYM, combining the weak coupling limit and  large imaginary values of the \(\gamma\)-parameters. It gives rise to a new 4D non-unitary CFT where the gauge interactions decouple and only chiral 4-scalar and Yukawa interactions are left. They are also expected to inherit the integrability properties of their  ``mother" theory -- the \(\gamma\)-deformed  \({\cal\ N}=4\) SYM.     In the simplest case, when only one double-scaling coupling is kept non-zero, it becomes a simple  theory of two interacting complex scalars (referred to in what follows as the ``bi-scalar theory"). Nevertheless, it is still a non-trivial interacting CFT but in addition it is integrable in planar limit! Its integrability, unlike the integrability of its ``mother" theory,  has a clear origin: its perturbation theory for various correlation functions is dominated by the ``fishnet" Feynman graphs. This means that  sufficiently large planar graphs have in the bulk  the shape of regular square lattice. It was noticed long ago\cite{Zamolodchikov:1980mb} that such a graph defines an integrable 2D statistical-mechanical spin system with   \(SU(2,2)\sim SO(4,2)\) symmetry, which is  a four-dimensional conformal group.  Thus the integrability of the bi-scalar theory is tightly related with the integrability of the conformal, non-compact \(SU(2,2)\) Heisenberg spin chain.  The  theory of integrable non-compact  spin chains has a long history~\cite{Izergin:1982ry,Faddeev-Volkov:1995,Bazhanov:2007mh,Lipatov:1993qn,Faddeev:1994zg,Derkachov:2001yn,Chicherin:2012yn} following the fundamental works of L.D.~Faddeev and the Leningrad school  (see \cite{Faddeev:1996iy} and references therein).   It had also   a few important applications, such as BFKL approximation in high-energy, Regge limit in QCD\cite{Lipatov:1993qn,Faddeev:1994zg}. Many of these and other old results on non-compact integrable spin chains  appear to be very helpful in the study of  non-perturbative dynamics of the bi-scalar theory and the other CFT's from the family of chiral CFT obtained in the DS limit from  \(\gamma\)-deformed  \({\cal N}=4\) SYM~\cite{Caetano:2016ydc}.

In this work, we will review the formalism of quantum spectral curve (QSC) for the spectrum of anomalous dimensions of the planar \({\cal\ N}=4\) SYM theory, including its  \(\gamma\)-deformed  version (named below as QSC\(\gamma\)). We will concentrate ourselves on the minimal set of basic propositions when formulating the  QSC\(\gamma\) equations, leaving aside its derivation. The algebraic part of QSC\(\gamma\), entirely dictated by the global \(PSU(2,2|4)\)  symmetry (broken to \(SU(2,2)\times U(1)^3\) by \(\gamma\)-deformation) and quantum integrability, is most conveniently formulated in terms of the \(Q\)-system - a set of \(2^8\) Baxter's \(Q\)-functions of spectral parameter \(u\). \(Q\)-functions are organized into the Hasse diagram, reflecting the fact that they are Grassmannian coordinates and they obey certain  Pl\"ucker relations. The analytic part of  QSC\(\gamma\) construction consists of description of the structure of Riemann surfaces of \(Q\)-functions, where the main element is an infinite ``ladder" of equally spaced ``Zhukovsky" quadratic cuts at \(u\in (-2g+i\mathbb{Z}\,\,,2g+i\mathbb{Z})    \), where \(g\) is the 't~Hooft coupling.  The large \(u\) asymptotics of \(Q\)-functions and their specific monodromy properties around the Zhukovsky cuts
conclude  the formulation of spectral problem. Roughly speaking,  QSC\(\gamma\)            represents a system of non-linear Riemann-Hilbert equations. We will start with the  section \ref{sec:spin_chains} where we demonstrate the similar formalism and the emergence of the supersymmetric \(Q\)-system on the example of Heisenberg \(GL(K|M)\) spin  chain where the analyticity constraints look much simpler. Then, in section \ref{sec:QSCgamma}, we will describe the QSC\(\gamma\) for \(\gamma\)-deformed  \({\cal\ N}=4\) SYM (AdS\(_5\)/CFT\(_4\) duality).  In section \ref{sec:fishnet}, we will  describe the CFTs emerging in the DS limit of this theory, and in particular the so-called bi-scalar theory dominated by integrable ``fishnet" graphs.  We will  review the results obtained for these models from  QSC\(\gamma\), such as the exact computation of certain multi-loop ``wheel" graphs and discuss the equivalence of the bi-scalar theory to the conformal integrable Heisenberg spin chain. The section \ref{sec:conclusions} is devoted to conclusions and unsolved problems.

\section{Spectrum of Heisenberg \(GL(K|M)\) spin chain from Baxter\\
\(Q\)-functions}\label{sec:spin_chains}

In this section, we  will give an alternative formulation of the well known solution for the spectrum of compact,  Heisenberg \(GL(K|M)\) spin  chain which will be useful for the generalization to AdS/CFT integrability. We will avoid the direct use of standard Bethe equations since in the  sigma-model on \(AdS_5\times S^5\) background, which describes the string side of the duality,   the notion of Bethe roots is a tricky and not very invariant issue.  We will rather rely on  the  full system of Baxter \(Q\)-functions, forming a Grassmannian, spectral parameter dependent structure   in the \(K+M\) dimensional space. The \(Q\)-functions are most naturally classified by the vertices of Hasse diagram -- the \((K+M)\)-dimensional hypercube.       Specifying the analyticity properties of these \(Q\)-functions w.r.t.  the spectral parameter \(u\) one can classify and efficiently study the spectrum. The supersymmetric generalization of this picture in terms of Hasse diagram, from \(GL(K+M)\) bosonic group to  to \(GL(K|M)\) supersymmetric group will be essentially a ``rotation" of the hypercube when imposing specific determinant relations (``determinat" flow) and analyticity conditions.
Many of the details, missing in this short overview of spin chains from the point of view of \(Q\)-functions, can be found in~\cite{Kazakov:2007na,Kazakov:2010iu,Kazakov:2015efa,Alexandrov:2011aa} and in the recent review~\cite{Gromov:2017blm}.

\subsection{Spectrum of \(GL(N)\) spin chain via \(Q\)-functions on\\ Hasse diagram}

Let us start from the  Heisenberg spin chain with spins belonging to the bosonic group \(GL(N) \) and twisted boundary conditions.  This spin-chain is defined through the \(1D\) hamiltonian
\begin{align}\label{Hamiltonian}
\hat H=\sum_{i=1}^{L-1}\CP_{i,i+1} +\CP_{L,1}\,\,G_{_L}G_1^{-1}
\end{align} where the spin at each site \(i\) takes the values \(s_i=1,2,\dots,N\), the permutation acts on a pair of spins as \(\CP_{i,j}(s_1,\dots,s_i,\dots s_j,\dots s_L)=(s_1,\dots,s_j,\dots s_i,\dots s_N)\) and the twist \(G={\rm diag}\{x_1,x_2,\cdots,x_N\}\in SU(N)\) is a fixed  element of Cartan subgroup.\footnote{All eigenvalues are assumed to be different, and we impose the unimodularity \(\prod_{j=1}^N x_j=1\).} Explicitly, in components, various terms in~\eqref{Hamiltonian} mean\begin{align} &\CP_{i,i+1}= \delta_{s_1}^{s_1'}\delta_{s_2}^{s_2'}\dots\delta_{s_i}^{s_{i+1}'}\delta_{s_{i+1}}^{s_i'}
\dots\delta_{s_L}^{s_L'}\,, \\
 &\CP_{L,1}\,\,G_{_L}G_1^{-1}=\delta_{s_2}^{s_2'}\delta_{s_3}^{s_3'}\dots\delta_{s_L-1}^{s_L'-1}\,\,
\delta_{s_L}^{s_1'}\delta_{s_1}^{s_L'}\frac{x_{s_L}}{x_{s_1}}.
\end{align}

To formulate the solution for the spectrum of this hamiltonian we introduce a set of \(N\)  Baxter \(Q_{ k}(u)\)-functions of spectral parameter \(u\) with  a single index
 \begin{align}\label{Qpolyn}
Q_{ k}(u)=x_k^{iu}\prod_{j=1}^{R_k}(u-u_j^{( k)}),\qquad k=1,2,\dots,N,
\end{align} each of them being a polynomial of spectral parameter \(u\) times  a twist-dependent exponential factor.  The positive integers \(\{R_1,R_2,\cdots,R_N\}\) are in fact the Cartan charges of the \(U(1)^N\) residual symmetry left after breaking the original \(GL(N)\) symmetry by twisting. 

Let us also define a natural object -- the multi-index \(Q\)-functions -- by the following Wronskian formula:  \begin{align}\label{QI}Q_I(u)\equiv Q_{j_1,\dots,j_k}(u)=\frac{\det_{1\le m,n\le k}\,\,Q_{j_m}^{[-1-k+2n]}}{\prod_{j=0}^{k-1}Q_\emptyset^{[-k+1+2j]}}, \end{align} where we denoted by capital letter \(I=\{j_1,\dots,j_k\}\in  \bar\emptyset\equiv \{1,2,\dots,N\} \)  a subset of the full set \(\bar\emptyset\) of indices. By definition, all indices in this subset are different and ordered from left to right. Any permutation \(P\) of indices in \eqref{QI} can only change the overall  sign by factor \((-1)^P\).  It was also natural to introduce the ``empty set" \(Q\)-function  \(Q_{\emptyset}(u)\) in denominator of  \eqref{QI}, but the reasons which will be clear below, when we will discuss the Pl\"ucker relations  \eqref{Plucker}.   In total, we have \(2^N\) different \(Q\)-functions, but they are obviously interrelated since they are given in terms of only \(N\) single index functions.

To fix all  the roots of these \(Q\)-functions, and thus to find all the eigenvalues of the above hamiltonian, it is enough to find all solutions of the following equation\cite{Pronko:1998xa,Pronko:1999gh} \begin{align}\label{spectralE}
Q_{12\cdots N}(u)=\Delta(x_1,\cdots,x_N)\,u^L\,,\qquad  Q_\emptyset(u)=1,
\end{align}
where, according to  \eqref{QI}, \begin{align}\label{WronskianN}
Q_{12\cdots N}(u)\equiv \det_{1\le k,j\le N} Q_k^{[-1-N+2j]}
\end{align} is  the Wronskian of the full set of single index
\(Q\)-functions and \(\Delta(x_1,\dots,x_N)=\prod_{k>j}(x_k-x_j)\) is the Vandermonde determinant of twist eigenvalues.
 Here and below we use the notations for standard shifts of arguments of the functions: \(f^{[n]}\equiv f(u+\frac{in}{2})\).

Once one   finds a solution of \eqref{spectralE}, the corresponding energy -- the eigenvalue of the hamiltonian \eqref{Hamiltonian} -- is given by the familiar formula \begin{align}\label{Energy}
E=L+i\,\partial_u\log\frac{Q_{\bar k}(u-i/2)}{Q_{\bar k}(u+i/2)}\left|_{u=0}\right. \,,
\end{align} where we used the  \(N-1\) index \(Q\)-functions   \(Q_{\bar k}\equiv Q_{12\dots \hat k\dots,N}\) given by \((N-1)\times(N-1)\)-determinant according to the formula \eqref{QI}\footnote{In general, the bar over a subset \(I\), i.e. \(\bar I\), means a subset complementary to \(I\)  w.r.t. the full set \(\bar\emptyset\equiv\{1,2,\dots,N\}\), and \(\{\bar k\}\equiv \{1,2,\dots,\hat k,\dots,N\}\) denotes the set of \(N-1\) indices with \(k\) missing.}.
The answer does not depend on the choice of  \(Q_{\bar k}\)  (this symmetry is related to the so called particle-hole duality).

\begin{figure}[htb]  
\begin{center}
\includegraphics[width=\textwidth]{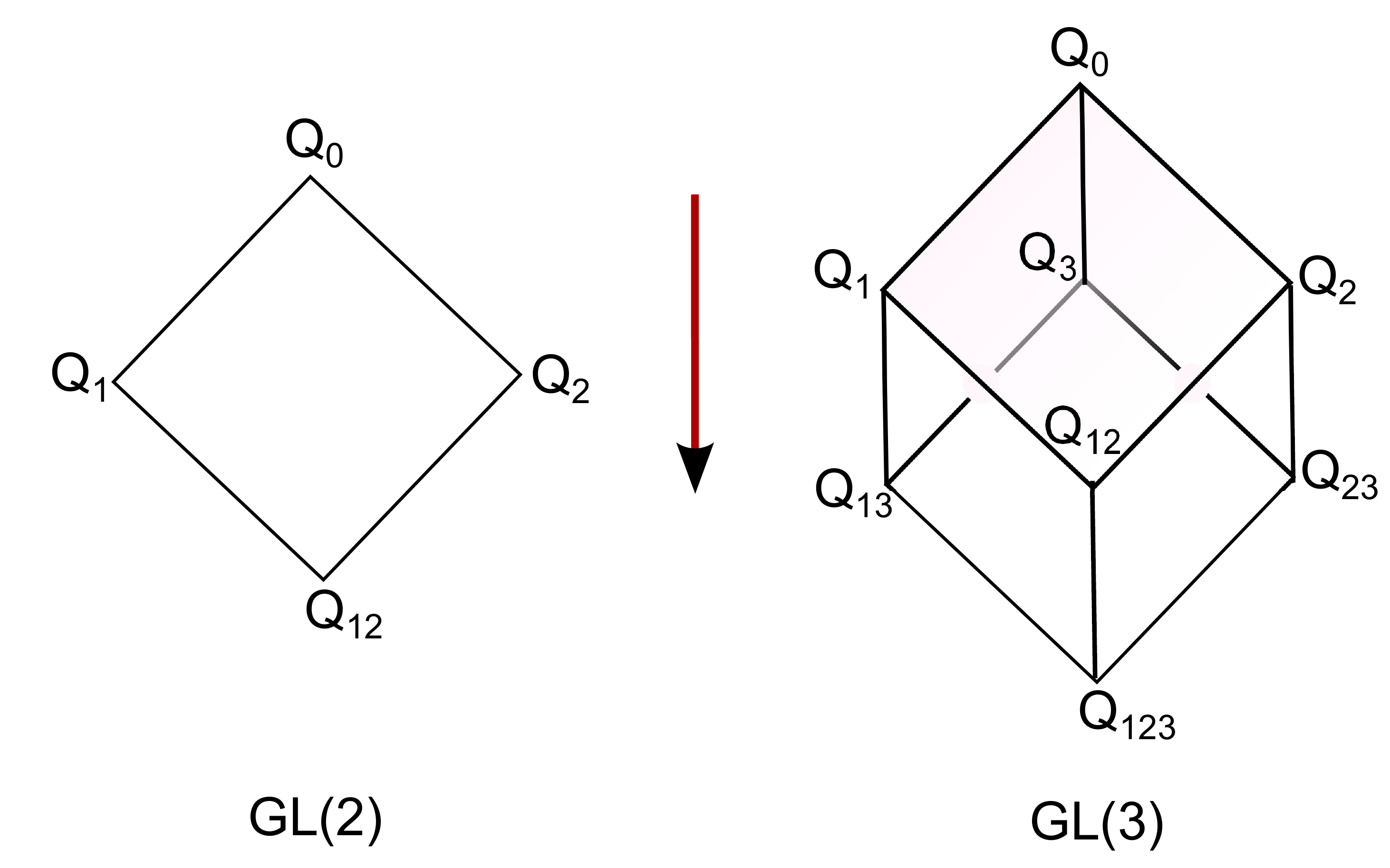}
\end{center}
\caption{Examples of Hasse diagrams for Q-system of Baxter functions of integrable models with \(GL(2)\) symmetry (on the left) and GL(3) symmetry (on the right).  The arrow shows the direction of ``determinant flow": the \(Q\)-functions with \(k\) indices are \(k\times k\) determinants \eqref{QI} of single index \(Q\)-functions, i.e. increasing in size with the increase of  the level \(k\).     }
\label{fig:Hasse23}
\end{figure}

It is natural to attach all these \(2^N\) \(Q\)-functions to the vertices of the \(N\)-dimensional hypercube which is called in this occasion the Hasse diagram. For example, in  the simplest case of \(GL(2)  \) spin chain (\(N=2\)) we have the set of 4 \(Q\)-functions: \(Q_\emptyset,Q_{1},Q_2,Q_{12}\) which we place at 4 vertices of the square, as shown
on Fig.\ref{fig:Hasse23}(left).
The upper vertex is occupied by \(Q_{\emptyset}\), which is connected by two edges with \(Q_1\) and \(Q_2\), which, in  turn, are connected by two edges with \(Q_{\bar\emptyset}\equiv Q_{12}=Q_1^+Q_2^--Q_1^-Q_2^+\)~\footnote{We used the notations \(f^{\pm}\equiv f(u\pm\frac{i}{2})\).}. The spectral equation \eqref{spectralE} takes the form
\begin{align}\label{Wronskian2}
Q_1^+Q_2^--Q_1^-Q_2^+=(x-1/x)u^L
\end{align} with the twist \(G=\{1/x,x\}\).  Imposing the specific analyticity condition -- the ``polynomiality'' of \(Q\)-functions  \eqref{Qpolyn} -- we obtain the usual Bethe equation for the roots of \(Q_1(u)\):   \begin{align}\label{Bethe2}
\frac{Q_1^{++}(u_j^{(1)})}{Q_1^{--}(u_j^{(1)})}=-\left(\frac{u_j^{(1)}+i/2}{u_j^{(1)}-i/2}\right)^L\,,\qquad j=1,2,\dots,R_1 .
\end{align} We used for that two relations \eqref{Wronskian2} at the roots of \(Q_1\), shifted from the original one by \(\pm i/2\), and divided one over another. A similar equation for the roots of \(Q_2\) leads to the same spectrum given by \eqref{Energy}.
 In the formula for energy \eqref{Energy} we can use either \(Q_{\bar 1}=Q_2\) or \(Q_{\bar 2}=Q_1\).

For \(N=3\), the Hasse diagram is  3D cube Fig.\ref{fig:Hasse23}(right). It is convenient to orient the cube in such a way that two of the vertices connected by the main diagonal  appear to be the upper and the lower ones. We place again \(Q_\emptyset\) at the top vertex, the single-indexed \(Q_{1},Q_2,Q_{3}\) -- at the vertices adjacent to it, then, say, \(Q_{13}\) --  on the next level, at the vertex adjacent to both \(Q_1\) and \(Q_3\), etc.
 One could pictorially think of the Hasse diagram as of the ``globe",   where the  level of \(Q\)-functions with given number of indices is like a  ``latitude", the \(\emptyset\)-vertex and the \(\bar\emptyset\)-vertex -- like the ``north and south poles", respectively.

The generalization to any \(N \) is straightforward.  All \(N+1 \) levels from top to bottom are ordered w.r.t. the number of indices \(|I|\equiv \text{Coordinality}(I)\) in the corresponding \(Q_I\) functions. This induces a natural direction in parameterization \eqref{QI} which we will call ``determinant flow". At a given \(u\), the collection of  functions  \(Q_{I}(u)\) on a   particular \(n\)-level with \(|I|=n\) forms an \(n\)-dimensional linear subspace \(V_{(n)}(u)\) representing the Pl\"ucker coordinates of a point on the Grassmannian \({\bf G}^n_N\) defined on the linear space \(\mathbb{C}^N\).
As was pointed out in \cite{Kazakov:2015efa}, the quantum integrability, constraining the spectra of various integrable models, from spin chains to \(2D\)  quantum field theories,  is based on the following abstract relation between these Pl\"ucker coordinates:
\begin{equation}
V_{(n)}(u + i/2)\cap\  V_{(n)}(u - i/2) = V_{(n-1)}(u)\,,\qquad  \forall n\in\{1, 2, . . . , N -1\},
\end{equation} following of course from \eqref{QI}. The Grassmannian structure of quantum integrability was first pointed out in \cite{Krichever:1996qd} on the example of transfer-matrices and Hirota bi-linear finite difference equations.

\begin{figure}[htb]  
\begin{center}
\includegraphics[width=\textwidth]{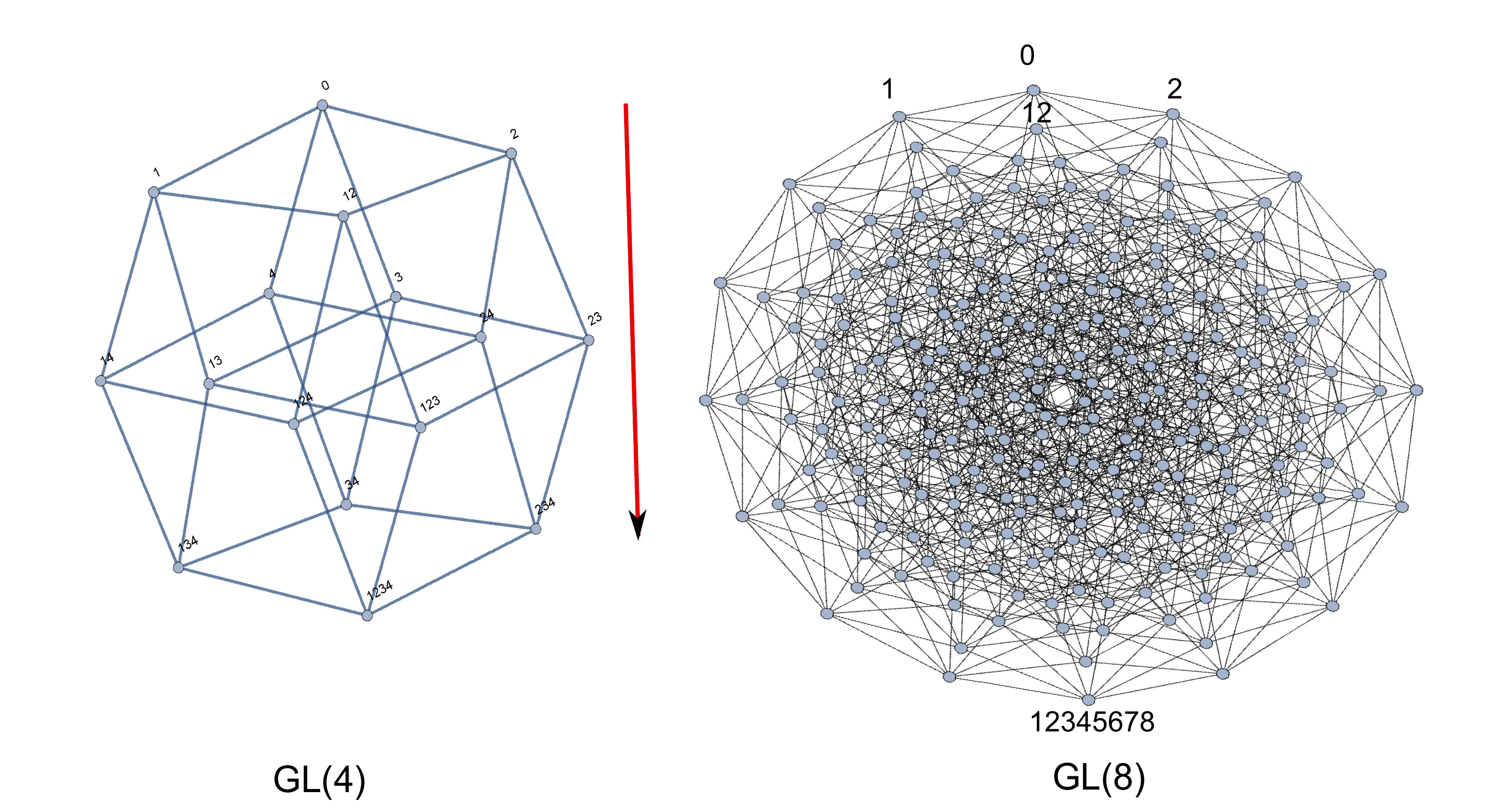}
\end{center}
\caption{Hasse diagrams for \(Q\)-system of Baxter functions of integrable models with \(GL(4)\) symmetry (on the left) and \(GL(8)\) symmetry (on the right) are hypercubes with the dimension of the rank of the symmetry of integrable quantum system. The arrow shows the direction of determinant flow according to \eqref{QI}.  The \(GL(8)\) Hasse diagram will be the same as for the \(PSU(2,2|4)\) super-group of  AdS\(_5\)/CFT\(_4\) duality (up to a certain ``rotation" of direction of the determinant flow and the details of analyticity structure).    }
\label{fig:Hasse48}
\end{figure}

We also present on Fig.\ref{fig:Hasse48} two other important examples. On Fig.\ref{fig:Hasse48}(left) we depicted the Hasse diagram for the \(GL(4)\) system, relevant for the conformal or for the R-symmetry subgroups of  \(PSU(2,2|4)\). On Fig.\ref{fig:Hasse48}(right) the Hasse diagram for the \(GL(8)\) system is presented.  As we will see, the last one is closely related to the full \(PSU(2,2|4)\) symmetry group of \({\cal\ N}=4\) SYM theory.

The determinant flow \eqref{QI} leads to the following Pl\"ucker relation (which is also called the QQ-relation in the AdS/CFT integrability literature) between the four \(Q\)-functions adjacent to the same two-dimensional face of the Hasse diagram (shown on Fig.\ref{fig:Plucker}):
\begin{figure}[htb]  
\begin{center}
\includegraphics[scale=.3]{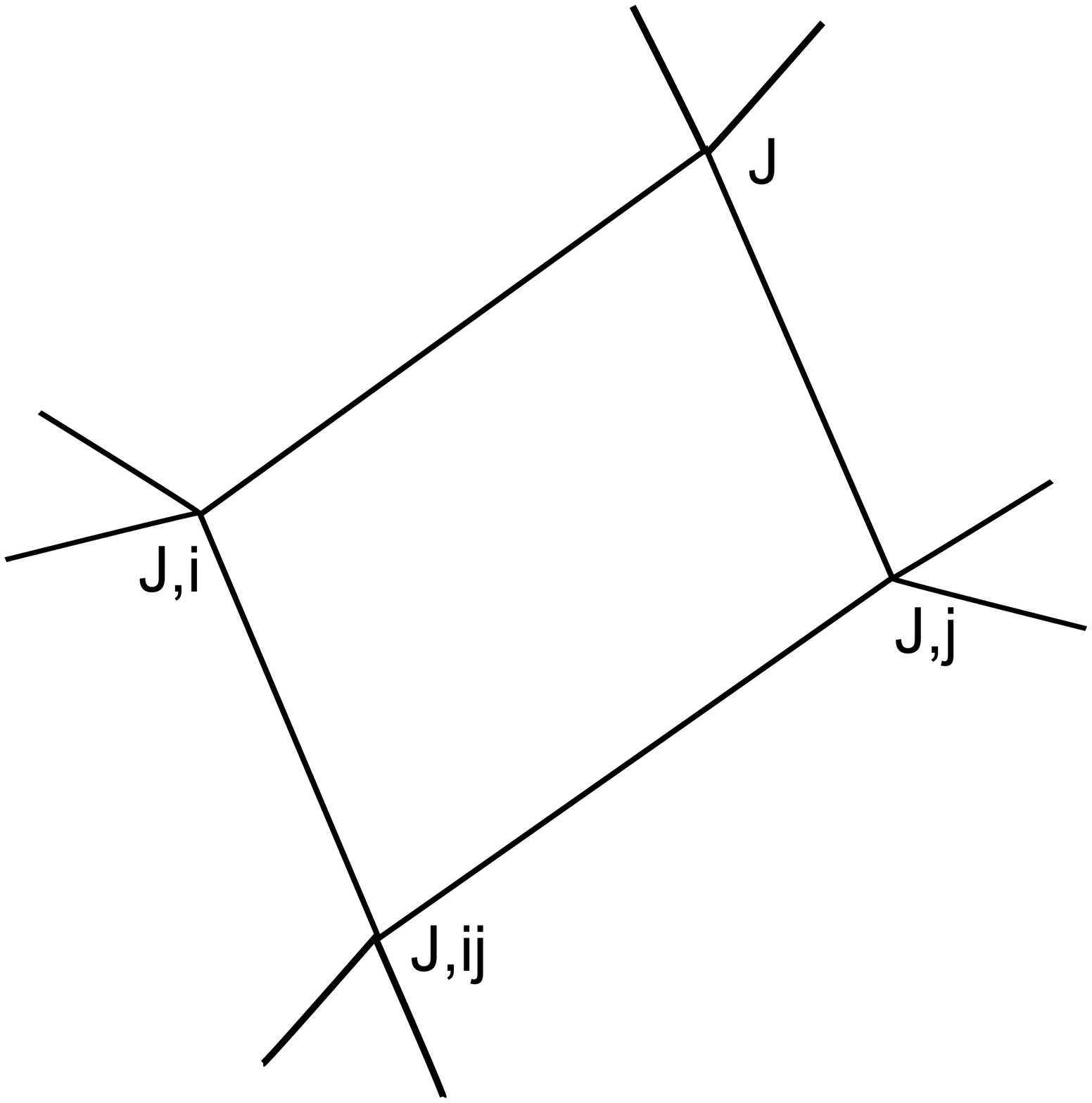}
\end{center}
\caption{QQ-relations (Pl\"ucker identities for the Grassmannian) \eqref{Plucker}. They emerge at any 2-dimensional face of  the hypercube of Hasse diagram  as a consequence of the determinant relations \eqref{QI}.    }
\label{fig:Plucker}
\end{figure}
\begin{align}\label{Plucker}
Q_{J}Q_{J,jk}=Q_{J,j}^+Q_{J,k}^--Q_{J,j}^-Q_{J,k}^{+}
\end{align}
where  \(J=\{j_1,\dots,j_k\}\in  \bar\emptyset\equiv \{1,2,\dots,N\} \) is a particular vertex on Hasse diagram. Notice, that the introduction of arbitrary function \(Q_\emptyset\) in denominator of  \eqref{spectralE}  was necessary for satisfying the \(QQ\)-relations  \eqref{Plucker} on the whole Hasse diagram, including the \(\emptyset\)-vertex. 

To obtain the standard nested Bethe ansatz equations, one has to choose a set of \(Q\)-functions along a ``meridian" of the Hasse diagram, say, \(Q_\emptyset ,Q_1,Q_{12},Q_{123},\dots,Q_{123\dots N}\).  Then one can use the above Pl\"ucker relations along the faces adjacent to this ``meridian" (on the same side of it)  and exclude all other \(Q\)-functions
 at the roots of the ``meridional" \(Q\)-functions by the trick similar to the one which led us to the Bethe equation \eqref{Bethe2}~\cite{Kazakov:2007fy,Leurent:2012xc,Kazakov:2010iu}.

\subsection{Spectrum of supersymmetric \(GL(K|M)\) spin chain via\\ \(Q\)-functions}

The Hamiltonian of the \(GL(K|M)\) super-spin chain has  the following form \begin{align}\label{superHamiltonian}
\hat H=\sum_{i=1}^{L-1}\CP^{K|M}_{i,i+1} +\CP^{K|M}_{L,1}\,\,G_{_L}G_1^{-1}
\end{align}   where the super-spin at each site \(i\) takes two kinds of values   \begin{equation} s_i=1,2,\dots,K|K+1,\dots,N.\end{equation} They correspond to two different gradings:   \(p_s=1,\,\text{if}\, s\le K,\,\,p_s=0,\,\text{if}\, s> K\). The super-permutation acts on a pair of spins as \(\CP_{i,j}(s_1,\dots,s_i,\dots s_j,\dots s_L)=(-1)^{p_{s_i}+p_{s_j}}(s_1,\dots,s_j,\dots ,s_i,\dots s_N)\) and the twist \(G={\rm diag}\{x_1,\dots,x_K|y_1,\cdots,x_M\}\in gl(K|M)\) is a fixed group element~\footnote{All eigenvalues are assumed distinct here.}.

\begin{figure}[htb]  
\begin{center}
\includegraphics[width=\textwidth]{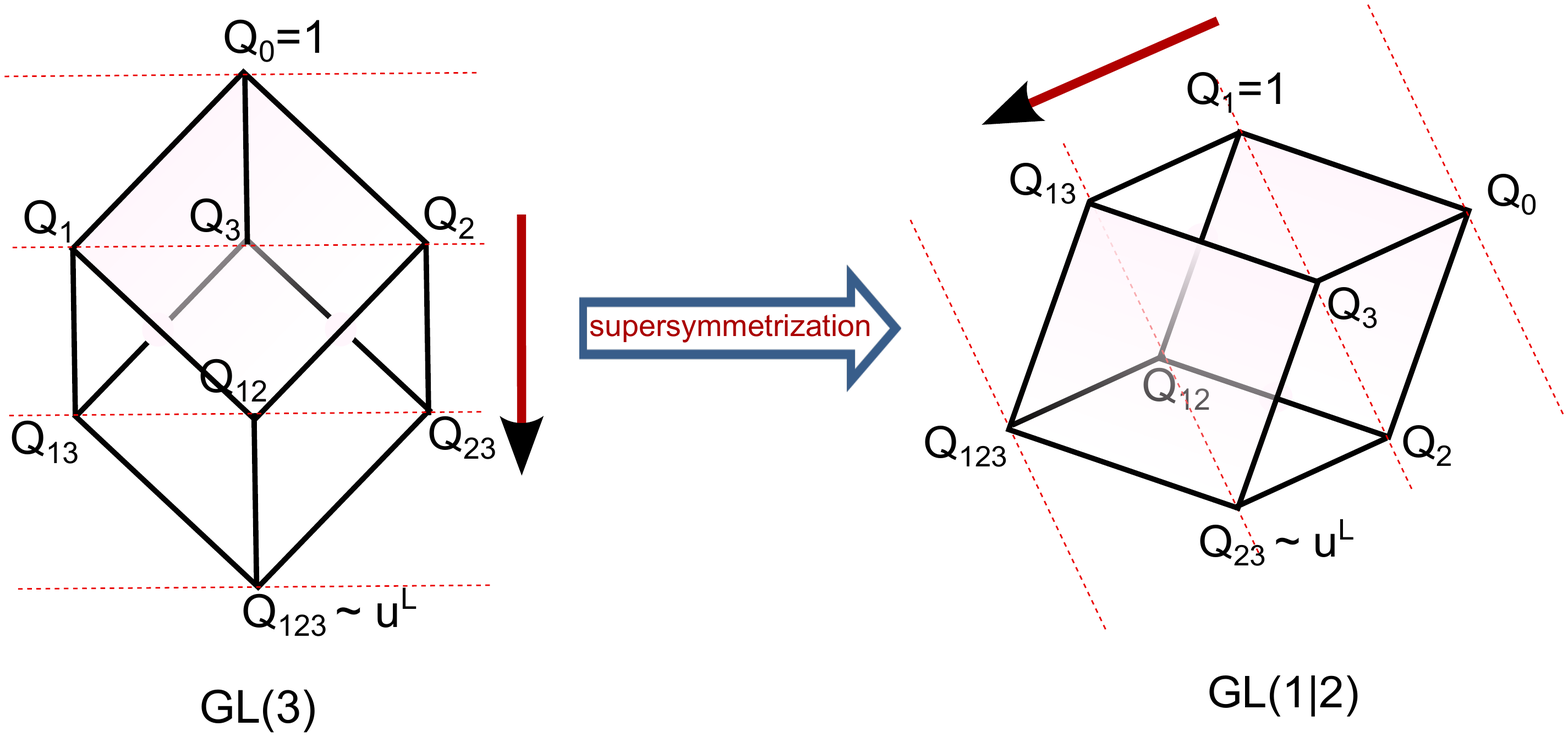}
\end{center}
\vspace*{-24pt}
\caption{Supersymmetrization of Hasse diagram of a QQ system of rank-3 on the  example of rational  Heisenberg spin chains: we pass from  \(GL(3)\) spin chain to \(SU(2|1)\) spin chain by rotating the direction of the determinant flow, i.e. imposing  different  analyticity (polynomiality) conditions (\eqref{QpolynSuper}-\eqref{QKKm} instead of \eqref{Qpolyn}) and fixing   a different pair of diametrally opposed \(Q\)-functions different in each case. The determinant flow is the same in both cases.}
\label{fig:superHasse}
\vspace*{12pt}
\end{figure}

The supersymmetric generalization of the above picture in terms of \(Q\)-functions can be nicely and easily presented as a specific ``rotation" of  Hasse diagram, when imposing the analyticity (``polynomiality") conditions. Namely, for \(GL(K|M)\) case  we can preserve the same determinant flow \eqref{QI} along the Hasse diagram as for \(GL(K+M)\) bosonic case. But to fix the analyticity conditions we choose, instead of \(Q_\emptyset,Q_{\bar\emptyset}\),  a pair of \(Q\)-functions at the extremes of a different main diagonal of the hypercube, one on the level \(K\), another on the level \(M.\)  For example, we can pick \(Q_{12\dots K}\) and \(Q_{K+1,K+2,\dots M}\).\footnote{Other choices of two such functions at the same levels are simple re-labelings. } This supersymmetrization procedure is shown for the example \(GL(3)\to GL(1|2)\) on Fig~\ref{fig:superHasse}.

To find the spectrum of  the Heisenberg super-spin chain we impose the following analyticity (polynomial times exponential for twist) conditions on the \(Q\)-functions:
 \begin{align}\label{QpolynSuper}
&Q_{1,2\dots,\hat k,\dots,K}(u)=x_k^{iu}\prod_{j=1}^{R_k}(u-u_j^{(k)}),\qquad k=1,2,\dots,K;\\
&Q_{1,\dots,K,K+m}(u)=y_{m}^{iu}\prod_{j=1}^{\hat R_m}(u-u_j^{(m)}),\qquad m=1,2,\dots,M.
\label{QKKm}\end{align} where the hat over \(\hat k\) in the l.h.s. of the first equation means that the corresponding index is missing from the sequence.  For the example \(GL(2|1)\) of Fig.~\ref{fig:superHasse} such three \(Q\)-functions in \eqref{QpolynSuper},\eqref{QKKm} are \(Q_\emptyset\) and \(Q_{12}\,,Q_{13}\), respectively.  The positive integers \(\{R_1,R_2,\cdots,R_K|\hat R_1,\hat R_2,\dots,\hat R_M\}\) are  the Cartan super-charges of the \(U(1)^{K+M}         \) residual symmetry left after breaking the original \(SU(K|M)\) symmetry by twisting. To fix all  the roots of these \(Q\)-functions, and thus to find all the eigenvalues of the above hamiltonian of super-spin chain of length \(L\), it is enough to find all solutions of the \(Q\)-system with the following conditions imposed \begin{align}\label{spectralEsuper}
Q_{K+1,K+2,\dots ,K+M}(u)=1,\qquad Q_{12\cdots K}(u)=\text{const}\times u^L\,.
\end{align}

Once one finds a solution of \eqref{spectralEsuper}, the corresponding energy -- an eigenvalue of the hamiltonian \eqref{superHamiltonian} -- is given by the familiar formula, through the \(Q\)-functions neighboring  the ``momentum-carrying" \(Q\)-function \(Q_{12\cdots K}\) on Hasse diagram
\begin{align}   
\label{EnergyN}
E&=L+i\,\partial_u\log\frac{Q_{k,k+1,\dots,K+M}(u+i/2)}{Q_{k,k+1,\dots,K+M}(u-i/2)}\left|_{u=0}\right.\\[6pt]
&=L+i\,\partial_u\log\frac{Q_{K+1,K+2,\dots,K+\hat k,\dots,K+M}(u-i/2)}{Q_{K+1,K+2,\dots,K+\hat k,\dots,K+M}(u+i/2)}\left|_{u=0}\right. \,,
\end{align}
where the answer does not depend on the choice of \(k\) or \(\hat k\).

We see that the scheme of solution for supersymmetric case is almost identical to the previous,  bosonic case. But the ``rotation" of the Hasse diagram in such a way leads to the dramatic change of  analyticity properties. For example, the known function  \(Q_{12\cdots K}(u)\) of eq.\eqref{spectralEsuper} cannot be expressed through the basic functions \eqref{QpolynSuper},\eqref{QKKm} as a simple determinant, as in the bosonic case, but rather has to be found by solving a chain of Pl\"ucker relations \eqref{Plucker}, which leads to more complicated formulas.
 One can also derive the corresponding supersymmetric Bethe ansatz equations \cite{Kulish:1980ii,Kulish:1985bj} directly from the QQ-relations \eqref{Plucker} as it was done in section~5 of \cite{Kazakov:2007fy} in  less invariant notations.
\footnote{For the sake of a unified description of \(Q\)-systems of bosonic and supersymmetric models, we avoid here the use of super-symmetric notations of the original papers~\cite{Gromov:2014caa,Kazakov:2015efa} where we would denote \(Q_{I|J}\equiv Q_{I,\bar\emptyset_M\backslash J+K}\), where \(I\subset \{1,2,\dots,K\},\,\,\, J\subset \bar\emptyset_M \equiv \{1,2,\dots,M\}\) and by \(\bar\emptyset_M\backslash J+K\) we denoted the subset, complementary to the set \(\bar\emptyset_M\), with every index shifted by \(+K\).  We mean that on the r.h.s. of this definition we have the \(Q\)-function in our current notations \eqref{QI}.  }

The \(Q\)-functional approach has a long history
 \cite{Baxter:2007book,Kluemper:1990,Krichever:1996qd,Bazhanov:1996dr,Kazakov:2007fy,Tsuboi:1997iq,Tsuboi:2003ng,Bazhanov:2001xm,Kazakov:2010iu,Bazhanov:2010jq,Frassek:2011aa,Gromov:2007ky,Belitsky:2006cp} and it has been developed in the form described above in the series of papers  \cite{Gromov:2014caa,Kazakov:2015efa,Gromov:2010km,Kazakov:2010iu,Tsuboi:2009ud,Tsuboi:2011iz,Leurent:2012xc}    where the reader can find many more details.
This approach is not only aesthetically attractive, it  also appeared to be more efficient in certain explicit computations compared to more conventional Bethe equations for rational spin chains~\cite{Marboe:2014gma,Marboe:2014sya}.
The  construction in terms of \(Q\)-system, based on Hasse diagram, presented above, can be applied for more complicated quantum integrable systems, such as non-compact (super)-spin chains and 2D sigma models in a finite volume. In this case, the analyticity conditions should be modified, since the \(Q\)-functions, or at least a part of them, cannot  be parameterized by polynomials anymore.
However, the solutions for the spectrum in such problems can be still formulated in terms of certain analyticity conditions on the set of \(Q\)-functions for which the algebraic structure of \(Q\)-system  is entirely defined by the symmetry group.  This approach was successfully applied for example for the study of spectrum of the principal chiral field model on a finite space-circle~\cite{Gromov:2008gj,Kazakov:2010kf}.~~
In the next section, we will use the   Q\(\)-system approach to formulate, in the most concise and general way, the Quantum Spectral Curve (QSC) equations~ \cite{Gromov2014a,Gromov:2014caa} -  a system of non-linear functional equations for computation of    anomalous dimensions of arbitrary local operators, at any coupling, in the planar limit of \({\cal\ N}=4\) Super-Yang-Mills (SYM) theory.

\section{Quantum spectral curve for twisted N=4 SYM }\label{sec:QSCgamma}

In this section we will give a concise formulation of the quantum spectral curve (QSC) for the spectrum of anomalous dimensions of local operators in planar \({\cal\ N}=4\) Super-Yang-Mills (SYM) theory, first introduced in~ \cite{Gromov2014a,Gromov:2014caa}, including its twisted version~\cite{Kazakov:2015efa,Gromov:2015dfa}.
It will be  based on the  \(Q\)-system\footnote{Called also AdS/CFT \(Q\)-system, referring to the duality between \({\cal N}=4 \) SYM and the string sigma model on \(AdS_5\times S^5\) background. For other recent reviews on integrability methods for this system see~\cite{Beisert:2010jr,Gromov:2017blm}. } approach described above. We will first make precise the algebraic structure of this \(AdS_5\)/CFT\( _4\) \(Q\)-system, based on the super-conformal \(PSU(2,2|4)\) symmetry of the model. Then we will describe the analyticity properties of the underlying \(Q\)-functions and the Riemann-Hilbert ``sewing" conditions allowing to completely fix the system of equations for the physical solutions.

Let us stress that we don't give here any derivation of the  \(AdS_5\)/CFT\( _4\)   QSC. We only formulate the final mathematical formalism, ready for further applications. Until the last chapter devoted to a particular  application of QSC to the chiral double limit of \(\gamma\)-twisted \({\cal\ N}=4\) SYM,  we    avoid, on purpose, the discussion of any consequences of QSC equations  and of secondary details, concentrating only on the basic foundations of  QSC  construction.
 For the derivation,  details and numerous consequences, the reader can turn to the original papers \cite{Gromov2014a,Gromov:2014caa,Kazakov:2015efa},  to the recent review~\cite{Gromov:2017blm} as well as to the already rich literature of its generalisations and applications \cite{Bombardelli:2017vhk,Cavaglia:2014exa,Cavaglia:2016ide,Gromov:2015dfa,Gromov:2015vua,Gromov:2015wca,Gromov:2016rrp,Gromov:2017cja,Kazakov:2015efa,Klabbers:2017vtw,Marboe:2014gma,Marboe:2014sya,Marboe:2017dmb}.

\subsection{Algebraic  structure of the AdS\(_5\)/CFT\(_4\) \(Q\)-system}

The Hasse diagram for the AdS/CFT \(Q\)-system is similar to  the one for the \(gl(4|4)\) super-spin chain described in the previous section. It represents an 8-dimensional hypercube with the \(2^8=256\) \(Q\)-functions attached to its vertices, as shown on Fig.\ref{fig:Hasse48}(right). The \(Q\)-functions have the same determinant flow as described by eq.\eqref{QI}.  Let us note that the \(Q\)-system obeys a certain residual \(sl(4)\times sl(4)\)  symmetry corresponding to two bosonic subgroups of the \(PSU(2,2|4)\) symmetry.  This algebraic symmetry refers to the linear transformations of, separately, 4 functions with 3 indices \(Q_{1,\dots,\hat k,\dots,4}(u),\quad k=1,2,3,4\) and 4 functions with 5 indices \(Q_{1,\\\dots,4,j}(u),\quad j=1,2,3,4\).  Another, ``gauge" symmetry of the \(Q\)-system, due to the homogeneity of \(QQ\)-relations,  consists of the rescalings of \(Q\)-functions (there are two such rescaling parameters, see~\cite{Gromov:2014caa}). 

But we should demand for QSC even more: we impose
two  to unit value\begin{align}\label{gauge}
Q_{1234}(u)=1=Q_{5678}(u)\qquad
\end{align} 
 at any spectral parameter \(u\).
The first of these conditions can be achieved by rescalings. But the second one, the   \(Q\)-function diametrally opposed on Hasse diagram (i.e. \(Q_{1234}\) and \(Q_{5678} \) are Hodge dual to each other)\footnote{A \(Q\)-function diametrally opposed to a given \(Q\)-function is usually called its  Hodge dual. Their index sets are complementary to each other w.r.t. the full set. All the algebraic relations of AdS/CFT \(Q\)-system are invariant w.r.t. the Hodge transformation, i.e. w.r.t. exchange of upper and lower  indices, once \eqref{gauge} is imposed.},     
this is an additional condition which replaces  \eqref{spectralEsuper} for the super-spin chain.\footnote{We could have imposed the condition \eqref{gauge} also for the super-spin chain. However, it would have spoiled the polynomial ansatz \eqref{QKKm} and significantly complicate the analyticity properties. The 2nd condition \eqref{gauge} is not a simple normalization, but a dynamical restriction, for the case of \((N|N)\)  \(Q\)-systems~\cite{Tsuboi:2011iz}.} It actually reflects the projectivity and super-unimodularity of the \(PSU(2,2|4)\) symmetry of the system.

This gauge appears to be the most suitable for the formulation of  analyticity properties of the whole \(Q\)-system.
But these properties are more complicated then the polynomial ansatz \eqref{QpolynSuper} since we deal not with the rational super-spin chain (which however occurs to be the case in the weak-coupling, one-loop limit of \({\cal\ N}=4\) SYM~\cite{Minahan:2002ve,Beisert:2003yb,Beisert:2005di}) but with the integrable string sigma-model on \(AdS_5\times S^5\) coset.

\quad\  We introduce special notations for the most useful ``near-equator" \(Q\)-functions mentioned above: \begin{align}\label{Pnotations}
&\bP_a\equiv Q_{1,\dots,\hat a,\dots,4}(u),\quad a=1,2,3,4;\\
&\bQ_{j}\equiv Q_{1,2,3,4,j+4}(u),\qquad j=1,2,3,4.
\label{Qnotations}\end{align} where by \(\hat a\) we denote again the missing index from the set \(\{1,2,3,4\}\). For example, \(\bP_2=Q_{134}\) and \(\bQ_3=Q_{12347}\).  Hence \(\bP_a\) have \(3\)  indices and \(\bQ_j\) has \(5\) indices in the standard notations for \(Q\) functions, as in \eqref{QI}.

   Their Hodge dual \(Q\)-functions have the same, but upper indices: \begin{align}\label{PdualNotations}
&\bP^a\equiv Q_{a,5,6,7,8}(u),\quad a=1,2,3,4;\\
&\bQ^j\equiv Q_{5,\dots,\widehat{j+4},\dots,8}(u),\qquad j=1,2,3,4.
\label{Qdualnotations}\end{align} For example, \(\bP^2=Q_{25678}\) and \(\bQ^3=Q_{568}\).
The positions of these functions on \((4|4)\) Hasse diagram are pictorially presented  on Fig~\ref{fig:singleIndexPQ}:
\begin{figure}[htb]  
\begin{center}
\includegraphics[width=\textwidth]{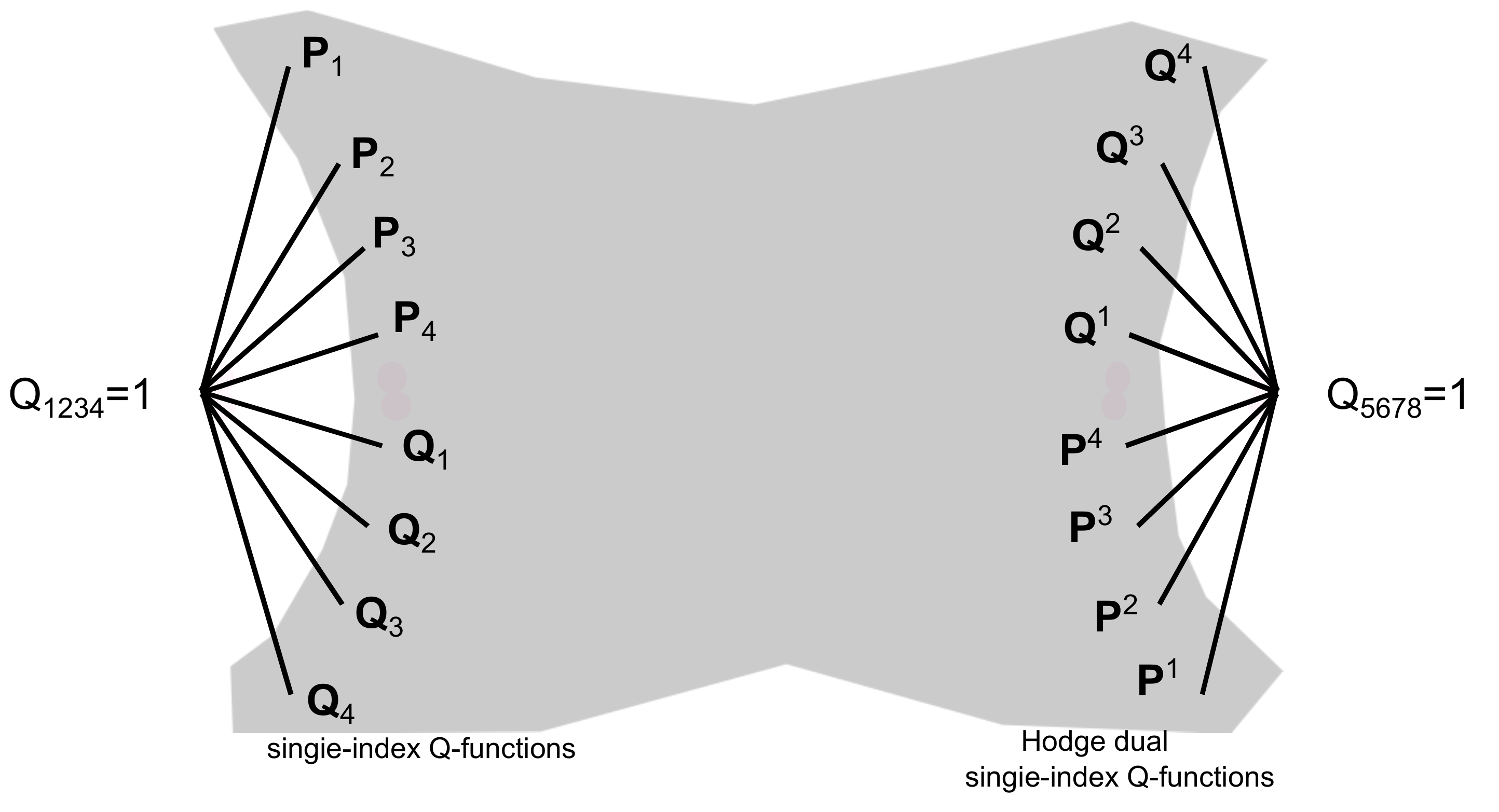}
\end{center}
\caption{Schematic presentation of positions of \(\bP\) functions and \(\bQ\) functions \eqref{Pnotations},\eqref{Qnotations},\eqref{PdualNotations},\eqref{Qdualnotations},  within the \((4|4)\) Hasse diagram. They are  neighboring the two ``poles" of Hasse diagram corresponding to empty-set and full-set labels. Each pair of functions  \(\bP_a,\bP^a\) or \(\bQ_j,\bQ^j\),  with the same label, are placed at the diametrally opposite vertices of Hasse diagram, i.e. they are Hodge dual to each other w.r.t. the Grassmannian structure of the \(Q\)-system. These 16 functions have the simplest analytic structure on the physical sheet.  The gray ``cloud" signifies all the elements of 8D hypercube missing on the picture.   }
\label{fig:singleIndexPQ}
\end{figure}

  The \(\bP\) and \(\bQ\) functions are, roughly, responsible for the dynamics of string fields on \(S^5\) (related to R-symmetry)  and   on \(AdS_5\) (related to the conformal symmetry) projections of the dual string sigma-model,   as will be seen from their large \(u\) asymptotics.

Another useful set   of 16  \(Q\)-functions and of their 16 Hodge duals deserves a special notation:  \begin{align}\label{Qajnotations}
&\bQ_{a|j}\equiv Q_{1,\dots,\hat a,\dots,4,j+4}(u),\quad a=1,2,3,4\,;\quad j=1,2,3,4,\\
&\bQ^{a|j}\equiv Q_{a,5,\dots,\widehat{ j+4},\dots,8}(u),\qquad a=1,2,3,4\,;\quad j=1,2,3,4.
\end{align}  where by ``hat" we denote again the missing member from consecutive integers. For example, \(\bQ_{2|3}=Q_{1347}\) and \(\bQ^{2|3}=Q_{2568}\).

Due to the determinant flow \eqref{QI}, together with the gauge conditions \eqref{gauge}, these functions satisfy a useful set of algebraic identities: \begin{align}
&\bQ_j=\bQ_{a|j}^{\pm}\bP^a,\qquad \bP_a=\bQ_{a|j}^{\pm}\bQ^j  &\text{(``metric" property)}, \label{PQid}\\
&\bQ_{a|j}^+-\bQ_{a|j}^-=\bP_a\bQ_j    &\text{(special QQ-relation)}.
\label{QQid}\end{align}  Notice that we can raise and lower the indices by the rules similar to the standard tensor algebra.
A useful automatic consequence of the Grassmannian structure of \(Q\)-system and of the gauge \eqref{gauge} is the orthogonality relations: \(\bP_a\bP^a=\bQ_j\bQ^j=0\).

The QSC formalism is based on a set of 256 \(Q\)-functions, out of which only a few are algebraically independent. The rest of them can be deduced from the determinant flow or from the Pl\"ucker QQ relations \eqref{Plucker}. The choice of the most convenient  algebraically independent subset of \(Q\)-functions depends on the problem being solved, i.e. on the type of studied operators and on the chosen approximations (weak coupling, strong coupling, numerics, etc).
Thus there exist many useful forms of QSC equations.   Let us mention one particularly important, especially for various weak coupling limits - the coupled system of 4th order Baxter equations on the functions \eqref{Pnotations},\eqref{Qnotations},\eqref{PdualNotations},\eqref{Qdualnotations}.  Namely, excluding the functions \(\bQ_{a|j}\)  from the  relations \eqref{PQid}-\eqref{QQid}\footnote{One uses for that a linear system of 5 dependent equations on the same \(\bQ_{a|j}\) obtained from \eqref{PQid} by 5 consecutive shifts of \(u\) by integers\(\times i\)  and a multiple use of \eqref{QQid}  for bringing various functions \(\bQ_{a|j}\) to the same argument \(u\) (see \cite{Alfimov:2014bwa} for details).}  one gets the following linear  4th order finite difference Baxter  equation\cite{Alfimov:2014bwa}
\begin{eqnarray}
a\,\,\bQ^{[+4]}-b\,\,\bQ^{[+2]}
+
c\,\,\bQ
^{[0]}-\bar b\,\,\bQ^{[-2]}+\bar a\,\,\bQ^{[-4]}=0
\end{eqnarray}
where the coefficients are explicit  functionals of \(\bP_a,\bP^a\)-functions:
\begin{eqnarray}  
\label{BaxterPQ}
&&a(u)=d_0,\quad\!
b(u)=d_1-\bP_a^{[+2]}\bP^{a[+4]}d_0,\quad\! c(u)=d_3+\bP_a\bP^{a[+4]}d_0+\bP_a\bP^{a[+2]}d_1\notag\\
&&\kern-5pt
\text{where}\quad\!\! d_m=\underset{1\le a,k\le 4}{\det}(\bP^a)^{[4-2k+2\theta_{k,m}]},\quad\!\!
(\theta_{m,k}=1\text{~if}\, m\geq k~\text{and}~\theta_{m,k}=0\text{~if}~m<k).\nonumber\\
\label{coeffBaxter}\end{eqnarray}

Four solutions of this equation give the functions \(\bQ_j\). Of course, any independent linear combinations of these 4 \(\bQ_j\)-functions with \(i\)-periodic coefficients \footnote{i.e. the coefficients are functions of \(u\) periodic w.r.t. the shift \(u\to u+i\).} are also algebraically admissible \(\bQ_j\)-functions.

For a general state/operator, this Baxter equation should be supplemented by three similar equations. One of them, for \(\bQ^ j\)-functions, uses  the Hodge duality  of the \(Q\)-system,  obtained from the above equation  by exchange of all upper\(\Leftrightarrow \)lower indices of the \(\bP_a\)-functions in the coefficients.  Two other 4th order Baxter equations, on \(\bP_a\)- and \(\bP^a\)-functions, can be obtained from the previous two by simply exchanging all \(\bP\)- and \(\bQ\)-functions. The existence of the last two equations is a simple consequence of the algebraic symmetry within the full \(Q\)-system between \(\bP\)- and \(\bQ\)-functions.

Let us note that the most frequent cases of the \({\cal N}=4\) SYM operators studied in the literature are those which obey the so called left-right (LR) symmetry w.r.t. to the exchange of two  subgroups of the full superconformal group: \(SU(2|2)_L\times SU(2|2)_R\in PSU(2,2|4)\). This symmetry has direct algebraic consequence for the underlying AdS/CFT \(Q\)-system. Namely, due this symmetry  we can  raise and lower the indices of \(\bP\)-functions and \(\bQ\)-functions, i.e. \(a\)-type or \(j\)-type, by means of a ``metric" whose role is played by a fixed constant matrix~\cite{Gromov:2011cx,Gromov:2014caa}: \begin{equation}
\bP_a=\chi^{ab}\bP^b,\quad\ \bQ_j=\chi^{jk}\bQ^k,\quad \text{where }\chi=\begin{pmatrix}0 & 0 & 0 & 1 \\
0 & 0 & -1 & 0 \\
0 & 1 & 0 & 0 \\
-1 & 0 & 0 & 0 \\
\end{pmatrix}.
\label{LRsymmetry}\end{equation}  Obviously, in this case only two 4th order Baxter equations are algebraically independent: one for \(\bQ_j\) and one for \(\bP_a\), which significantly simplifies the problem. In addition,  in various weak coupling limits, such as one-loop\cite{Beisert:2003jj,Beisert:2005di} or BFKL\cite{Alfimov:2014bwa} approximations, or the double scaling (DS) limit of \(\gamma\)-deformed \({\cal\ N}=4\) SYM\cite{Gurdogan:2015csr} described in the next section, the analytic properties of \(\bP\)-functions simplify even further: they can have only  finite order poles at \(u=0\) and \(u=\infty\) and thus they can be parameterized by a finite number of coefficients in the corresponding polynomials. Then the Baxter equation \eqref{BaxterPQ} on \(\bQ\)-functions starts to remind the one for the integrable \(SU(2,2)\) spin chain reflecting the 4D conformal symmetry of the problem.

\subsection{Analyticity:  quantum spectral curve as a Riemann-Hilbert problem}

The QSC formalism is based on two fundamental ingredients: the first is the algebraic structure of the underlying \(Q\)-system, entirely based on the superconformal symmetry \(PSU(2,2|4)\) of the model, and the second is the analyticity properties of the underlying  \(Q\)-functions. The analyticity is greatly, but not completely dictated by the algebraic structure of \(Q\)-system.  It was established in the original papers~\cite{Gromov2014a,Gromov:2014caa}. It was extracted from the exact solution of the AdS/CFT spectral problem, first proposed in the form of the AdS/CFT Y-system~\cite{Gromov:2009tv} and then via the TBA approach~\cite{Gromov:2009bc,Bombardelli:2009ns,Arutyunov:2009ur}. The papers \cite{Cavaglia:2010nm,Gromov:2011cx} have been  important steps towards the  discovery  of the QSC formalism.

In the rest of this section, we will describe the analytic properties of \(Q\)-functions.  The main ingredients of their analytic structure are i)~Infinitely branching Riemann surface due to branch cuts at fixed positions -- ``Zhukovsky cuts"~\footnote{The name originating from the Zhukovsky conformal map, inverse of \(u=g(x+1/x)\) function.};  ii)~Asymptotics at large values of spectral parameter fixing the \(SU(2,2|4)\) representation of state/operator; iii) Riemann-Hilbert ``sewing" conditions relating various \(Q\)-functions via monodromies around Zhukovsky cuts; iv) Absence of any other singularities anywhere on the Riemann surface of any \(Q\)-function, except mentioned above. Let us inspect these properties in detail.
\begin{figure}
\centerline{\includegraphics[scale=0.4]{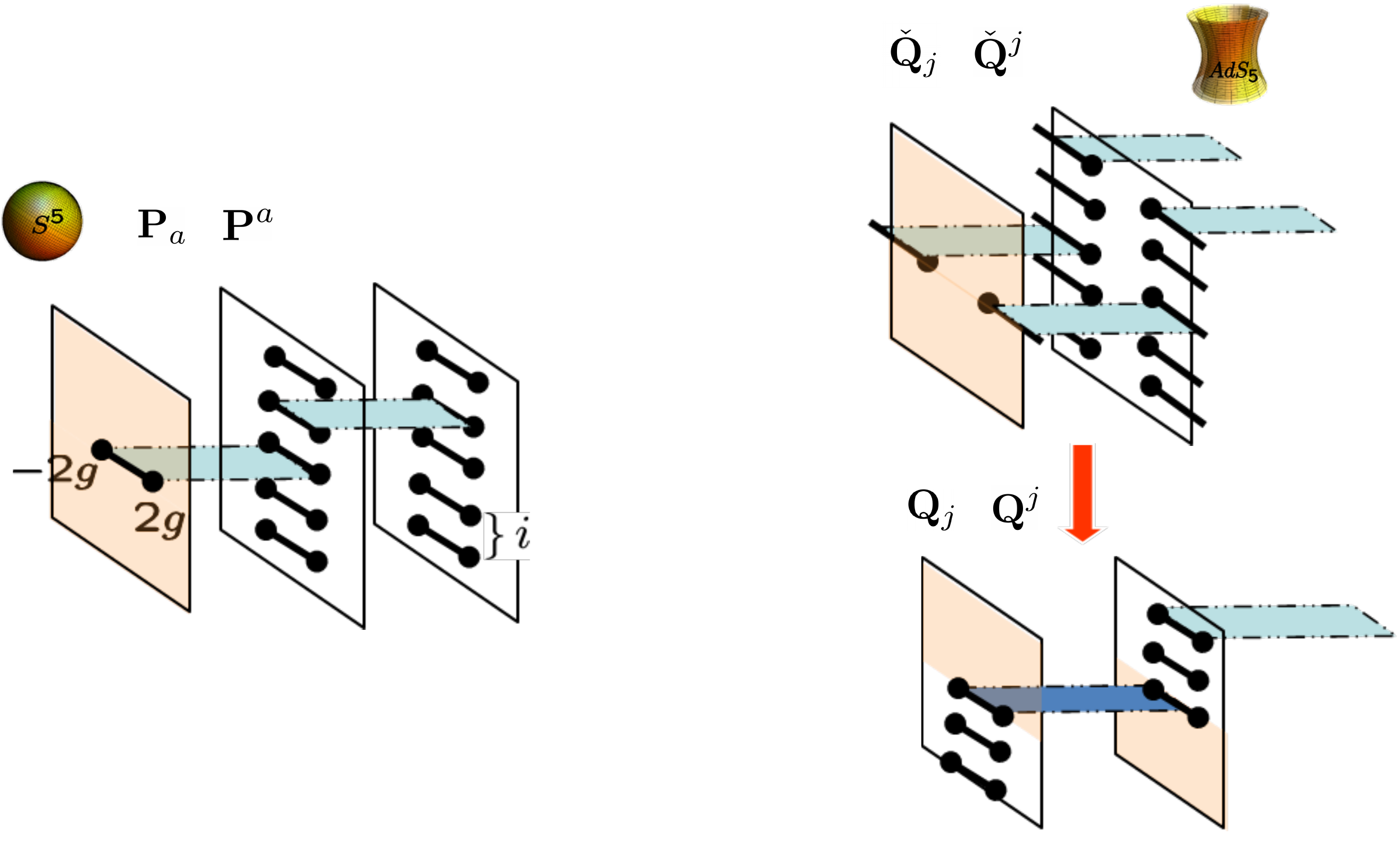}}
\caption{Schematic depiction of analytic structure of Riemann surfaces of  \(\bP\) and \(\bQ\)  functions defined by \eqref{Pnotations},\eqref{Qnotations},\eqref{PdualNotations},\eqref{Qdualnotations}.   On the left, \(\bP_a\) and \(\bP^a\)  have, each, a special, physical sheet of the Riemann surface  where it has only one Zhukovsky cut for the range of spectral parameters \(u\in (-2g,2g)\), where \(g^2=\frac{1}{16\pi^2}N_c g_{_{YM}}^2\) is the 't~Hooft coupling. This cut is connected to the next sheet which has a ladder of equidistant Zhukovsky cuts spaced by \(i\), at positions \(u\in (-2g+i\mathbb{Z}\,\,,2g+i\mathbb{Z})\), along the imaginary axis. On the right(up), the same picture of Riemann surface is true for the \(\check\bQ_j\) and \(\check\bQ^j\) functions, with an important difference: short Zhukovsky cuts should be replaced by long Zhukovsky cut, passing through \(u=\infty\), i.e. for \(u\in (-\infty,-2g)\,\cup\,(2g,\infty)\). On the next sheets we have an infinite ladder of such cuts spaced by \(i\).    On the right(down) we show the rearrangement of Riemann surface, by re-gluing the upper-half plane of the physical sheet with the lower-half-plane of the next sheet, and vice versa for the other two halves.  This flips the long Zhukovsky cut on the real axis to short cut, but also creates a sequence of cuts in the lower half plane (which can be made short by the same re-gluing procedure for the next sheets).
 }
\label{fig:cutsPQ}
\end{figure}

\subsubsection{Zhukovsky branch cuts and Riemann surface for \(Q\)-functions}

The main analyticity observation in QSC formalism is about the particular subset of 16 \(Q\)-functions, precisely the ones listed in \eqref{Pnotations},\eqref{Qnotations},\eqref{PdualNotations},\eqref{Qdualnotations} and shown on Fig.\ref{fig:singleIndexPQ}.  Namely, the 8 functions \(\bP_a\) and \(\bP^a\)  have,
each, a special sheet of the Riemann surface (which will be called physical) where it has only one Zhukovsky cut for the range of spectral parameters \(u\in (-2g,2g)\) and \(g^2=\frac{1}{16\pi^2}N_c g_{_{YM}}^2\) is the 't~Hooft coupling\footnote{\(g_{_{YM}}\) is the original Yang-Mills coupling and \(N_c\) is the number of colors of the  \(U(N_c)\) gauge group. From now on, for brevity, we will rather  call \(g\) the 't~Hooft coupling.}. The physical sheet is depicted on the left of  Fig~\ref{fig:cutsPQ}(left).   Similarly, the other 8 \(Q\)-functions, \(\check\bQ_j\) and \(\check\bQ^j\), have a special, physical sheet where they have only one cut with the same branch-points but passing through \(u=\infty\), i.e. for \(u\in (-\infty,-2g)\,\cup\,(2g,\infty)\)~.  The physical sheet is depicted on  Fig~\ref{fig:cutsPQ}(upper-right). It is natural call the first type of cuts as ``short cuts" and the second one as ``long cuts".  The positions of the branchpoints of these cuts are actually the only place in the QSC formalism where the 't~Hooft-Yang-Mills coupling constant is encoded.\footnote{From now on, we will distinguish, by ``check" or its absence, two functions \(\check\bQ_j\) and \(\bQ_j\). In fact, it is the same function but the Riemann sheets are organized differently in two cases:  \(\check\bQ_j\) has only long cuts and the notation \(\bQ_j\) will be reserved for the same function with Riemann sheets re-glued in such a way that all its cuts are short. See the text below and the Fig.\ref{fig:cutsPQ} for more explanations.}

{\it There are no other singularities on the physical sheets of these \(Q\)-functions except the one   at \(u=\infty\)}, described in the next subsection.

Next, we want to know what happens under the cut, on the next sheet of the Riemann surface. In fact, the structure of the \(Q\)-system, and in particular of the  QQ-relations \eqref{Plucker}, dictates that for \(\bP_a\) and \(\bP^a\) functions, apart from the same short cut \(u\in (-2g,2g)\), we find on the second sheet an infinite ``ladder" of its periodically\footnote{We call this periodicity the \(i\)-periodicity since the cuts  are spaced by the distance \(i\) in complex plane. }  repeating replicas at \(u\in (-2g+i\mathbb{Z}\,\,,2g+i\mathbb{Z})    \), as shown on the right of the Fig~\ref{fig:cutsPQ}(left). If we pass through any of these cuts we will encounter another sheet, with the same infinite ladder  of short cuts repeating periodically along the whole imaginary axis. Passing through any of these cuts we discover the other sheets with the same infinite ladder of cuts. Consequently, each \(\bP\)-function lives on an infinitely branching Riemann surface of the topology of sphere with a puncture at \(u=\infty\).\footnote{We conjecture the spherical topology of the Riemann surface since we see no obvious reasons for the existence of any non-contractible closed paths on this surface (except those encircling  \(u=\infty\)). }

As for the functions
 \(\check\bQ_j\) and \(\check\bQ^j\), the picture of cuts on the sheets next to the physical one is exactly the same as for the functions \(\bP_a\) and \(\bP^a\), except that all cuts are long, i.e. they are \(i\)-periodic, at positions \(u\in (-\infty,-2g+i\mathbb{Z})\,\cup\,(2g+i\mathbb{Z},\infty)\), as shown on the right of the Fig~\ref{fig:cutsPQ}(upper-right). Of course the fact that these cuts  are accumulated at  \(u=\infty\) leads to an infinite branching at infinity and allows for asymptotics with arbitrary power law w.r.t. spectral parameter. As we will see below, this is the way how the parameter \(\gamma(g)\) -- the anomalous dimension of the operator -- arises in the QSC formalism as a power in the  large \(u\) asymptotics of \(\bQ\)-functions.

\subsubsection{Large \(u\) asymptotics}

Now we  describe the behavior of \(\bP\)- and \(\check\bQ\)-functions at \(u\to\infty\) on the physical sheet. To avoid complications with degeneracy of solutions we first consider the case of the totaly deformed superconformal symmetry
of the model: \(PSU(2,2|4)\to [U(1)^2\times\mathbb{R}]_{conf.}\times[U(1)^3]_R\). This is done by introduction of special twist on the \(AdS{_5}\times\)CFT\(_4\) duality~ parameterized by a fixed Cartan group element: \(\text{diag}\{x_1,x_2,x_3,x_4|y_1,y_2,y_3,y_4\}\in PSU(2,2|4) \), with the group constraint \(\prod_{a=1}^4x_a=1=\prod_{j=1}^4y_j\).
This deformation is easy to perform directly in the  \({\cal N}=4\) SYM
 action~\cite{Leigh1995,Lunin:2005jy,Beisert:2005if,Fokken:2013aea} for the case of so called \(\gamma\)-twist, when the conformal part of the superconformal symmetry is not twisted \(y_1=y_2=y_3=y_4=1\) and the \(x_a\) are  parameters  of  \(SU(4)\) \(R\)-symmetry deformation.\footnote{Twisting the conformal part of the  supergroup leads to a non-commutative generalization of \({\cal N}=4\) SYM theory~\cite{Beisert:2005if}.}

The picture  for \(\bP\)-functions is very simple: since the only singularity at the finite part of the \(u\)-plane is a short Zhukovsky cut, we can approach the \(u=\infty\) by any path, and the asymptotics is completely fixed by the global R-symmetry charges \(\{J_1,J_2,J_3\}\in so(6)\sim su(4)\) and the value of the twist: \begin{align}\label{assfulltwistP}
&\bP_b\sim  x_b^{i\,u} \,u^{-\lambda_b}\left(1+\frac{p^{(b)}_1}{u}+\frac{p^{(b)}_2}{u^2}+\dots\right),\qquad \text{where}\quad {\{x_1,x_2,x_3,x_4\}\in SU(4),} \\
&\lambda_b=\frac{1}{2}\{+J_1+J_2-J_3,\quad +J_1-J_2+J_3,\quad -J_1+J_2+J_3,\quad -J_1-J_2-J_3\}.
\label{lambdaass}\end{align}  So we see that  these asymptotics can have only integer or half-integer powers\footnote{In fact, in  more ``physical" quantities, such as transfer-matrix eigenvalues, the \(\bP\)-functions enter bi-linearly, so that the asymptotics becomes single-valued at \(u=\infty\). So in some sense the the quadratic branch-point  of \(\bP\) at  \(u=\infty\)  is not an important issue for the analyticity properties of the \(Q\)-system, see~\cite{Gromov:2014caa} for the details. }.

The situation with large \(u\) asymptotics on the physical sheet of  \(\bQ\)-functions is slightly more involved: due to the presence of the long Zhukovsky cut we should speak in principle separately of the large \(u\) asymptotics in the upper-half plane (UHP) and in the lower-half plane (LHP). However one can easily argue that those two asymptotics can be different only by an overall constant (see \cite{Gromov:2014caa} for the calculation of this constant).  We can thus impose that, for example in the UHP, far away from the real axis (to avoid the vicinity of the long cut) their  exponential and power-like  parts are defined, respectively, by \(SU(2,2)\) twists and  the Cartan charges of conformal group \(\{\Delta,S_1,S_2\}\in so(4,2)\sim su(2,2)\), as follows \begin{align}\label{assfulltwistQ}
&\check\bQ_j\sim  y_j^{-i\,u} \,u^{-\nu_j}\left(1+\frac{q^{(j)}_1}{u}+\frac{q^{(j)}_2}{u^2}+\dots\right),\qquad \text{where}\quad {\{y_1,y_2,y_3,y_4\}\in SU(2,2),} \\
&\nu_j=\frac{1}{2}\{+\Delta-S_1-S_2,\quad +\Delta+S_1+S_2,\quad -\Delta-S_1+S_2,\quad -\Delta+S_1-S_2\}  .
\end{align}   Here \(S_1,S_2\) are integer conformal spins and  \(\Delta(g)\equiv \Delta(0)+\gamma(g)\) is the dimension of the studied operator (energy of the state on the string  side of duality) which is the main quantity under study in QSC formalism. Generically,  \(\Delta(g)\) is a complicated function of the 't~Hooft coupling \(g\), of conserved charges \(J_1,J_2,J_3|S_1,S_2\) and of the twist parameters \(\{x_1,x_2,x_3,x_4|y_1,y_2,y_3,y_4\}\). With all these parameters fixed, we should have a finite or infinite  discrete  set of operators/states with different  anomalous dimensions fixed by the values of the other conserved charges present in this integrable model. The presence of an arbitrary (if we vary \(g\)) power \(\pm\Delta/2\) in the asymptotics means the presence of, in general, infinite branching at \(u=\infty\). This is a natural consequence of the presence of a long cut passing through \(u=\infty\) point. Notice that on the next sheets of \(\check\bQ\)-functions it is hardly possible to speak about such power-like\(\times\)exponential asymptotics due to the accumulation of long cuts forming an infinite ladder. On the contrary, one can define this kind of asymptotics at large \(u\) for the \(\bP\)-functions if we avoid approaching \(u=\infty\) along the imaginary axis, in the vicinity of infinite ladder of short cuts.

Notice that we did not impose separately the asymptotics of Hodge dual \(\bP^a\) and \(\bQ^j\)-functions since those are not independent of  \(\bP_a\) and \(\bQ_j\). They are completely constrained by the structure of the \(Q\)-system (with an important role of the gauge condition \eqref{gauge}) and the leading asymptotics are  inverse powers w.r.t. the original \(\bP_a\) and \(\bQ_j\), namely,
\begin{align}
\label{assPcheckQ}
\bP^b\sim x_b^{-i\,u} \,u^{\lambda_b}\!\left(1+\frac{p_1^{''(b)}}{u}
+\frac{p^{''(b)}_2}{u^2}+\dots\right)\!,\quad\!
\check\bQ^j\sim  y_j^{i\,u} \,u^{\nu_j}\!\left(1+\frac{q^{''(j)}_1}{u}+\frac{q^{''(j)}_2}{u^2}
+\dots\right)
\end{align}

It is important to notice that all these asymptotics are multiplied by the expansion in integer powers w.r.t. \(1/u\). This is a special choice of the \(Q\)-functions, since any linear combination of them would spoil this property and mix up  different combinations of twists and charges. We call our choice ``pure" asymptotics, and this choice will be important for the rest of analyticity properties given below in the form of Riemann-Hilbert conditions.

We can partially   remove the deformations by making some of the twist parameters
\(\{x_1,x_2,x_3,x_4|y_1,y_2,y_3,y_4\}\) equal to each other, i.e. restoring some subgroups of \(PSU(2,2|4\)) symmetry. Then we have to modify the asymptotics  \eqref{assfulltwistP},\eqref{assfulltwistQ} by shifting the exponent by certain integers because, asymptotically,  certain determinant formulas for \(Q\)-functions will become ambiguous and will not render the right asymptotics. The whole classification of various twist configurations and of the corresponding asymptotics is given in \cite{Kazakov:2015efa}.  We will discuss a couple of the most important cases.  One of them, used in the next section, is the so called \(\gamma\)-deformation which preserves the entire conformal subgroup \(SU(2,2)\), i.e. \(y_1=y_2=y_3=y_4=1\), and leaves arbitrary  twists \(\{x_1,x_2,x_3,x_4\},\quad \prod_j x_j=1\), thus breaking \(R\)-symmetry \(SU(4)\to U(1)^3\).   Then the asymptotics \eqref{assfulltwistQ},  should be modified as follows \begin{align}\label{gammatwist}
\check\bQ_j\sim \,u^{-\nu'_j}\left(1+\frac{q^{(j)}_1}{u}+\frac{q^{(j)}_2}{u^2}+\dots\right)
\end{align}
where \begin{align}
\nu_j'=\frac{1}{2}\{+\Delta-S_1-S_2, +\Delta+S_1+S_2-2, -\Delta-S_1+S_2-4, -\Delta+S_1-S_2-6\}  .
\label{nuprime}\end{align} where as the leading asymptotics of  \(\bP_b\) remain  as given by \eqref{assfulltwistP},\eqref{lambdaass}.

Finally, the most studied case is of course the fully untwisted, completely \(PSU(2,2|4)\) symmetric SYM theory (or the equivalent dual  superstring sigma-model on \(AdS_5\times S^5\) background). In this case, the above asymptotics of \(\check\bQ\) are the same as in \eqref{assfulltwistQ}, but for \(\bP\)they  look now as follows\cite{Gromov2014a,Gromov:2014caa} \begin{align}\label{untwisted}
\bP_b\sim \,u^{-\lambda_b'}\left(1+\frac{p^{(b)}_1}{u}+\frac{p^{(b)}_2}{u^2}+\dots\right),
\end{align} with \begin{align}
&\lambda_b'=\frac{1}{2}\{+J_1+J_2-J_3,+J_1-J_2+J_3-2,-J_1+J_2+J_3-4, -J_1-J_2-J_3-6\} .
\end{align}

Using these asymptotics and the Grassmannian structure of the QQ system we can even compute  a few  leading coefficients of all these asymptotics, which appear to depend only on the global charges, not on particular solutions~\cite{Gromov2014a,Gromov:2014caa}. The classification of the coefficients of the leading asymptotics can be found  in \cite{Kazakov:2015efa}. We don't give here   explicit formulas since we limit ourselves only to the formulation of basic rules of QSC construction, leaving aside its consequences.

\subsubsection{Riemann-Hilbert sewing conditions}

Finally, we have to describe how one can move among the sheets of the Riemann surface for the \(\bP\) and \(\check\bQ\)-functions. In other words, one should detail the properties of monodromy around the branch-points of Zhukovsky cuts of these functions.

\begin{figure}[htb]
\begin{center}
\includegraphics[scale=.4]{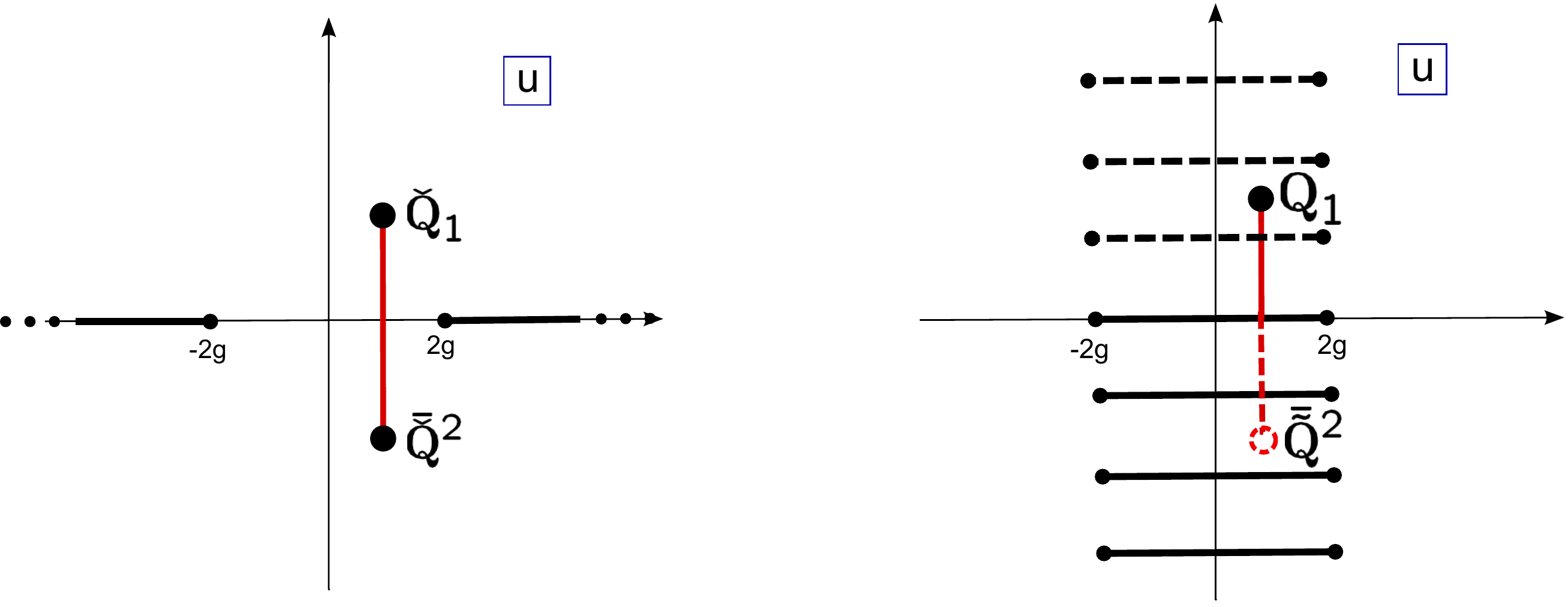}
\end{center}
\caption{Demonstration of the Riemann-Hilbert sewing relations: on the left, the complex conjugation relation \eqref{RH} between a pair of functions \({\check\bQ}_1=\bar\beta_1\bar{\check \bQ}^2\) on the physical sheet with a single long cut is presented. Notice that the path connecting them should go between the branch-points. On the right, the same relation is demonstrated on the sheet with short cuts. It takes the form  \({\tilde\bQ}_1=\bar\beta_1\bar{\bQ}^2\). Dotted cuts,  are situated on the second sheet. The conjugation path  is now passing through the short cut at the real axis, i.e. the conjugation involves now the monodromy (denoted by tilde) as well.      }
\label{fig:sew}
\end{figure}

It was noticed in\cite{Gromov:2015wca}  that, after having imposed the ``purity" of asymptotics, as discussed after  eq.\eqref{assPcheckQ},  one can fix completely the system of spectral equations, by demanding within the QSC formalism the following Riemann-Hilbert sewing conditions~\cite{Gromov:2014caa}\footnote{For the origin and explanations of these sawing conditions see the section 4.4.2 in \cite{Gromov:2014caa}, and in particular eq.(4.63). In \cite{Gromov:2015wca} these relations are called ``gluing conditions", but we call them here ``sewing conditions" which seems to be a more frequent terminology for Riemann-Hilbert problems. }
\begin{equation}\label{RH}
 \begin{pmatrix}
\bar{\check\bQ}_{ 1} \\
\bar{\check\bQ}_{2}\\
\bar{\check\bQ}_{ 3} \\
\bar{\check\bQ}_{ 4}\\
\end{pmatrix} \simeq
\begin{pmatrix}0 & \beta_1 & 0 & 0 \\
\bar\beta_1 & 0 & 0 & 0 \\
0 & 0 & 0 & \beta_2 \\
0 & 0 & \bar\beta_2& 0 \\
\end{pmatrix} \begin{pmatrix}
{\check\bQ}^{1} \\
{\check\bQ}^{2}\\
{\check\bQ}^{ 3} \\
{\check\bQ}^{ 4}\\
\end{pmatrix},
\end{equation}
where  \(\bar{\check\bQ}_j\) means the complex conjugation, i.e. reflection of the main sheet w.r.t. the real axis where the long cut is present, see Fig.\ref{fig:sew}(left),
 and \(\beta_1,\beta_2    \) are  constants non-trivially depending on the parameters of the operator/state, to be defined self-consistently in the process of solution of QSC equations. The origins of this sewing condition originate already from the properties of quasi-momenta of classical  finite gap solution of the string dual -- the sigma model on \(AdS_5\times S_5\) coset \cite{Gromov:2010kf,Kazakov:2004qf,Beisert:2005bm}.\footnote{See section 2.5 and  the equations (2.59) and (3.3) in \cite{Beisert:2005bm}.  The parameter \(x\) is related to the spectral parameter used here by Zhukovsky map  \(u=x+1/x\) (the 't~Hooft coupling \(g\) plays the role of the ``Planck constant" for the string sigma model and can be scaled out in classical limit). }

These conditions mean that the  \(\check\bQ\)-functions are not all independent but  rather glued together into a smaller number of analytic functions.
This sewing condition  is the finite element of QSC construction which  locks completely the  QSC relations into a closed system of equations for spectrum. Their solution renders a discrete set of dimensions/energies of all the operators/states with the given 't~Hooft coupling \(g\), the  global charges \(J_1,J_2,J_3|S_1,S_2\) and twist parameters \(\{x_1,x_2,x_3,x_4|y_1,y_2,y_3,y_4\}\).

For various applications, especially related to the weak coupling approximations \(g\to 0\),  it is very convenient to reshuffle the sheets of the  Riemann surface of each \(\check\bQ\) function in such a way that it would have only short cuts on the whole Riemann surface.  For example,  the function \begin{equation} x(u)=\frac{1}{2g}(u+\sqrt{u-2g}\sqrt{u+2g}), \label{Zhuk}\end{equation}  inverse to Zhukovsky map \(\frac{u}{g}=x+1/x\), has by definition a short cut. But  its analytic continuation, the function \(\check x(u)=\frac{1}{2g}(u+   i\sqrt{4g^2-u^2})\), already has a long cut\footnote{what can be immediately seen if one plots real and imaginary parts of these two functions on Mathematica using Plot3D function.  } and it corresponds to  regluing halves of the  two sheets of the first functions along the real axis. As was already mentioned, we will reserve from now on the notation \(\bQ_j\) (without ``check") for the configuration of Riemann surface with short cuts.  This transition  \(\check\bQ\to\bQ\)  is shown on Fig~\ref{fig:cutsPQ}(right-down). Re-gluing in this way the halves of the first and the second sheets we obtain the first, upper sheet of the new Riemann surface which is free of singularities in the UHP, inheriting this analyticity from the UHP of the original physical sheet with the long cut; but on the real axis and below it we will have now a half-infinite ladder of Zhukovsky cuts inherited from the second sheet of the original Riemann surface. All these cuts can be made short by the same operation, involving the sheets next to the second one. So we can always work only with the short cuts, but the analyticity on the main sheet becomes more involved.

The sewing Rienamm-Hilbert relations \eqref{RH} now look as follows
\begin{equation}\label{sew}
 \begin{pmatrix}
\tilde\bQ_{ 1} \\
\tilde\bQ_{2}\\
\tilde\bQ_{ 3} \\
\tilde\bQ_{ 4}\\
\end{pmatrix} \simeq
\begin{pmatrix}0 & \bar\beta_1 & 0 & 0 \\
\beta_1 & 0 & 0 & 0 \\
0 & 0 & 0 & \bar\beta_2 \\
0 & 0 & \beta_2& 0 \\
\end{pmatrix} \begin{pmatrix}
\bar\bQ^{1} \\
\bar\bQ^{2}\\
\bar\bQ^{ 3} \\
\bar\bQ^{ 4}\\
\end{pmatrix},
\end{equation} where tilde sign corresponds to the  monodromy of a function around the Zhukovsky branchpoint on the  real axis.  That means that   \(\tilde\bQ_j\)  is identical to    \(\check\bQ_j\) on the second sheet on Fig.\ref{fig:cutsPQ}. Notice that on both sides of \eqref{sew}  the sequence of short cuts goes from the real axis upwards.

In fact, it turns out that any particular one of the conditions \eqref{sew}
\begin{align}
\tilde \bQ_1\sim  \bar \bQ^2\,,\quad \tilde \bQ_2\sim \bar \bQ^1\,,\quad
 \tilde \bQ_3\sim  \bar \bQ^4\,,\quad \tilde \bQ_4\sim \bar \bQ^3\,,
\label{sewQ}\end{align}  imposed on  functions with pure asymptotics is enough to fix completely the set of physical solutions for energies/dimensions of states with given global charges of the superconformal symmetry\cite{Gromov:2015vua}~.
The first of them is demonstrated on Fig.~\ref{fig:sew}(right).
In fact, it was also observed in
\cite{Gromov:2015vua} that, in the case of complete twisting, as in \eqref{assPcheckQ}, only one  of these conditions  is enough to fix completely such a set of solutions. The other three will follow from it.  If we have only a partial twisting, or coinciding twist parameters, the situation is more complex since the  asymptotics for some groups of \(Q\)-functions will be different only by integer powers and  the choice   of pure solutions becomes ambiguous.\footnote{The discussion and classification of asymptotics of all possible generations of twists can be found in~\cite{Kazakov:2015efa}.}  This is the case for example in the case of \(\gamma\)-deformation \eqref{assPcheckQ} where there is no twisting for \(\bQ\)-functions. In this case, a pair of relations \eqref{sewQ}, containing asymptotics with different sign of \(\Delta\)  is enough (the first and the third, or the second and the forth).

To conclude, in this and preceding sections, we gave the general scheme of the QSC formalism, concentrating on its universal and the most general defining features.  We specially avoided so far  any secondary details and consequences of this construction. In the next section, we will discuss an interesting physical application of QSC\(\gamma\), related to conformal filed theory following from the \(\gamma\)-twisted  \({\cal\ N}=4\) SYM in a specific double scaling (DS) limit of large (imaginary) \(\gamma\) twists and weak coupling.

\section{Double scaling limit of \(\gamma\)-twisted \({\cal N}=4\) SYM and fishnet\\
Feynman graphs}\label{sec:fishnet}

In this section, we demonstrate the force of QSC on a particular example of the study of the double scaling (DS) limit of \(\gamma\)-deformed   \({\cal\ N}=4\) SYM theory proposed in~\cite{Gurdogan:2015csr}.  The resulting non-unitary chiral CFTs  inherit the integrability properties of the full \(\gamma\)-twisted  \({\cal\ N}=4\) SYM. However it  demystifies to some extent the, still hypothetic though always properly working, AdS/CFT integrability: at least in its simplest, bi-scalar version the theory in DS limit is dominated by so called ``fishnet" Feynman graphs, explicitly related to the integrable conformal, \(SU(2,2)\) Heisenberg spin chain~\cite{Zamolodchikov:1980mb,Gurdogan:2015csr,Gromov:2017cja}.

\subsection{Lagrangian and conformal properties of \(\gamma\)-deformed \({\cal\ N}=4\) SYM}

The QSC formalism described in the previous section is deeply  rooted in the  bootstrap solution of  the string \(\sigma\)-model on \(\gamma\)-deformed \(AdS_5\times S^5\) background. It is the result of a long development of integrability methods, such as Y-system, TBA and Destri-De~Vega-type equations, applied to this two-dimensional string \(\sigma\)-model.  However, the AdS/CFT correspondence, as applied to the \(\gamma\)-deformed
case\cite{Lunin:2005jy,Frolov:2005dj,Dymarsky:2005uh,Beisert:2005if}~, states that the energy spectrum of this \(\sigma\)-model  is in one-to-one correspondence with the  spectrum of anomalous dimensions of
 the \(\gamma\)-deformed   \({\cal\ N}=4\) SYM theory with the Lagrangian  given in the following explicit form (see \cite{Sieg:2016vap,Gurdogan:2015csr}) \begin{align}\label{Lagrangian}   {\cal L}=&N_c\tr\biggl[
  -\frac{1}{4} F_{\mu\nu}F^{\mu\nu}
  -\frac{1}{2}D^\mu\phi^\dagger_iD_\mu\phi^i
  +i\bar\psi^{\dot\alpha}_{ A}D^\alpha_{\dot\alpha}\psi^A_{\alpha }
+\notag\\
& +g^2\,\left(\frac{1}{4} \{\phi^\dagger_i,\phi^i\}
     \{\phi^\dagger_j,\phi^j\}-\,e^{-i\epsilon^{ijk}\gamma_k}
     \phi^\dagger_i\phi^\dagger_j\phi^i\phi^j\right)+\notag \\
&+g\,\Big(-e^{-\frac{i}{2}\gamma^-_{j}}\bar\psi^{}_{ j}\phi^j\bar\psi_{ 4}
     +e^{+\frac{i}{2}\gamma^-_{j}}\bar\psi^{}_{ 4}\phi^j\bar\psi_{ j}
     + i\epsilon_{ijk}e^{\frac{i}{2} \epsilon_{jkm}
     \gamma^+_m} \bar\psi^k \phi^i \bar\psi^{ j}\,
\notag\\&+\,\,{\rm  conjugate \,\, terms} \big)\hfill\, \biggr],
\end{align} where \(A=1,2,3,4\), and we  sum up over all   doubly repeated or (abusing the standard tensorial notations)  triply repeated indices \(i,j,k,m=1,2,3.\)\footnote{Such a summation over triply repeated indices occurs since the \(SU(4)\sim SO(6)\) symmetry is broken by \(\gamma\)-twists.} \setcounter{footnote}{0}\def\thefootnote{\alph{footnote}\alph{footnote}} Here  \(\phi_{j=1,2,3}\)  are complex scalar fields and \(\psi_j^\alpha\) are Majorana-Weyl fermions.\footnote{We also suppressed the spinorial indices in the second and third lines in the above formula, always assuming that they are contracted in the standard way: \((\dots\psi\dots\psi)\to(\dots\psi^\alpha\dots\psi_\alpha)\) and \((\dots\bar\psi\dots\bar\psi)\to(\dots\bar\psi^{\dot\alpha}\dots\bar\psi_{\dot\alpha})\).} We also  used  the shorthand notations \(\gamma^\pm_{1}=\frac{\gamma_3\pm\gamma_2}{2},\,\,\gamma^\pm_{2}=\frac{\gamma_1\pm\gamma_3}{2},\,\, \gamma^\pm_{3}=\frac{\gamma_2\pm\gamma_1}{2}\).  The three parameters   \(\gamma_j\) are related to the twist parameters \(x_k\) of \eqref{gammatwist} by the following formulas~\cite{Kazakov:2015efa}  \begin{align}
&x_1= e^{\frac{i}{2}\left[(\gamma_2-\gamma_1)J_3-(\gamma_1+\gamma_3)J_2+(\gamma_2+\gamma_3)J_1\right]}\,,\qquad x_2= e^{\frac{i}{2}\left[(\gamma_1+\gamma_2)J_3-(\gamma_2+\gamma_3)J_1+(\gamma_1-\gamma_3)J_2\right]}\,, \notag \\
&x_3= e^{\frac{i}{2}\left[-(\gamma_1+\gamma_2)J_3+(\gamma_1+\gamma_3)J_2+(\gamma_3-\gamma_2)J_1\right]}\,,\qquad x_4= e^{\frac{i}{2}\left[(\gamma_3-\gamma_1)J_2+(\gamma_1-\gamma_2)J_3+(\gamma_2-\gamma_3)J_1\right]}\,.
\end{align}

Notice that if we put \(\gamma_1=\gamma_2=\gamma_3=0\) we obtain the standard case of the superconformal,   \({\cal\ N}=4\) SYM - a  CFT with the unbroken \(PSU(2,2|4)\) symmetry~\footnote{Unless it is spontaneously broken in Coulomb branch.}. The twisting corresponds to the following rule. In each term of the   Lagrangian \eqref{Lagrangian}, the deformation factors depend on the order of the fields under the trace.  For two arbitrary fields \(A\) and \(B\) the matrix product \(AB\) is replaced by a star product: \begin{align}A\,B\to A\star B\equiv  q_{A,B}\, A\, B\,,\quad  \text{where}\,\,\, q_{A,B}= e^{-\frac{i}{2}\epsilon^{mjk}\gamma_m\, J_j^A\, J_k^B}=(q_{B,A})^{-1}
\end{align} and \(J_1^A,\, J_2^A,\,J_3^A\in SO(6)\) are the three Cartan charges of \(A\)-field. Using this rule, it is easy to recover the \(\gamma\)-deformed Lagrangian \eqref{Lagrangian} from the undeformed one. On the classical level, the full superconformal symmetry appears to be explicitly broken: \(PSU(2,2|4)\to SU(2,2)\times U(1)^3\) but the conformal symmetry remains.

Strictly speaking, on the quantum level the \(\gamma\)-deformation breaks the conformal symmetry, even in the large \(N_c\) limit~\cite{Fokken:2013aea}~: although the 't~Hooft coupling \(g\) does not run with RG flow,  a few new, double-trace terms of the type  \(\tr(\phi_j\phi_k^{\dagger })\tr(\phi_k\phi_j^{\dagger})\) or \(\tr(\phi_j\phi_k)\tr(\phi^\dagger_j\phi_k^{\dagger })\)\cite{Tseytlin:1999ii,Dymarsky:2005uh,Fokken:2013aea} are generated, whose couplings do run. In particular,  for the double-trace interaction term
\(\alpha_{jj}^2\tr(\phi_j\phi_j)\tr(\phi^\dagger_j\phi_j^{\dagger })\) the one-loop beta-function
 is given by \cite{Fokken:2014soa}
\begin{equation}
 \beta_{\alpha^2_{jj}}=\frac{g^4}{ \pi^2}\sin^2\gamma_j^+\sin^2\gamma_j^-+\frac{\alpha_{jj}^4}{  4\pi^2}\,+{\cal O}(g^6)),\qquad (\text{no sum over}\, j),
\end{equation}
so that  at a complex fixed points \begin{equation}\label{fp}
\alpha_{jj}^{2}=\pm 2i g^2 \sin\gamma_j^+\sin\gamma_j^-+O(g^4)
\end{equation} the
  \(\gamma\)-deformed  \({\cal\ N}=4\)   SYM theory  becomes again a true, though non-unitary, CFT!\cite{Sieg:2016vap,Grabner:2017pgm}. This statement was demonstrated in~\cite{Grabner:2017pgm}~ in a few orders of perturbation theory in the specific double scaling limit described below. It was also claimed in~\cite{Grabner:2017pgm}~, and checked in many different ways, that the \(\gamma\)-deformed QSC of the previous section solves the problem of planar spectrum of
 the \(\gamma\)-deformed   \({\cal\ N}=4\) SYM precisely at this fixed point.
\subsection{Double scaling limit and fishnet graphs}

The
  \(\gamma\)-deformed  \({\cal\ N}=4\)   SYM theory admits an interesting DS limit\cite{Gurdogan:2015csr} which significantly clarifies the origins of integrability of  \({\cal\ N}=4\)  SYM  itself. The DS limit can be explicitly performed on the  level of Lagrangian \eqref{Lagrangian}.  Namely, it combines the weak coupling limit\(\) and big imaginary \(\gamma_j\) parameters: \begin{align}
g\to 0, \qquad e^{-i\gamma_j/2}\to\infty,\qquad \xi_j=g\,e^{-i\gamma_j/2}\,\,
   -\,{\rm  fixed},\qquad (j=1,2,3).
\end{align} The resulting non-unitary  CFT directly follows from \eqref{Lagrangian}. It  depends  on three DS couplings \(\xi_j\) and is defined~\footnote{Up to the already discussed double-trace  scalar interactions, tuned to the conformal point.} by the following  Lagrangian\cite{Gurdogan:2015csr}   \begin{align}\label{fullDS}
{\cal L}=&N_c\tr\biggl[
  -\frac{1}{2}\p^\mu\phi^\dagger_i\p_\mu\phi^i
  +i\bar\psi^{\dot\alpha}_{ A}\p^\alpha_{\dot\alpha}\psi^A_{\alpha }
+\notag\\
&+\xi_1^2\,\phi_2^\dagger \phi_3^\dagger \phi_2\phi_3+
\xi_2^2\,\phi_3^\dagger \phi_1^\dagger \phi_3\phi_1+\xi_3^2\,
\phi_1^\dagger \phi_2^\dagger \phi_1\phi_2+
\notag\\
 &+i\sqrt{\xi_2\xi_3}(\psi^3 \phi^1 \psi^{ 2}+ \bar\psi_{ 3} \phi^\dagger_1 \bar\psi_2 )
 +i\sqrt{\xi_1\xi_3}(\psi^1 \phi^2 \psi^{ 3}+ \bar\psi_{ 1} \phi^\dagger_2 \bar\psi_3 )
 \notag\\ &+i\sqrt{\xi_1\xi_2}(\psi^2 \phi^3 \psi^{ 1}+ \bar\psi_{ 2} \phi^\dagger_3 \bar\psi_1 )\,\biggr].
\end{align} Notice that the gauge field and the 4th component of fermion are completely decoupled in DS limit.

This theory obeys a certain chirality property which shows up if we consider the Feynman perturbation theory for graphs with fixed topology appearing in \(1/N_c\) expansion.  Namely, each  term in the last two lines of~\eqref{fullDS}, with quartic scalar or Yukawa interaction, does not have its Hermitian conjugate counterpart. This means that, only a certain   order of propagators around each  vertex of a planar graph is possible, and the vertex with opposite order, which would correspond to the Hermitian conjugated term in the Lagrangian,  does not appear.  We will show in the next subsection, along the lines of \cite{Gurdogan:2015csr,Caetano:2016ydc}, that this chiral property leads to
the  RG independence of \(\xi_j\) couplings and the absence of mass generation, since the corresponding Feynman graphs would necessarily include the vertices of both chiralities.

\subsection{Bi-scalar model and integrable fishnet graphs}

\begin{figure}[htb]  
\begin{center}
\includegraphics[scale=1.05]{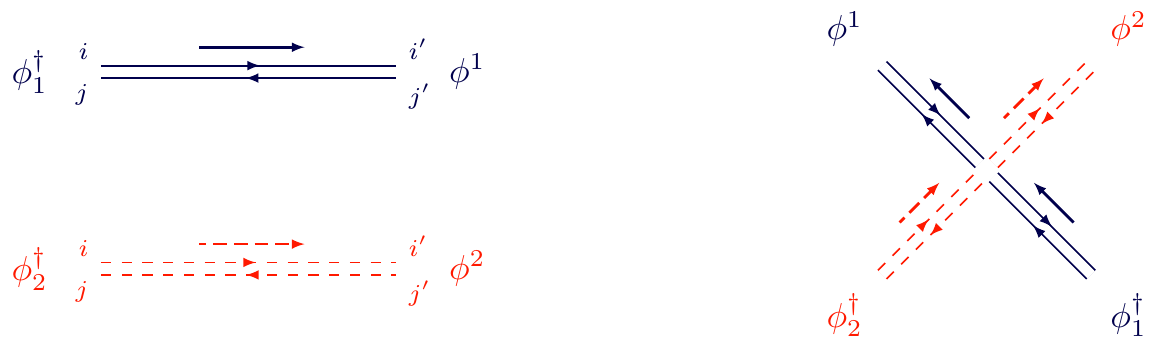}
\end{center}
\caption{Propagators and the vertex of bi-scalar model in double-line 't~Hooft notations. The external arrows show the direction from a field to its Hermitian conjugate. The model has a particular orientation of this arrows for two different fields (solid lines for \(\phi_1\) and dotted lines for  \(\phi_2\)) around the vertex, fixing its chirality. The vertex of opposite chirality, corresponding to the complex conjugate interaction term, is absent.   }
\label{fig:vertex}
\end{figure}

Let us discuss a particular case of the model \eqref{fullDS}, taking \(\xi_1=\xi_2=0\) and keeping only \(\xi\equiv \xi_3\ne 0\). Its Lagrangian appears to be extremely simple\cite{Gurdogan:2015csr}
 \begin{equation}
{\cal L}_{\text{bi-scalar}}[\phi_1,\phi_2]= \frac{N_c}{2}\tr\,\,
    \left(\p^\mu\phi^\dagger_1 \p_\mu\phi_1+\p^\mu\phi^\dagger_2
    \p_\mu\phi_2+2\xi^2\,\phi_1^\dagger \phi_2^\dagger \phi_1\phi_2\right)\,.
\label{bi-scalarL}\end{equation} The planar Feynman graphs are built of  two types of scalar propagators:
\begin{equation}
\left\langle \phi_1^{*ij}(y)\phi_1^{kl}(x)\right\rangle_0 \quad=
\quad\left\langle \phi_2^{*ij}(y)\phi_2^{kl}(x)\right\rangle_0\quad=\frac{1}{N}\delta
^{ik}\delta ^{jl}\,\frac{1}{(x-y)^2}
\end{equation} and a scalar vertex
\begin{equation}
V=\xi^2\tr\,(\phi_1^\dagger \phi_2^\dagger \phi_1\phi_2).
\end{equation}
These elements are presented
in double-line notations on Fig.\ref{fig:vertex}.

\begin{figure}[htb]  
\begin{center}
\begin{subfigure}{.5\textwidth}
  \centering
\includegraphics[scale=1.1]{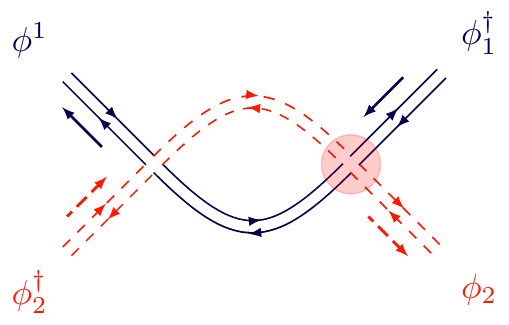}
\end{subfigure}%
\begin{subfigure}{.5\textwidth}
\includegraphics[scale=1.1]{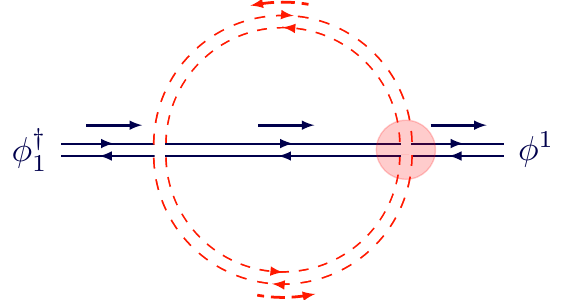}
\end{subfigure}%
\end{center}
\caption{These elements of graphs, renormalizing the vertex (on the left) and the mass (on the right), are absent from the planar Feynman diagrams of bi-scalar model due to the wrong chirality of vertices marked by a spot, absent from the action. This is an illustration of a general phenomenon leading to the all-loop conformality of the model.    }
\label{fig:nonchiral}\end{figure}

The  perturbative expansion for this theory   appears to contain very limited set of Feynman graphs, with very specific structure.  Notice for example that, in the lowest  order of perturbation theory, the Feynman diagrams renormalizing the coupling \(\xi\)  (on the left of Fig.\ref{fig:nonchiral}) and
the mass of scalars (on the right of Fig.\ref{fig:nonchiral})   are absent since they can be built only from two vertices of opposite chirality. This property persists in higher orders as well.

\begin{figure}[htb]  
\begin{center}
\begin{subfigure}{.5\textwidth}
\includegraphics[scale=0.87]{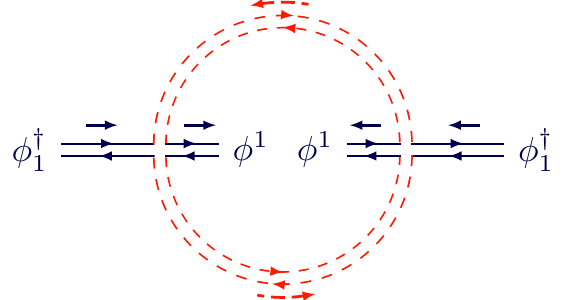}
\end{subfigure}%
\begin{subfigure}{.5\textwidth}
\includegraphics[scale=0.86]{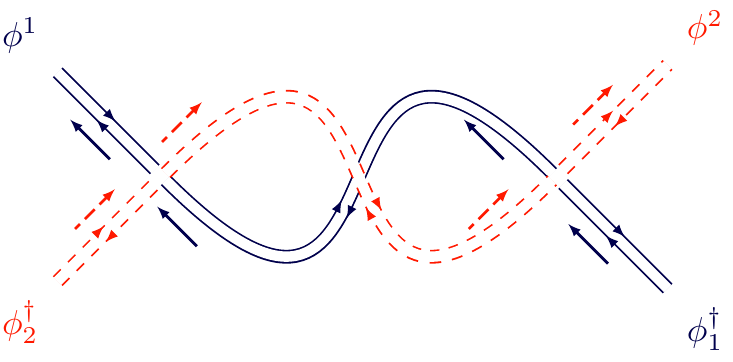}
\end{subfigure}%
\end{center}
\caption{Feynman diagrams generating the double-trace interactions in
  \(\gamma\)-deformed  \({\cal\ N}=4\)   SYM theory in general, and in the bi-scalar model in particular.  They survive even in the planar limit, but with the appropriate fine-tuning of these (complex) couplings the theory remains conformal.  }
\label{fig:doubletracegen}\end{figure}

However,  double-trace vertices will still be generated by RG, similarly to the full
  \(\gamma\)-deformed  \({\cal\ N}=4\)   SYM. For example, the diagrams of the types depicted on Fig.\ref{fig:doubletracegen} will generate the double-trace  terms \begin{align}
\label{double-tr-L}
 \alpha_1^2\sum_{i=1}^2\tr(\phi_i\phi_i)\,\tr(\phi_i^{\dagger}\phi_i^{\dagger})
 - \alpha_2^2\,\tr(\phi_1\phi_2)\tr(\phi_2^{\dagger }\phi_1^{\dagger })
-\alpha_3^2\tr(\phi_1\phi_2^{\dagger })\tr(\phi_2\phi_1^{\dagger })\, .
\end{align}
The couplings \(\alpha_i\) can be again adjusted, as functions of the non-renormalized coupling \(\xi\), to the fixed point where the model  becomes a non-unitary CFT\cite{Sieg:2016vap,Grabner:2017pgm}. The last two couplings appear to be only  one-loop renormalizable, so that at the critical point they are fixed to \(\alpha_2^2=\alpha_3^2=\xi^2\), whether as the critical coupling \(\alpha_1(\xi)\) is more complicated and can have two complex conjugate values\cite{Grabner:2017pgm}, given in dimensional regularization scheme by   \begin{align} \label{alpha1}
\alpha_{1,\pm }^2 = \pm \frac{i \xi ^2}{2}-\frac{\xi ^4}{2}\mp \frac{3 i \xi ^6}{4}+\xi ^8\pm\frac{65 i \xi
   ^{10}}{48}-\frac{19 \xi ^{12}}{10}+O\left(\xi ^{14}\right)\,.
\end{align}

To study the spectrum of anomalous dimensions of local operators in such a theory, we have to be able to compute the mixing matrix among various operators which is given in terms of two-point correlation functions. Generically, such operators are of single trace type (in the planar limit) and can be presented as linear combinations of ``words" (with cyclic symmetry) built out of  four fields \(\phi_1,\phi_1^\dagger,\phi_2,\phi_2^\dagger\) (taken at the same space-time point \(x\)) and two light-cone derivatives \(\p_\pm\) applied any number of times to any of the scalar fields constituting a local operator. Not all of these operators are independent: some of them are descendants (full derivatives) of simpler ones and some can be excluded by equations of motion. Still their amount is quickly increasing with the number of constituent fields.
Diagonalizing the mixing  matrix with particular linear combinations of such operators   we obtain the anomalous dimensions as its eigenvalues.

\begin{figure}[htb]  
\begin{center}
\begin{subfigure}{.5\textwidth}
  \centering
\includegraphics[scale=0.9]{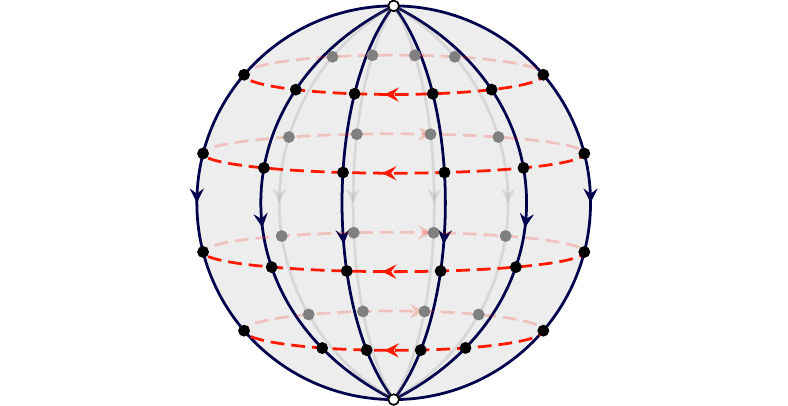}
\end{subfigure}%
\begin{subfigure}{.5\textwidth}
\includegraphics[scale=0.9]{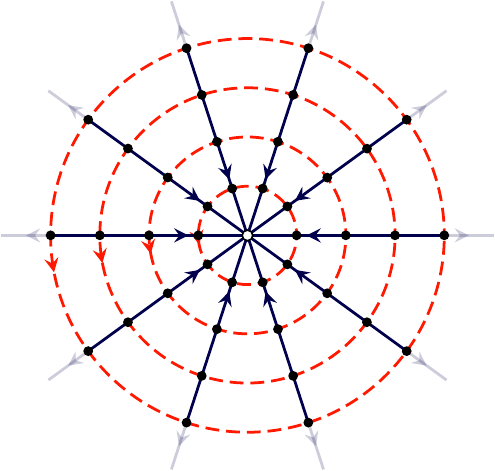}
\end{subfigure}%
\end{center}
\caption{The ``globe" graphs (on the left) are the only type of Feynman graphs (for \(L>2\)) contributing to the pair correlation function of the vacuum operator \(\tr(\phi_1)^L(x)\). In the bulk, such a Feynman diagram has the structure of fishnet: the regular square lattice of propagators with \(\phi^4\) interactions in the vertices. Due to the conformal invariance of these graphs we can send the coordinate  at one of the poles to infinity and, using the fact that we compute UV divergent expression, factor out the propagators around this pole as \(|x|^{-2L}\). We will be thus left with the calculation of the ``wheel" graph,  on the right of the picture. To compute the corresponding anomalous dimension we need to know only the coefficient of the simple pole \(1/\epsilon\) of the wheel graph. The wheel graphs are integrable. They can be studied by means of QSC\(\gamma\) (see next section) or using its \(SU(2,2)\) Heisenberg spin chain interpretation.      }
\label{fig:globe}\end{figure}

Let us consider one type of such operators -- the multi-magnon operators built only from \(L-M\) fields   \(\phi_1\) and \(M\) fields \(\phi_2\) (with no derivatives):
\footnote{In the description of these operators and the related Feynman graphs we closely follow the paper\cite{Caetano:2016ydc} and most of  the figures are also borrowed from there.  }
\begin{align}
{\cal O}_{L,M}(x)={\rm tr} \left(\underset{L\,\,
{\rm fields}}{\underbrace{\phi_2\phi_1\phi_1\phi_1\phi_2\dots\phi_1}
}\right)(x)\,\, + \,\, {\rm permutations.}
\label{magnonOp}\end{align} There exist linear combinations of such operators with given \(L,M\), diagonalizing the mixing matrix, i.e. leading to the standard conformal two-point correlation functions \begin{align}
\langle{\cal O}(x)\,\,{\cal O}(0)\rangle
  \,\, \sim |x|^{-2L-2\gamma(\xi)}
\label{vacuumcorr}\end{align} where \(\gamma(\xi)\) is the anomalous dimension of such an operator -- typically a complicated function of \(\xi\) which we want to compute. \setcounter{footnote}{0}\def\thefootnote{\alph{footnote}\alph{footnote}}
The simplest of such operators is the ``vacuum" operator\footnote{This operator is protected in the undeformed \({\cal N}=4\) SYM theory, i.e. it has the dimension \(\Delta=L\) for any value of coupling. The corresponding string state is usually called BMN vacuum. We keep calling it the ``vacuum" operator, though it gets non-trivial corrections in \(\gamma\)-deformed case, produced by  so called wrapped Feynman graphs. } without magnons (\(M=0\)) \begin{align}
{\cal O}_{L}(x)={\rm tr\,}[\phi_1(x)]^L\,.
\end{align} It does not mix with  any other operator and hence it has a particular anomalous dimension \(\gamma_L(\xi)\). If we  try to compute the pair correlation function of such operator \(\langle {\cal O}_{L}^\dagger(x){\cal O}_{L}(0)\) by the   Feynman perturbation technique we quickly realize that a single non-zero Feynman graph contributes    at each  \(\xi^{2L}\) order of perturbation theory and it has the shape of a ``globe" with meridians consisting of only \(\phi_1\)-type propagators and the parallels consisting of only \(\phi_2\)-type propagators, as depicted on Fig.\ref{fig:globe}(left). Due to the conformal invariance, we can send in this two-point correlator \(x\to\infty\) without the loss of information. In this case, the trivial divergent contribution \( |x|^{-2L}\) will factor out from the correlator \eqref{vacuumcorr} and  we can chop off \(L\) propagators adjacent to the  ``north pole" of the globe responsible for that contribution and   reduce the globe graph to a wheel graph shown on Fig.\ref{fig:globe}(right).  We will call this procedure a UV reduction.
That means that if we were able to compute such a Feynman integral at any loop order at a given \(L\) we would calculate a very non-trivial quantity in this CFT -- the anomalous dimension of the ``vacuum" operator.  Remarkably, the bulk of this graph looks like a regular square lattice -- ``fishnet" -- and is known to define an   integrable statistical-mechanical lattice model\cite{Zamolodchikov:1980mb}. The problem of vacuum anomalous dimension is exactly solvable due to integrability and QSC is a very efficient approach for that. In the next subsection we will describe these results.

\begin{figure}[htb]  
\begin{center}
\begin{subfigure}{.5\textwidth}
  \centering
\includegraphics[scale=0.9]{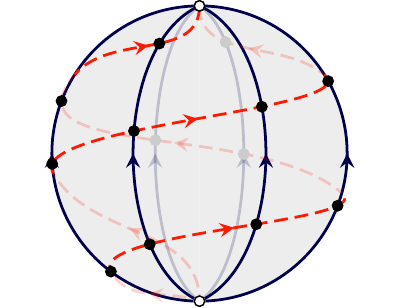}
\end{subfigure}%
\begin{subfigure}{.5\textwidth}
\includegraphics[scale=0.9]{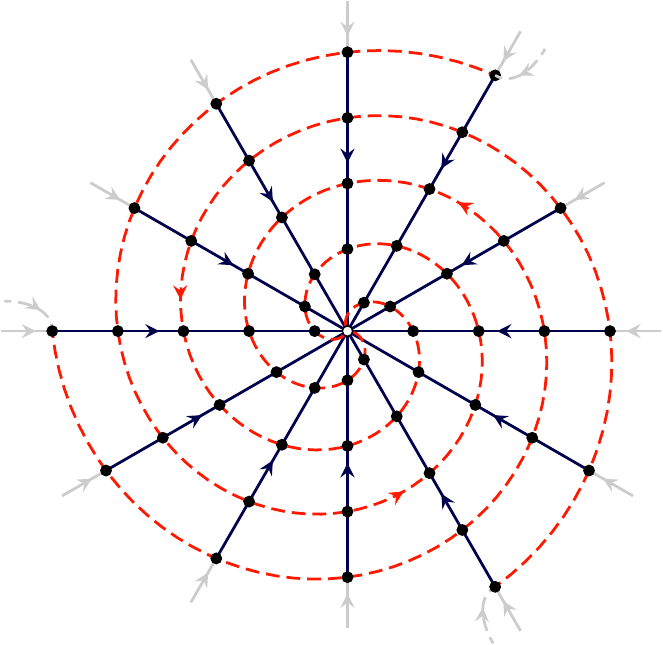}
\end{subfigure}%
\end{center}
\caption{The spiral graphs, such as one on the left -- an example with 2 magnons, dominate the operators with magnons of the type \eqref{magnonOp}. The UV reduction, similar to the one leading from globe graphs to wheel graphs for vacuum operators, gives a spiderweb-type graph on the right, obtained from the spiral graph by chopping off the propagators around one of the poles, such as the 3-magnon graph on the right. All those graphs are integrable and thus, in principal calculable, at least for the simple pole contributions in dimensional regularization.    }
\label{fig:spiral}\end{figure}

\begin{figure}[htb]  
\begin{center}
\includegraphics[scale=1.1]{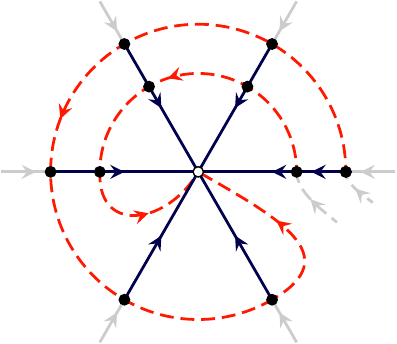}
\end{center}
\caption{An unwrapped magnon graph (spiderweb), calculable by the asymptotic Bethe ansatz methods developed in \cite{Caetano:2016ydc}. The absence of wrapping concerns the topology of planar graph. It can be seen as existence of a path connecting the center (fixed) node with a point outside (``infinity") without crossing any propagators.     }
\label{fig:wrapping2mag}
\end{figure}

Let us also consider a more general case of the operators \eqref{magnonOp}, in the presence of magnons, i.e. \(M\ne 0\). They are also dominated by very particular Feynman graphs of a spiral type, as shown on Fig.\ref{fig:spiral}(left).  The  UV reduction, similar to the previous globe/wheel case, brings us to a graph  on Fig.\ref{fig:spiral}(right) which can be called a ``spiderweb" graph.\footnote{This is precisely the way  the spiders weave their web.} These graphs also have a fishnet structure in their bulk. However, on the boundary of this fishnet the structure of the spiderweb graph is very different from the wheel graph. We can see that these are the same integrable lattice systems but with different boundary conditions. QSC allows to compute the anomalous dimensions of such multi-magnon operators as well. Such a calculation  at arbitrary coupling, or arbitrary loop order, is  yet to be done.
But for  low orders of perturbation theory, corresponding to    ``unwrapped" magnon graphs, such as shown on Fig.\ref{fig:wrapping2mag}  can be computed by means of the asymptotic Bethe ansatz (ABA)\cite{Caetano:2016ydc}  -- a doubly scaled version of Beisert-Staudacher equations\cite{Beisert:2006ez}.

 Let us note that, unlike the vacuum operators, the multi-magnon operators mix with each other for \(M>1\) magnons. QSC approach automatically solves the problem of finding the true anomalous dimensions - the eignevalues of the mixing matrix.

The limited Feynman graph content
of correlation functions in bi-scalar theory brings us to the idea of using integrability for exact computation of fishnet graphs (or at least of their specific UV singularity) with various integrable boundaries, specified by the appropriate conformal operators.  This program was significantly advanced in\cite{Gurdogan:2015csr} where the double wheel graphs (with double wrapping) have been computed using the TBA results of \cite{Ahn:2011xq}~,  in
\cite{Caetano:2016ydc}~\footnote{Tomake it precise, the explicit results for each 2-magnon 5-loop graph come about from a combination of ABA approach of \cite{Caetano:2016ydc} and explicit computation of one of these graphs in~\cite{Georgoudis:2018olj}. } where the 5-loop unwrapped  2-magnon  graphs have been computed using the doubly scaled ABA equations,  and finally in \cite{Gromov:2017cja} where the problem of \(L=3\) wheel graphs (i.e. with 3 spokes) is reduced, by the double scaling procedure applied to QSC formalism of the previous section, to a Baxter equation with specific quantisation conditions described in the next subsection. The last result gives essentially the full solution of the problem since it is very easy to generate from this equation the results for UV \(1/\epsilon\) divergency of such a graph at very high loop orders (12 loops are reached by a laptop mathematica program),  in terms of explicit  multiple \(\zeta\)-value (MZV) expressions.  Numerical solution for the anomalous dimension at finite couplings, with very high precision, is available as well.

We will describe the results of \cite{Gromov:2017cja} for the vacuum operator in the next section.  We will conclude this section by an interesting observation which can have important consequences for the non-perturbative study of this model.

\subsection{Feynman graphs of bi-scalar model and \(SU(2,2)\) conformal Heisenberg spin chain}

\begin{figure}[htb]  
\begin{center}
\includegraphics[scale=0.4]{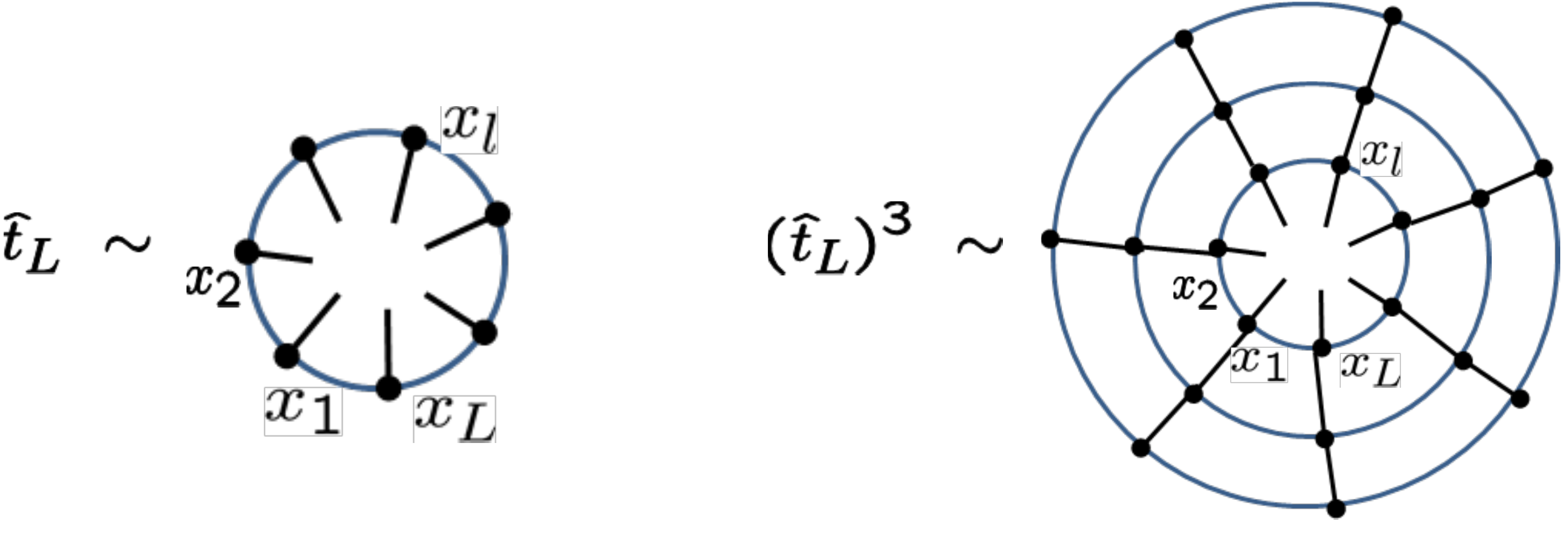}
\end{center}
\caption{On the left, we give a schematic representation of the graph-building operator \(\hat t_L\) \eqref{graph-building}(transfer-matrix), with a row of propagators around the circle and the radial propagators given by inverse laplacians. On the right, the 3rd power of this operator is depicted as a consecutive action of the kernel of this operator, producing the  integrations w.r.t. the  4D variables in the intermediate vertices.  These graphs give the main contribution to a certain \(2L\)-point correlation function of the bi-scalar model.    }
\label{fig:graphBuild}
\end{figure}

It was noticed in \cite{Gurdogan:2015csr} that the problem of computation of the wheel graphs can be formulated in terms of the ``graph-building" operator in the space of space-time coordinates \(x_l\) \begin{eqnarray}
{\hat t}_L=\xi^{2L}\prod_{l=1}^{L}\frac{1}{(x_{l+1}-x_{l})^2}\prod_{l=1}^{L}\Delta_{x_l
}^{-1}
\label{graph-building}\end{eqnarray} which represents one row of a wheel graph, as depicted in Fig.\ref{fig:graphBuild}(left).
Here the  first factor represents propagators of field \(\phi_2\), placed along the circular frame in angular direction, and  the second factor uses the standard property of inverse Laplacian \begin{eqnarray*}
 \Delta_{x}^{-1}\,\,\delta^{(4)}(x-y)=\frac{1}{(x-y)^2}
\end{eqnarray*} to insert the propagators in radial direction.

The wheel graph with \(n\) frames is given by the following formal expression: \begin{align}
W_{L,n} =\int d^4x_1\dots \int d^4 x_L\,\,\langle x_1,\dots,x_L|\left(\hat t_L\right)^n|0,\dots,0\rangle.
\end{align}
A power of graph-building operator \(\hat t_L\) is illustrated by the Fig.\ref{fig:graphBuild}~(on the right we show the example of \(\hat (t_L)^3\)). Summing up  matrix elements of powers  of  \(\hat t_L\)  we obtain a certain \(2L\)-point correlation function \begin{align}\label{Kkernel}
K_L( x_1,\dots,x_L|y_1,\dots,y_L)&=\prod_{l=1}^{L}(x_{l+1}-x_{l})^2\times\sum_{n=1}^\infty\langle x_1,\dots,x_L|(\hat t_L)^n|y_1,\dots,y_L\rangle= \notag\\
&=\prod_{l=1}^{L}(x_{l+1}-x_{l})^2\times\langle x_1,\dots,x_L|\frac{\hat t_L}{1-\hat t_L}|y_1,\dots,y_L\rangle
\end{align}  given by the sum of cylindric fishnet graphs, as the one  on  Fig.\ref{fig:graphBuild}(right).

Of course these expressions are UV divergent and need to be regularized. The standard dimensional regularization introduces the dimension as the regularization parameter \(\epsilon=4-D.\) To extract the anomalous dimension \(\gamma_L(\xi)\) of the vacuum operator \(\tr(\phi_1)^L\), we only  need to know the residue of the lowest, \(1/\epsilon\) pole of \(W_{L,n}\) for each \(n\).

\begin{figure}[htb]  
\begin{center}
\includegraphics[scale=0.4]{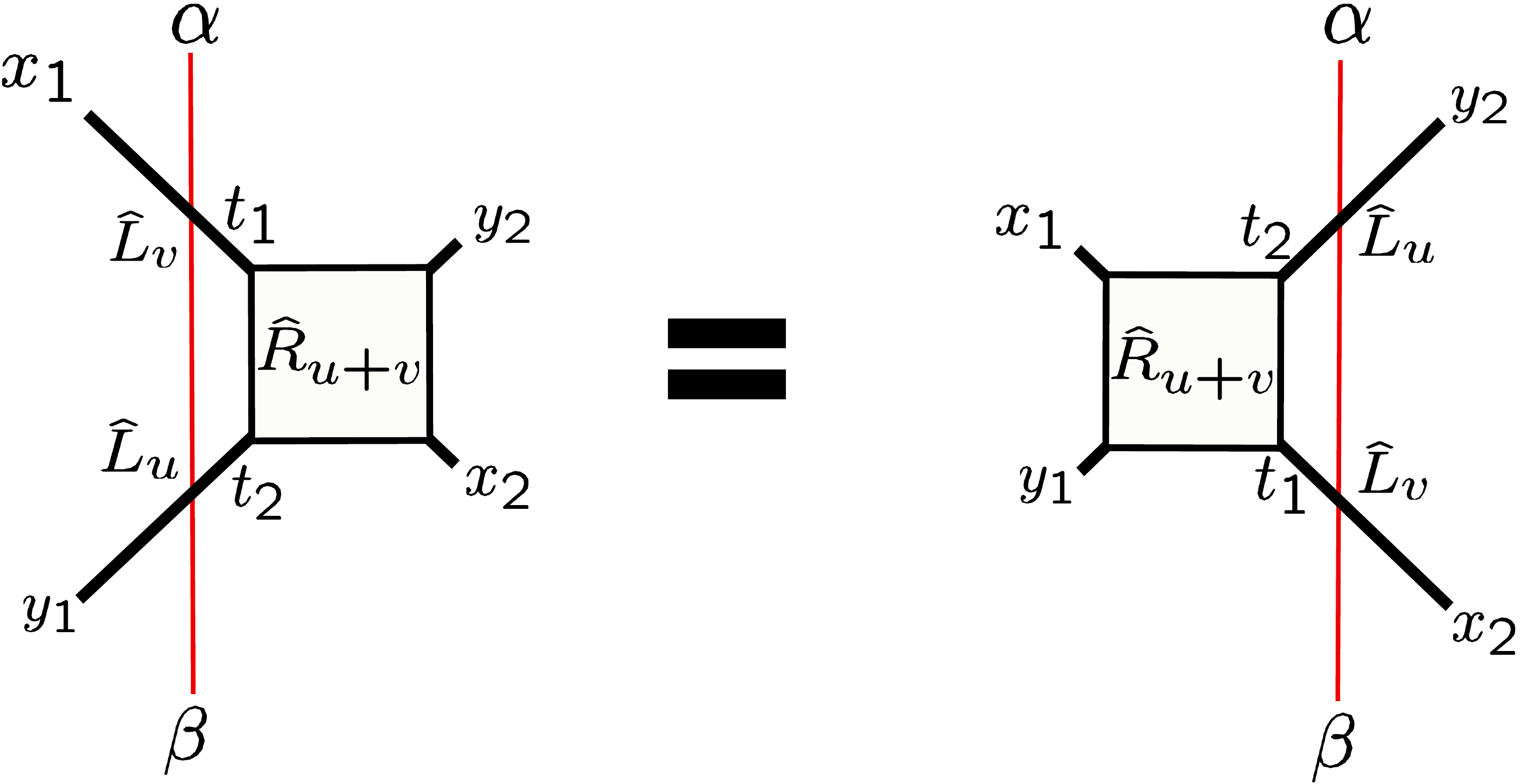}
\end{center}
\caption{Yang-Baxter relations for the Lax operators \eqref{Lax}, including the intertwiner    R-matrix \eqref{R-matrix} leaving in the product of principal series representations of the conformal group \(SU(2,2)\) realized by 4D coordinates \(x_j,y_j,t_j\). Lax operator leaves in the product of principal series representation (thick black lines) and fundamental representation (thin red lines), with indices \(\alpha,\beta=1,2,3,4\). The integrations over coordinates \(t_1,t_2\) are assumed.     }
\label{fig:YangBaxter}
\end{figure}

\begin{figure}[htb]  
\begin{center}
\includegraphics[scale=0.4]{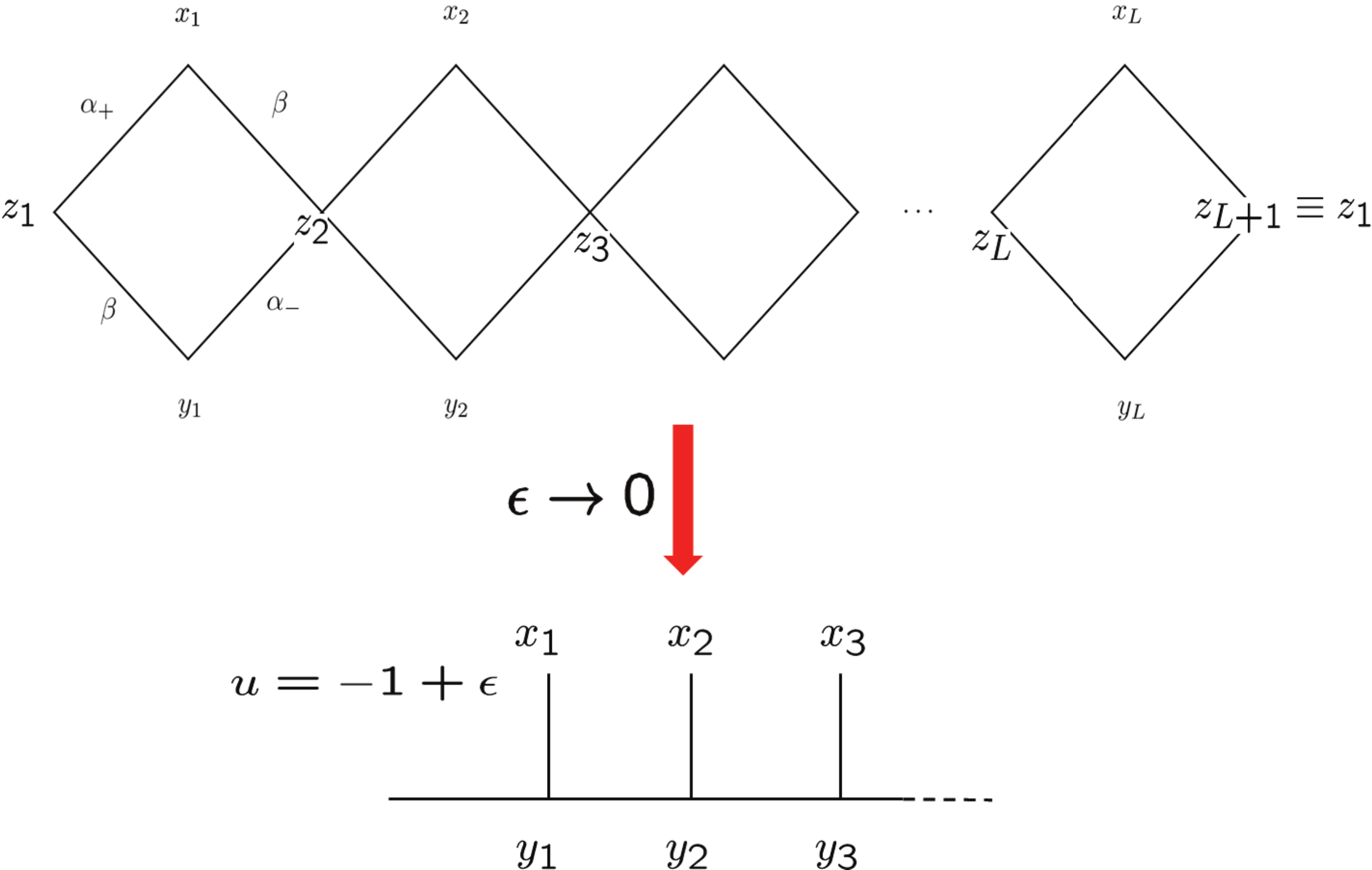}
\end{center}
\caption{Graphical demonstration of relation between the R-matrix \eqref{R-matrix}(upper chain of squares, with integrations over variables \(z_j\)) and the graph-building operator \eqref{graph-building}. The latter one emerges from the former for a particular limit of spectral parameter \(u\to -1\), when \(\alpha_+\to 0,\,\,\beta,\alpha_-\to 1, \) and  \(\delta^{(4)}(z_i-y_i)\) occur.  This demonstrates the fact that the graph-building operator is in involution with all the conserved charges of the integrable conformal \(su(2,2)\) spin chain.       }
\label{fig:Ttot}
\end{figure}

We will now argue that  the graph-building operator \eqref{graph-building} is a nontrivial conserved charge of the noncompact  Heisenberg spin chain based on the conformal group \(su(2,2)\). Indeed, let us define a Lax operator \begin{align}
\hat L_{\alpha\beta}(u) = u \,\delta_{\alpha\beta}+ \frac12 s^{ab}_{\alpha\beta}\,
\rho_{ab}
\label{Lax}\end{align} where \(u\) is the spectral parameter,   \(s^{ab}_{\alpha\beta}=\delta^a_\alpha \delta^b_\beta-\frac{1}{4}\delta^{ab} \delta_{\alpha\beta}\) are the standard adjoint \(su(4)\) generators and \(\rho_{ab}=\{{P}_{i,\mu},{L}_{i,\mu \nu},{D}_i, {K}_{i,\mu}\}  \in su(2,2)\) is a \(4\times 4\)matrix of  generators of conformal group in  representation \((S_1=0,h=1,S_2=0)\), i.e. with zero conformal spins and a unit weight (dimension) corresponding to that of the scalar field. This Lax operator satisfies the Yang-Baxter equations graphically represented on Fig.\ref{fig:YangBaxter},
 with the intertwining R-matrix living in the direct product of principal series representations and given by the following expression\cite{Chicherin:2012yn} \begin{eqnarray}
\!\!\!\!\!R_u(z_1,x_1|y_1,z_2) =
 \frac{c(u) }{ [(x_1-z_1)^2]^{-u-1} [(x_1-z_2)^2 (z_1-y_1)^2]^{u+2} [(y_1-z_2)^2]^{-u+1}}\,.
\label{R-matrix}\end{eqnarray}
where \(c(u)=\frac{2^{4u}}{\pi^4}\frac{\Gamma^2(u+2)}{\Gamma^2(-u)}\) is a useful normalization factor. This  R-matrix can be used to construct a transfer-matrix by taking the trace of their matrix product in auxiliary space (with the 4D variables \(z_1,\dots,z_L\) as labels):
\begin{align}  
&\hat T_{L}(u)\notag\\
&=\int d^4z_1\int d^4z_2\dots \int d^4z_L\,\,R_u(z_1,x_1|y_1,z_2) R_u(z_2,x_2|y_2,z_3) \dots R_u(z_L,x_L|y_L,z_1).
\end{align}
This quantity is depicted on Fig.\ref{fig:Ttot}(upper part).

Remarkably,
 when tuned to a particular value of the spectral parameter \(u= -1+\epsilon,\,\,\, (\epsilon\to 0)\) this transfer-matrix becomes exactly the graph-building operator \eqref{graph-building}!\cite{Gromov:2017cja} Namely,
 \begin{eqnarray*}
\langle x_1,\dots,x_L|T_{L}(-1+\epsilon)|y_1,\dots,y_L\rangle &=
\frac{1}{ (16\pi^2\epsilon)^{L}} \prod_{i=1}^{L}
\frac{1}{   (y_{i}-y_{i+1})^2\,\,(y_i-x_i)^2  } \\ &\sim \langle x_1,\dots,x_L|\hat t_L|y_1,\dots,y_L\rangle,
\end{eqnarray*} which can be also easily seen from the Fig.\ref{fig:Ttot}(lower part).

With the help of Lax operator  \eqref{Lax} we can also construct another \(u\)-dependent transfer-matrix, \begin{align}
\hat {\cal T}_L(u)&=\hat L_{\alpha_1\alpha_2}(u)\hat L_{\alpha_2\alpha_3}(u)\dots\hat L_{\alpha_{L}\alpha_1}(u)
\end{align}  Both transfer-matrices represent  generating functions of  quantum integrals of motion for the conformal, \(SU(2,2)\) Heisenberg spin chain.  In virtue of the Yang-Baxter relations they should commute:
\begin{align}
[\hat {\cal T}_L(u),\hat  T_L(u')]=0,
\end{align}
and, consequently, they both commute at any \(u\) with the
 graph-building transfer matrix \eqref{graph-building}: \begin{align}
[\hat T_L(u),\hat  t_L]=
[\hat {\cal T}_L(u),\hat  t_L]=0.
\end{align}
 This means that many problems of computation of physical quantities in the bi-scalar theory, such as OPE data, correlation functions,  etc,  given by fishnet Feynman graphs, can be formulated and studied within   the relatively well developed formalism of  integrable  non-compact Heisenberg spin
chains~\cite{Izergin:1982ry,Faddeev-Volkov:1995,Bazhanov:2007mh,Lipatov:1993qn,Faddeev:1994zg,Derkachov:2001yn,Chicherin:2012yn}. This promising approach to the study of bi-scalar model and its generalizations, based on the   conformal spin chain, is  still at its very early stage, though some important observations have been done on this way in\cite{Gromov:2017cja}. This Lax approach appeared to be very fruitful in  application to the scattering amplitudes of the bi-scalar model where  the Yangian symmetry has been discovered and explicitly demonstrated~\cite{Chicherin:2017cns,Chicherin:2017frs}, though in the original \({\cal N}=4\) SYM this symmetry remains still an open issue.

We will turn now to a more developed QSC approach which is however limited so far to the computation of spectra of local  operators.\footnote{And some non-local operators as well, such as a cusped Wilson loop\cite{Gromov:2015dfa}.}

\section{QSC solution for wheel graphs}\label{sec:wheels}

In this section, we will consider, as an application of QSC method, the calculation of anomalous dimensions for the vacuum operators \(\tr(\phi_1)^L\) in the bi-scalar CFT \eqref{bi-scalarL}, with a particular stress on the \(L=3\) case.  We will give only the main ideas of calculations. The interested reader can  find all   the details
in the original papers\cite{Gromov:2017cja,Grabner:2017pgm} where these results have been obtained.\footnote{Most of the figures of this section are also borrowed from\cite{Gromov:2017cja}.}

Our computation will be based on three main ingredients:
\begin{itemize}
\item  Analyticity properties of \(\bP\) and \(\bQ\) functions in the DS limit,
 as described in the previous section;       \item The DS limit of general QSC\(\gamma\) Baxter equation \eqref{BaxterPQ} for \(\bQ\)-functions;
       \item
Quantization condition for solutions of Baxter equation as a consequence of RH sewing relations~\eqref{sew}.
\end{itemize}

\subsection{DS Baxter equations}

Let us first notice that,  the operators \(\tr(\phi_1)^L\)  belongs to representation \((J_1=L,J_2=0,J_3=0)\) of \(SO(6)\to U(1)^3\) broken R-symmetry group of the full \(\gamma\)-deformed SYM theory. This state, as well as more general  multi-magnon operators \eqref{magnonOp}, obey the RL-symmetry and hence the upper index \(\bP^a\) and \(\bQ^j\) functions are trivially expressed through  lower index \(\bP_a\) and \(\bQ_j\)  functions  the relations \eqref{LRsymmetry}, thus greatly simplifying the algebraic structure of QSC Baxter equation  \eqref{BaxterPQ}, even before the DS limit.

The asymptotic \(1/u\) expansions of  \(\bP_a\) and \(\bQ_j\)  functions follow from the values of Cartan charges for this state: \(\{J_1,J_2,J_3|\Delta,S_1,S_2\}=\{L,0,0|\Delta,0,0\}\). We have from \eqref{assfulltwistP},\eqref{lambdaass},\eqref{gammatwist},\eqref{nuprime}
\begin{align}
 \label{asymptoticsPa}
&\bP_a\sim  A_a x_a^{i u} u^{-\lambda_a}\,\left(1+{\cal O}(1/u)\right)\,,\qquad
&\bQ_i\sim B_i u^{-\nu_i'}\left(1+{\cal O}(1/u)\right)\,, \end{align}
where
\begin{align}
\lambda_a=\left\{\frac{L}{2},\frac{L}{2},-\frac{L}{2},-\frac{L}{2}\right\},
\qquad\nu_i'=\left\{-\frac{\Delta}{2},-1-\frac{\Delta}{2},-2+\frac{\Delta}{2},-3+
\frac{\Delta}{2}\right\}. \label{assq}\end{align}
The leading  coefficients of asymptotics \(A_a,B_j\),  as well as the subleading ones, can be  fixed, up to the normalization conventions, by plugging them into the QSC Baxter equation   \eqref{BaxterPQ}.

We can also make some  precisions on the  't~Hooft coupling \(g\) dependence of \(\bP\) functions, using their important analyticity property -- the presence of a single short Zhukovsky cut for \(x\in (-2g,2g)\) on its physical sheet. That is why we can
uniformize \(\bP\) functions by expanding them in powers of variable \(x(u)\) instead of the spectral parameter \(u\) itself.  Namely, we can write \begin{align}
\bP_a(u)=x_a^{i u} (g x(u))^{- \lambda_a}\bp_a(u)\;,
\label{Pexp}\end{align} where
\begin{align}
\bp_a=\left\{A_1 f_1(u),A_2 f_1(-u),A_3 f_2(u),A_4 f_2(-u)\right\}
\end{align}
  and
\begin{align}
&f_1=1+g^{2L} \sum\limits_{n=1}^{\infty}\frac{g^{2n-2}c_{1,n}}{(g x)^n},\label{f1exp}\\
&f_2=(g x)^{-L}\left(u^L+  \sum_{k=0}^{L-1} c_{2,-k}u^k+ \sum\limits_{n=1}^{\infty}\frac{g^{2n}c_{2,n}}{(g x)^n}\right).
\label{f2exp}\end{align}
where the coefficients \(c_{1,n}(g), c_{2,n}(g)\) are functions of \(g\) and \(\gamma_j\), yet to be defined. We use the Zhukovsky variable which can be also expanded in powers of \(g/u\)   \begin{align}
\label{sqrt-exp}
gx(u)=\frac{1}{2}\left(u+\sqrt{u^2-4g^2}\right)=u\left(1-\frac{ g^2}{u^2}-\frac{ g^4}{u^4}-\frac{2 g^6}{u^6}+O\left(g^8\right)\right)\;.
\end{align}
 The asymptotics \eqref{asymptoticsPa} are already incorporated into this expansion and  we
used here   natural assumptions about the symmetry of \(\bP\) functions of the vacuum state \(\tr (\phi_1)^L\) w.r.t. the refection \(u \leftrightarrow -u\).

 In the DS limit, we take\footnote{There is no \(\gamma_1\) dependence for this particular state.} \begin{equation}\kappa=e^{-\frac{i}{2}(\gamma_3+\gamma_2)}\to\infty,\,\, \hat\kappa=e^{-\frac{i}{2}(\gamma_3-\gamma_2)}\to\infty,\,\, g\to 0,\quad (\xi= g\,\kappa\to \text{fixed)},\end{equation} we have to make some natural assumptions about the \(g\)-dependence of  coefficients  \(c_{1,n}(g), c_{2,n}(g)\). We will assume that all of them  have regular expansion
around \(g=0\).   Notice that \(f_1=1+{\cal O}(g^{2L})\), which reflects the fact that the non-trivial, wrapping contributions to the main asymptotics  start here from \(g^{2L}\) terms\cite{Kazakov:2015efa}. The power \(g^{2n-2}\) in each term of expansion in \eqref{f1exp} is needed to make each of these terms  regular in the weak coupling limit \(g\to 0\) in the function \(\tilde \bP_a(u)\) obtained from \(\bP_a(u)\) by monodromy around the branchpoint. This monodromy \(\bP_a\to\tilde\bP_a\) is achieved by simply flipping everywhere \(x(u)\to1/x(u)\), so that, e.g. \(\frac{g^{2n}}{(g x)^n}\to \frac{g^{2n}x^n}{g ^n}\sim {\cal O}(g^0)\).  A similar reasoning applies to \(f_2\). Notice that positive powers of \( u\) can be always converted to expansion in \(x\) using the inverse map \(u=g(x+1/x)\).

Plugging the large \(u \) asymptotics of  expansion \eqref{Pexp}-\eqref{f2exp}
into the coefficients \eqref{coeffBaxter} of QSC Baxter equation we can immediately fix the following relation between the leading coefficients \(A_a\) of the asymptotics:  \cite{Kazakov:2015efa}
\begin{align}\label{A1A2}
&A_1=-A_2=\frac{\hat \kappa^L (\kappa^L-1)^3}{(1+\kappa^L)(\kappa^L-\hat \kappa^L)((\kappa \hat \kappa)^L-1)}\\
&A_3=-A_4=-\frac{\kappa^L (\hat\kappa^L-1)^3}{(1+\hat\kappa^L)(\kappa^L-\hat\kappa^L)((\kappa \hat \kappa)^L-1)}.
\label{A3A4}\end{align}
where we chose the twist parameters as \(x_a=\left\{\kappa^L,\kappa^{-L},\hat\kappa^L,\hat\kappa^{-L}\right\}\). Notice that for the bi-scalar limit    \(\hat\kappa\to\kappa\).\footnote{We consider the case \(\hat\kappa\ne\kappa\) for regularization of certain divergencies. Then the LR-symmetry relation \eqref{LRsymmetry}  between upper and lower index \(\bP\)'s  should be accompanied by the simultaneous exchange  \(\hat\kappa\leftrightarrow\kappa\).\cite{Kazakov:2015efa} At the end we will put them equal. }

We can use all this information to compute the coefficients  \eqref{coeffBaxter} of QSC Baxter equation  in the DS limit but for finite \(u\). To this end, we expand the coefficients \(c_{1,n}(g), c_{2,n}(g),\)   as well as the Zhukovsky
variable \(gx(u) \), in regular series in \(g^2\):\begin{align}
&c_{m,n}(g)=c_{m,n}(0)+c_{m,n}'(0)g^2+\frac{1}{2}c_{m,n}''(0)g^4+O\left(g^6\right)\;\quad (m=1,2), \label{coeff-exp}
\end{align}  plug these expansions into             \eqref{f1exp},\eqref{f2exp} and use the resulting expansions of \(f_1,f_2\)  to compute the   coefficients \eqref{coeffBaxter} of QSC Baxter equation. Notice that we assume the right DS scaling regime for these functions to be \( g\ll u\sim 1\). As we mentioned in formulating the basic `axioms" of QSC, we assumed that the short Zhukovsky cut is the only singularity of \(\bP\) functions at finite part of the physical sheet. Hence the expansion \eqref{sqrt-exp} is the only source of poles at the origin in the DS limit of \(\bP\) functions.

Now we can perform the DS limit in coefficients
  \eqref{coeffBaxter} by grouping  the  powers of \(g\) with the powers of twist parameters \(\hat\kappa,\kappa\) (appearing in coefficients \(A_a\) in \eqref{A1A2}-\eqref{A3A4})  into finite DS couplings \(\xi= g\,\kappa,\,\, \hat\xi= g\,\hat\kappa\) and dropping all the subleading terms \({\cal O}(g,1/\kappa,1/\hat\kappa)\). Remarkably, no terms which blow up in this limit occur during this DS procedure, which perfectly confirms our assumptions for the ansatz \eqref{f1exp},\eqref{f2exp}. Of course, only a finite number of expansion coefficients is retained, though their number increases with the length of operator \(L\), as is obvious from the form of the leading coefficients \eqref{A1A2}-\eqref{A3A4}.\footnote{Practically, extracting the DS limit from such expansions was done by using Mathematica program. Increasing \(L\) can be rather time-consuming, so that we have done the actual calculations only for \(L=2,3,4.\)\cite{Gromov:2017cja}.}.    Another positive sign is that all the  coefficients  \eqref{coeffBaxter} of QSC Baxter equation  appear to be of the same order in \(g\) and hence we have at the end a perfectly defined Baxter equation in DS limit for the ``vacuum" state, valid for the full chiral CFT \eqref{fullDS}. If we want to limit ourselves to the bi-scalar case we simply put \(\hat\xi=\xi\) which appears to be a smooth limit in the DS Baxter equation.

The resulting Baxter equation, for slightly modified definition of \(\bQ\)-function \begin{equation}\bQ_j=u^{L/2}q_j\,,\end{equation} takes a rather symmetric form\cite{Gromov:2017cja}
\begin{align}
&A\left(u+i\right)q\left(u+2i\right)-B\left(u+\frac{i}{2}\right)q\left(u+i\right)
+C\left(u\right)q\left(u\right)- \notag\\&-B\left(u-\frac{i}{2}\right)q\left(u-i\right)+A\left(u-i\right)q\left(u-2i\right)=0\;,
\label{DSBaxter}
\end{align}
where
 for \(L=3\) \begin{align} &A(u)=u^{3}\,,\qquad B(u)= u\left(4u^{2}-\frac{\alpha+5}{2}\right)\,,  \\
 &C(u)=6u^{3}-\left(\alpha+5\right)u+\frac{\left(\alpha-1\right)^{2}}{16u}+\frac{m^2}{u^{3}}\end{align}and \(\alpha=(\Delta-2)^2\). The higher conserved charge \(m(\xi)\), as well as the value of  dimension \(\Delta( \xi)\) for this state -- our main goal -- will be fixed from the auxiliary quantization condition following essentially from the QSC Riemann-Hilbert sewing conditions \eqref{RH}.

The Baxter equation \eqref{DSBaxter} is very suggestive as concerns already mentioned direct relation of the current problem to the periodic \(SU(2,2) \) Heisenberg spin chain described in~\cite{Chicherin:2012yn}~(see Appendix~A of \cite{Gromov:2017cja}). One can even justify its form  \eqref{DSBaxter},  with \(A(u)=u^{L}\),  which is completely fixed by the choice of spin representation \(\{0,1,0\}\), whereas
\(B(u) \)  and \(u^{L}C(u)\) are polynomials in $u$ of degree \({L}\) and \(2{L}\), respectively, obeying the symmetry (proper to this state):
 \(B(u)=(-1)^{L} B(-u)\,,\quad u^{L} C(u)=(-u)^{L}C(-u)\).    This information, as well as the asymptotic behavior of \(q\) functions: \begin{equation}\label{Csymmetry}
 q\sim u^\delta \,,\qquad \delta=\left\{\frac{\Delta-L }{2},\,\frac{\Delta-L }{2} +1,\, 2-\frac{\Delta +L}{2} ,\, 3-\frac{\Delta +L}{2}\right\}
\end{equation} following from  \eqref{asymptoticsPa} partially fix the coefficients of \eqref{DSBaxter} at any \(L\):\cite{Gromov:2017cja}
\begin{align}\label{BC}
 B(u) &= 4u^L-\frac{1}{2}(\alpha+3 L-4)u^{L-2}+bu^{L-4}+\sum_{k=3}^{[L/2]} d_ku^{L-2k}\,,
\\
C(u) &=  6u^L-(\alpha+3 L-4)u^{L-2}+\frac{(\alpha -4)^2+32 b+3 L^2+2 (\alpha -7) L}{16}u^{L-4}\notag\\
&+\sum_{k=3}^{L}c_ku^{L-2k}\,,
\end{align}
where  \(\alpha=(\Delta-2)^2\). These expressions depend on \(1+(L-2)+([L/2]-2)=L+[L/2]-3 \)  arbitrary constants
\(b, c_k,d_k\),  to be fixed  by  additional, yet to be derived, quantization conditions.

\subsection{DS quantisation condition }

The Baxter equation \eqref{DSBaxter} has a few unfixed coefficients, including the main quantity under study -- the dimension \(\Delta(\xi)\). In addition,  any linear combination, with \(i\)-periodic coefficients,  of  4 independent solutions of this equation is again a solution. We have to find such a set of 4 solutions, corresponding to \(q_i=u^{-L/2}Q_i\) that they are pure functions. But even the condition of purity does not fix them completely. We have to find an additional condition which fixes the solutions, as well as the yet unfixed coefficients in \eqref{BC}, completely. Such a quantisation condition should be based on the RH sewing relations \eqref{RH} which we did not   use so far.

Let us now concentrate on the case \(L=3\) for the bi-scalar model. First of all, we notice that the 4th order Baxter equation \eqref{DSBaxter} can be factorized in this case to two 2nd order equations: \((\hat L_+ u^3\hat L_-)q\equiv(\hat L_-u^3\hat L_+)q=0\), where \begin{eqnarray} \label{bax2-fact}
\hat L_\pm q(u)\equiv\left(\frac{(\Delta-1)(\Delta-3)}{4u^2}\pm\frac{m}{u^3}-2\right)q(u)+q(u+i)+q(u-i)=0.
\end{eqnarray} The asymptotics \eqref{assq} suggest that the pure solutions  of the first of these equations \(\hat L_-q(u,m)=0\) are \begin{align}
\label{as-inf1}
&q_2(u,m)= u^{\Delta/2-1/2}\left(1+\frac{a_1}{u}+O\left(\frac{1}{u^2}\right)\right),
\\
\label{as-inf2}
& q_4(u,m)= u^{-\Delta/2+3/2}\left(1+\frac{b_1}{u}+O\left(\frac{1}{u^2}\right)\right),
\end{align} Then \(q_1(u,m)=q_2(u,-m),\quad q_3(u,m)=q_4(u,-m). \)

From  solutions to the Baxter equation \eqref{bax2-fact}, we can now construct four \(\bQ\)-functions.
In the double scaling limit they are linear combinations of \(q\) functions
\begin{align}
&  {\bf Q}_1(u)=\frac{-is^6}{2m(\Delta-2)}u^{3/2}[q_2(u,m)-q_2(u,-m)]\,,
\notag\\
& {\bf Q}_2(u)=\frac{u^{3/2}}{2}[q_2(u,m)+q_2(u,-m)]\,,
\notag\\&  {\bf Q}_3(u)=\frac{is^6}{2m(\Delta-2)} u^{3/2}[q_4(u,m)-q_4(u,-m)]\,.
\notag\\&{\bf Q}_4(u)=\frac{u^{3/2}}{2}[q_4(u,m)+q_4(u,-m)]\,,
\notag\\\label{Qq}
\end{align}
where the coefficients on the left are chosen as a normalization and the other two are fixed from  compatibility with the   QSC Baxter equation   \eqref{BaxterPQ}. These functions obey a fixed parity w.r.t. \(u\to-u\), as assumed  in the initial ansatz, since the Baxter equations are invariant w.r.t. the simultaneous change  \(u\to -u,\,\,m\to-m\). Notice that this also means that \begin{equation}\bar\bQ_j(u)\sim\bQ_j(-u)\label{barQQ}\end{equation} since we will see below that  for physical solutions the constant \(m\) is purely imaginary.

The RH conditions \eqref{RH}, together with the RL-symmetry \eqref{LRsymmetry}, lead to an additional constraint \(\beta_1=1/\bar \beta_2\) and thus to the following analyticity constraints \begin{equation}
\quad\ \tilde \bQ_1(u)=\bar\beta_1\bar\bQ_3(u),\quad \tilde \bQ_2(u)=-\beta_1\bar\bQ_4(u).
\label{Qquant}\end{equation}Notice that if  we take  the argument at one of the  branch-points \(u^*=\pm 2g\) then we have an obvious  equality \(\tilde\bQ_j(u^*)\sim\bQ_j(u^*)\) since, by assumption, the only singularities of \(\bQ_j\) at finite \(u\) are the Zhukovsky cuts. That means that  keeping \(u=u^*\)  we  can extract from \eqref{Qquant},\eqref{as-inf1}, \eqref{as-inf2}  two expressions for  \(\beta_1 \):
\begin{equation}
 \beta_1=\frac{ \bQ_1(u^*)}{\bQ_3(-u^*)}=-\frac{ \bQ_2(u^*)}{\bQ_4(-u^*)}.
\label{Qquantbeta}\end{equation}
Since in the  DS limit \(u^*=\pm 2g\to 0\),
plugging here  \eqref{Qq} we obtain the final quantization condition:\footnote{I thank N.Gromov for sharing with me this and the next shortcuts for deriving the quantization condition and the formula for \(m^2\). See more rigorous  original derivation in our paper\cite{Gromov:2017cja}.} \begin{equation}
q_2(0,-m)q_4(0,m)+ q_2(0,m)q_4(0,-m)=0\,
\label{qquant}\end{equation} which fixes unambiguously the physical solution of Baxter equation \eqref{bax2-fact}.\footnote{This relation implies the cancellation of the proportionality constants appearing in \eqref{barQQ}. These constants can be computed by following the asymptotic behavior behavior of functions in \eqref{Qq} when going from \(u\to\infty\) to \(u\to-\infty\) along a big semicircle in the upper half-plane.   }

Recall that the \(\bQ\) functions, by the assumptions of QSC formalism, can only have the short cuts in the lower half-plane, starting from the real axis. In the DS limit these cuts can give only the poles at \(u=-i n, \quad n=0,1,2,\dots\).  Inspecting the equation \eqref{bax2-fact} around these poles we realize that, first of all, they cannot be of the order higher than \(\frac{1}{u^3}\), and second, \(q_2(u, m),q_4(u, m) \) are regular functions at \(u=0\).~\footnote{A pole at \(u=0\) would immediately produce a pole at \(u=i\), which is absent by assumption}  Similar argument applies to \(\bar\bQ\), except that the poles can be now found only in the upper-half plane, and hence \(q_2(u, -m),q_4(u, -m) \) are also regular at \(u=0\).

This quantization condition suffices to establish the relation between \(\Delta\) and \(m\). However, we don't know yet how any of them are related to the coupling constant \(\xi\).  The dependence of \(m\) on \(\xi\) was derived in\cite{Gromov:2017cja} from the QSC formalism and appeared to be very simple \begin{equation}
\quad  m^2=-\xi^6.
\label{m2xi3}\end{equation}  implying, as we already mentioned, that \(m\) is imaginary. We will not reproduce here these arguments but we note that it can be obtained entirely within the \(SU(2,2)\) Heisenberg spin chain formalism\cite{Chicherin:2012yn,Gromov:2017cja} by showing that the coefficient \(m^2\), an   eigenvalue of  the conserved charge \(\hat m^2\), can be directly related to the graph-building hamiltonian \(\hat  h_3\) defined by \eqref{graph-building}, namely  \begin{equation}
 \hat m^2=-\xi^{6}(\hat h_3)^{-1}.
\end{equation} Then we notice that the energy of this state  should be defined by the  position of the pole in  the expansion \eqref{Kkernel}, namely at \(\hat h_3\to 1 \). For the eigenvalue of \(\hat m\) this means precisely the relation \eqref{m2xi3}.\footnote{The last argument will appear in the forthcoming work\cite{GGKK}.}

In conclusion, we obtained the following formulation of solution of the problem of computation of anomalous dimension of the operator \(\tr(\phi_1)^3\) in the bi-scalar model: to fix \(\Delta(\xi)\), we find a pair of pure solutions of  Baxter equation  \eqref{DSBaxter} (choosing the lower sign ``-" there and fixing \(m=\pm i\xi^3\)) satisfying the quantization condition \eqref{qquant}, the asymptotic expansions \eqref{as-inf1}-\eqref{as-inf2} and having the poles up to the third order order for \(u=-i n,\quad n=1,2,3,\dots    \).
This will fix a discrete set of dimensions \(\Delta(\xi)\), the lowest of them corresponding to the operator \(\tr(\phi_1)^3\).

\subsection{Some results for \(J=3\) spectrum}

Let us briefly present some results of the analysis, perturbative and numerical, of equations obtained in the previous section.

\begin{figure}  
\center{\includegraphics[scale=0.4]{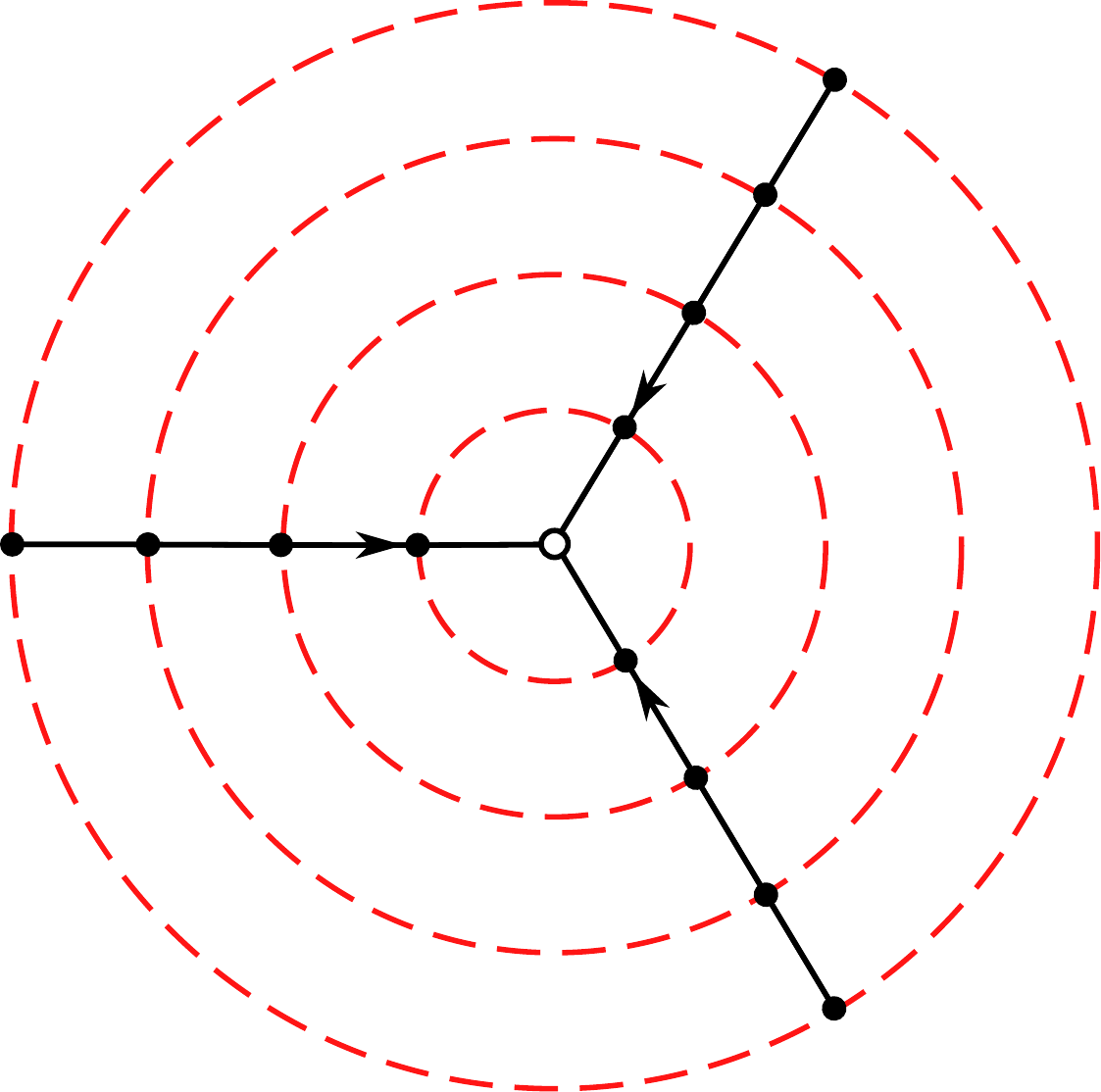}}
\caption{The ``wheel" Feynman graph corresponding to $O(\xi^{24})$ term in the weak coupling expansion \eqref{D3exp}
of anomalous dimension of the operator \(\Tr( \phi_1^3)\). \label{fig:L3graph}  }
\vspace*{-12pt}
\end{figure}

\paragraph{Weak coupling solution for the operator \(\tr(\phi_1)^3\): }
At zero order we have \(m=\,\,-i\xi^3=0\) and \(\Delta(0)=3\). We find thus a pair of independent solutions of \eqref{DSBaxter}: \(\;\;q_I(u,0)=u\;\;,\;\;q_{II}(u,0)=1\,\). Expanding \( \Delta=L + \sum_k \xi^{6k} \Delta^{(k)}  \) and, for each function, \(q(u,m)=w_1(u)+m w_2(u)+m^2 w_3(u)+\dots\) and solving  \eqref{DSBaxter} iteratively by means of varying coefficients, we can find the next term of \(m=\pm i \xi^3\) expansion with the right pole structure: \begin{align}
q_I&=u-i m \left(\eta _1-\eta _2 u\right)+\notag\\&+m^2 \left(-\eta _{1,2}+\eta _{2,1}+u \eta
_{1,3}-u \eta _{2,2}-\frac{i \delta }{2}+\frac{1}{2} i   \eta _1
u \delta+\frac{u \delta}{2}\right)+{\cal O}\left(m^3\right)\,,\notag\\
q_{II}&=1-i m \left(\eta _2-\eta _3 u\right)+\notag\\&+m^2 \left(-\eta _{2,2}+\eta _{3,1}+u \eta
_{2,3}-u \eta _{3,2}-\frac{1}{2} i\eta _1\delta +\frac{1}{2} i \eta _2
u\delta\right)+{\cal O}\left(m^3\right)\;,\notag
\end{align} where \(\Delta=3-m^2\delta + O(m^4)\,\) and we introduced a standard set of functions~\cite{Leurent:2013mr} \begin{equation}\label{eta}
\eta_{s_1,\dots,s_k}(u)=\sum_{n_1>n_2>\dots>n_k\geq 0}\frac{1}{(u+in_1)^{s_1}\dots (u+in_k)^{s_k}}\;.
\end{equation} Using these two \(q\)-functions we can find, order by order, the right linear combinations of them corresponding to two pure solutions \(q_2\) and \(q_4\) which fit the asymptotics \eqref{as-inf1}-\eqref{as-inf2}. Leaving aside the details, which can be found in\cite{Gromov:2017cja}, we give here the result for the dimension of \(\tr(\phi_1)^3\) up to 12 loops:
\begin{align}
&\Delta_3-3=-12 \zeta _3\xi^6+ \xi ^{12}\left(189 \zeta _7-144 \zeta _3{}^2\right)\notag\\
&+\xi ^{18} \left(-1944 \zeta _{8,2,1}-3024 \zeta
_3{}^3-3024 \zeta _5 \zeta _3{}^2+6804 \zeta _7 \zeta
_3+\frac{198 \pi ^8 \zeta _3}{175}\right.\notag\\
&\left.+\frac{612 \pi ^6 \zeta
_5}{35}+270 \pi ^4 \zeta _7 +5994 \pi ^2 \zeta _9-\frac{925911
\zeta_{11}}{8}\right)\notag\\
&+\xi^{24}\left(-93312 \zeta_3\zeta_{8,2,1}+\frac{10368}{5}\pi^4\zeta_{8,2,1}+5184\pi^2
\zeta_{9,3,1}+51840\pi^2\zeta_{10,2,1}\right.\notag\\
&\left.-148716\zeta_{11,3,1}-1061910\zeta_{12,2,1}+62208\zeta_{10,2,1,1,1}
-77760\zeta_3{}^4-145152\zeta_5\zeta_3{}^3\right.\notag\\
&\left.-\frac{576}{7}\pi^6\zeta_3{}^3-864\pi^4\zeta_5\zeta_3{}^2
-2592\pi^2\zeta _7\zeta_3{}^2+244944\zeta_7\zeta_3{}^2+186588\zeta_9\zeta_3{}^2\right.\notag\\
&\left.+\frac{9504}{175}\pi^8\zeta_3{}^2-2592\pi^2\zeta_5{}^2\zeta_3
+\frac{29376}{35}\pi^6\zeta_5\zeta_3+298404\zeta_5\zeta_7\zeta_3\right.\notag\\
&\left.+12960\pi^4\zeta_7\zeta_3+287712\pi^2\zeta_9\zeta_3
-5555466\zeta_{11}\zeta_3+\frac{2910394\pi^{12}\zeta_3}{2627625}
+57672\zeta_5{}^3\right.\notag\\
&\left.-71442\zeta_7{}^2+\frac{13953\pi^{10}\zeta_5}{1925}+\frac{7293\pi^8\zeta_7}{175}
-\frac{19959\pi^6\zeta_9}{5}+\frac{119979\pi^4\zeta_{11}}{2}\right.\notag\\
&\left.+\frac{10738413\pi^2\zeta_{13}}{2}-\frac{4607294013\zeta _{15}}{80}\right)
+O\left(\xi^{30}\right)\,,
\label{D3exp}
\end{align}
where $\zeta _{i_1,\dots, i_k}=\sum_{n_1> \dots>n_k >0} 1/(n_1^{i_1}\dots  n_k^{i_k} )$ are  multiple Riemann \(\zeta\)-values. Here the coefficients in front of \(\xi^{6M}\) give the residues
at simple pole \(1/\epsilon\) in dimensionally regularized Feynman integrals corresponding to  the \(L=3\) wheel graphs with \(M=1,2,3,4,\dots\) frames (see Fig. \ref{fig:L3graph}).
The first two terms of (\ref{D3exp}) (one and two  wrappings) coincide with  the known results \cite{Broadhurst:1985vq,Panzer:2013cha}.

\begin{figure}  
\includegraphics[scale=0.5]{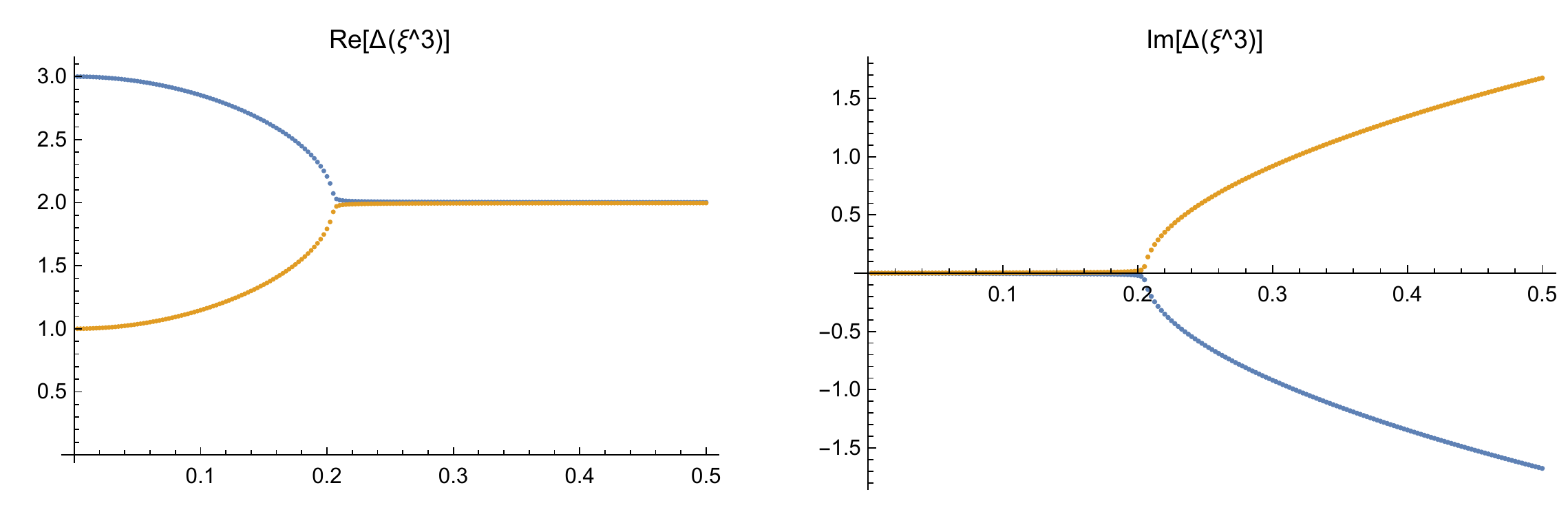}
\caption{Numerical results for  the scaling dimension of the operator  \(\tr(\phi_1^3)\) as a function of the coupling \(\xi^3\). At \(\xi^3\simeq 0.21\)  the scaling dimension hits the value $\Delta=2$ and becomes imaginary. This point defines the radius of convergency of the weak coupling expansion.  The second branch, starting from \(\Delta(0)=1\), arises due to the  symmetry of the Baxter equation (\ref{DSBaxter}) under \(\Delta\to 4-\Delta\).
 It corresponds to a non-local, ``shadow" operator.\label{fig:numericsL3}
}
\vspace*{-12pt}
\end{figure}

\paragraph{Numerical solution}  The equations of the previous subsections can be also solved, very  efficiently and with virtually unlimited accuracy, numerically at all interesting finite values of the coupling, using the methods developed in\cite{Gromov:2015dfa}~. The results for the dimension \(\Delta_3(\xi)\) are presented on Fig.\ref{fig:numericsL3}. This dimension is real for sufficiently small values of \(\xi\), but it becomes imaginary starting from a certain value   \(\xi^3\simeq 0.21\), which is not an abnormal behavior for a non-unitary theory, such as our bi-scalar model.

\begin{figure}[htb]  
\begin{center}
\includegraphics[scale=0.4]{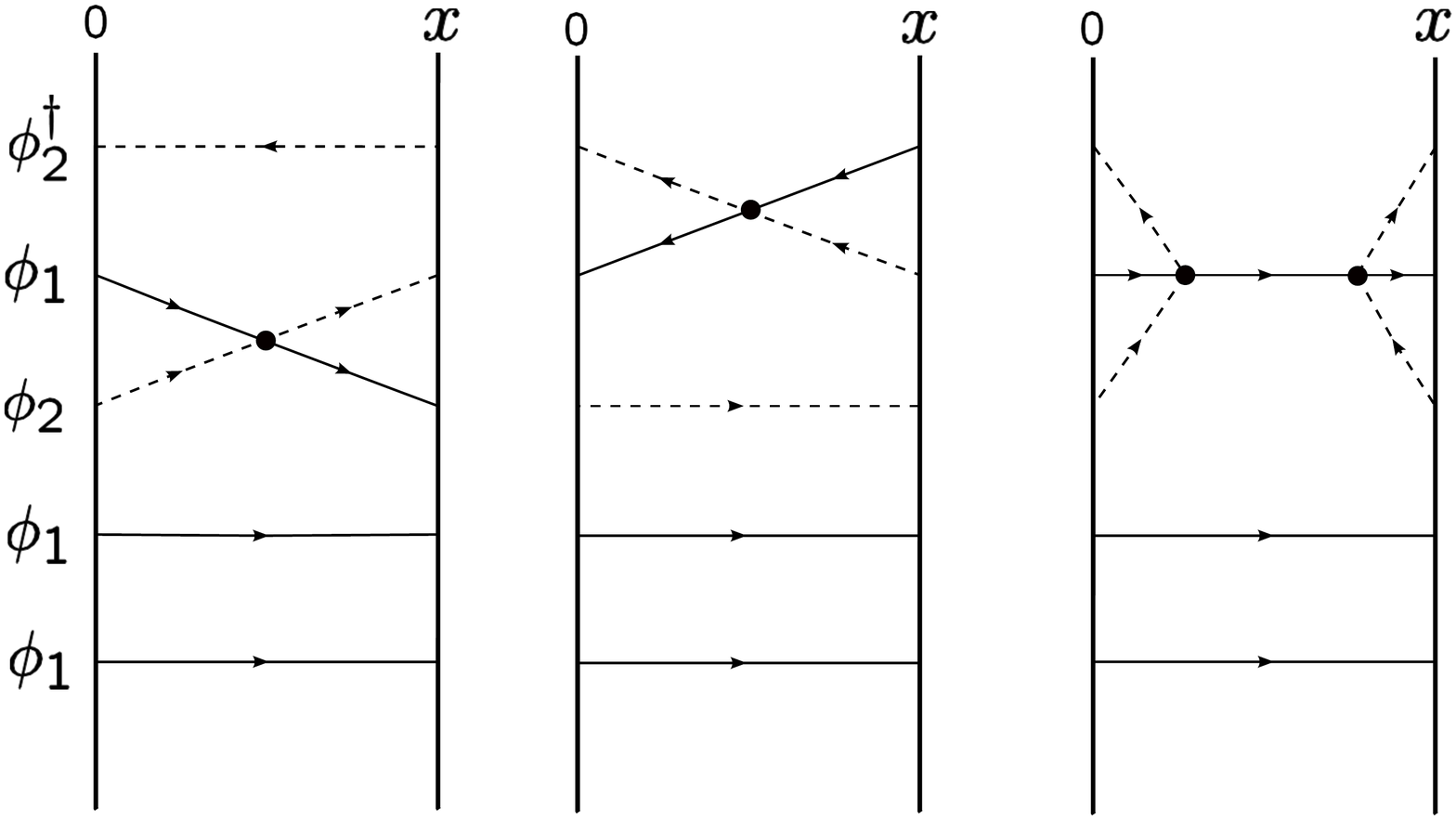}
\end{center}
\vspace*{-24pt}
\caption{The lowest order Feynman graphs contributing to the mixing matrix \eqref{Jordan} of operators of charge \(J_1=3, J_2=J_3=0\) and the length \(L=5\). Bringing this matrix to the Jordan form we encounter two  complex conjugate anomalous dimensions  \eqref{DeltaL5} as well as \(2\times 2\) a Jordan cell leading to logarithmic conformal correlator \eqref{log-corr}, typical for non-unitary CFTs.      }
\label{fig:Fgraphs}
\end{figure}

\begin{figure}[h]
\begin{center}
\includegraphics[scale=0.55]{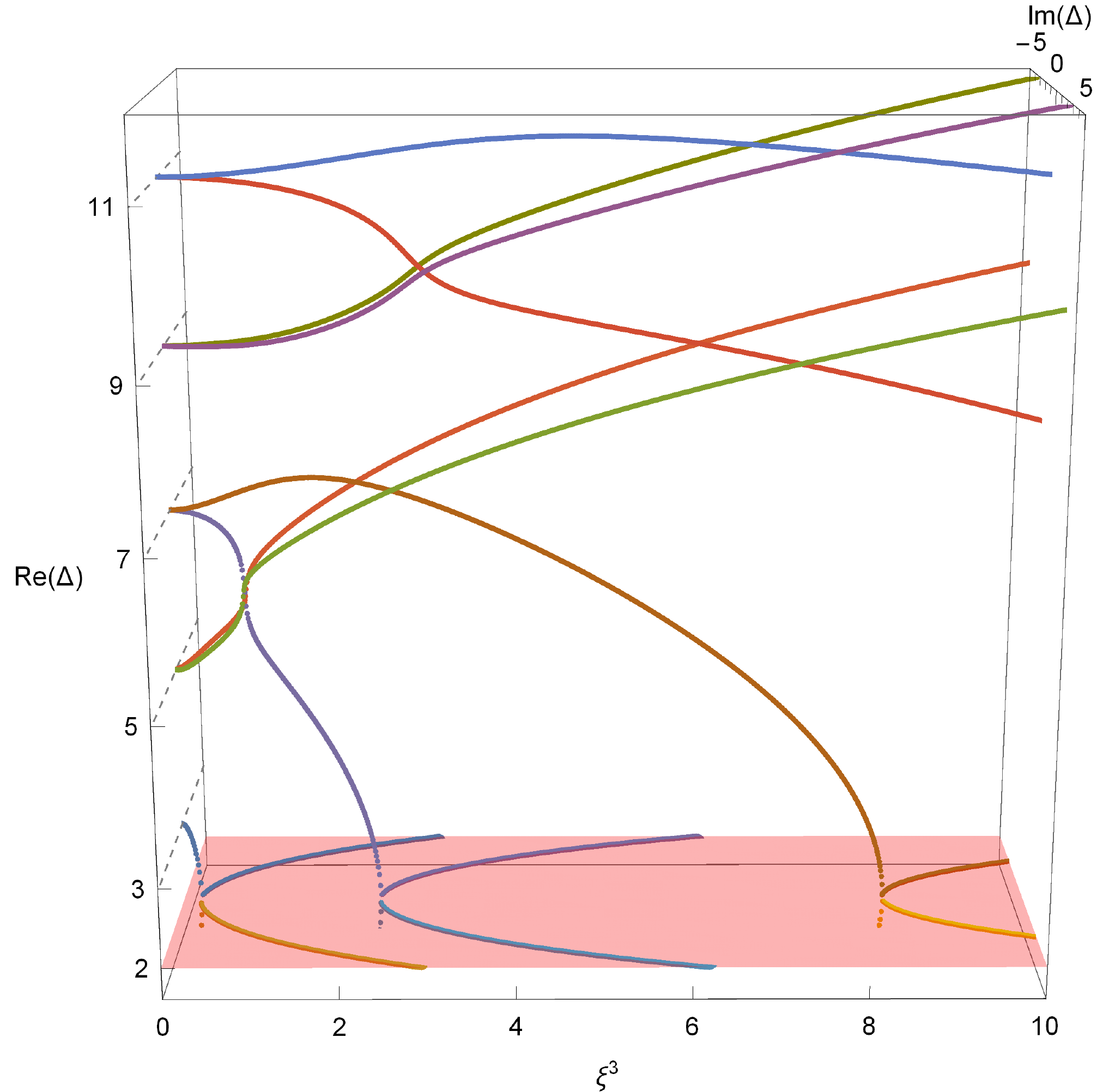}
\end{center}
\caption{Real and imaginary part of the scaling dimension of the nine lowest lying states with \(L=3\). The curve that starts at \(\Delta(0)=3\) corresponds to the operator \(\tr(\phi_1^3)\). The pair of states that start at \(\Delta(0)=3+2k\) with \(k=1,2,3,4\) correspond to the operators of the form \(\tr[(\phi_1)^3(\phi_2^\dagger\phi_2)^k]+\text{permutations}\).}
\label{fig:states}
\end{figure}

\paragraph{Higher twist solutions and Jordan cells for mixing matrix} Actually, our Baxter equation (\ref{DSBaxter}), together with the quantization conditions \eqref{qquant}, describes not only the operator \(\tr(\phi_1)^3\) but also all operators with the same \(R\)-charge \(J_1=L,J_2=J_3=0\). They can be represented as linear combinations of operators of  length\(=3+2n\), of the type: \(\tr[(\phi_1)^3(\phi_2^\dagger\phi_2)^n]\) with all possible permutations of fields there.\footnote{The operators with insertions of \(\phi_1^\dagger\) appear to be protected.} The weak coupling expansion, similar to the described above, leads to the following, complex conjugate values of two dimensions of length-5 operators:
\begin{eqnarray}  
\kern-2pt
\Delta_{5}^{\pm}=5\mp2 i \xi ^3+3 \xi ^6\pm\frac{31 i \xi ^9}{4}+\xi ^{12} \left(3 \zeta_3-\frac{97}{4}\right)\pm i\xi ^{15} \left(\frac{27 \zeta_3}{2}-\frac{5359 }{64}\right)+\dots.\quad
\label{DeltaL5}
\end{eqnarray}
In fact, they correspond to a multiplet formed by four operators
\begin{align}\label{O's}
&O_1=\tr(\phi_1^3 \phi_2 \phi_2^\dagger)\,,\qquad
&&O_2=\tr(\phi_1^2 \phi_2\phi_1 \phi_2^\dagger)\,,
\notag\\
&O_3=\tr(\phi_1 \phi_2\phi_1^2 \phi_2^\dagger)\,,\qquad
&&O_4=\tr(\phi_2\phi_1^3 \phi_2^\dagger)\,.
\end{align}
Their mixing matrix is not Hermitian, reflecting the non-unitarity of the theory. Computing it in the lowest order of perturbation theory by means of Feynman graphs presented on Fig.\ref{fig:Fgraphs}
 and bringing it it
Jordan form we obtain \begin{align}
V = \frac{1}{4\pi^2}\left[\begin{array}{cccc}0 & \xi^2  & 0 & 0 \\0 & 0 & \xi^2  & 0
\\0 & -\xi^4  & 0 & \xi^2  \\0 & 0 & 0 & 0\end{array}\right]=U\cdot \left(
\begin{array}{cccc}
 0 & 1 & 0 & 0 \\
 0 & 0 & 0 & 0 \\
 0 & 0 & -i \xi ^3 & 0 \\
 0 & 0 & 0 & i \xi ^3 \\
\end{array}
\right)\cdot U^{-1}.
\label{Jordan}\end{align}  Plugging it into Callan-Symanzik equation \(\mu \frac{d}{ d\mu} O_i(x) = -V_{ij} O_j(x)\) and solving it for  pair correlation functions we obtain a correlation matrix  consisting of two standard conformal correlation functions on the diagonal,
 corresponding to the dimensions
\eqref{DeltaL5} (at two lowest orders), as well as a Jordan block of correlators with the logarithmic behavior characteristic for the non-unitary CFTs\footnote{The presence of such logarithmic correlation functions in the bi-scalar theory, omnipresent in non-unitary CFTs\cite{Gurarie:1993xq}~, was fist noticed by J.Caetano (unpublished).}:
  \begin{align}
\langle \tilde{\cal O}^{\dagger}_\alpha(x) \tilde{\cal O}_\beta(0)
 \rangle_{\text{ren}} =\frac{1}{x^{10}}
 \left[\begin{array}{cc}0 & 1 \\ 1 & \log x^2\mu^2 \end{array}\right]_{\alpha\beta}.
\label{log-corr}\end{align}
We also performed in\cite{Gromov:2017cja} numerical calculations for this and a few higher multiplets, revealing a rich  structure of the spectrum of this conformal CFT (see Fig.\ref{fig:states}).

The behavior of these dimensions was also analysed in\cite{Gromov:2017cja} analytically in the strong coupling limit \(\xi\to\infty\). The results suggested an interesting interpretation of the system in terms of the classical dynamics of three non-compact spins. It is an interesting step in the direction of understanding whether the bi-scalar theory has a string dual description. Probably for a better understanding of this problem one has to analyze the operators of the type \(\tr(\phi_1)^L\) at \(L\to\infty\), in analogy with the Frolov-Tseytlin limit in standard \({\cal\ N}=4\) SYM theory.

\section{Prospects and unsolved problems}\label{sec:conclusions}

This   review has two main purposes: Firstly, to give the most general, albeit minimalist formulation of the quantum spectral curve (QSC) formalism for the spectrum of dimensions of local operators of the \(\gamma\)-deformed \({\cal\ N}=4\) SYM theory; Secondly,  to demonstrate the  power of QSC on the example of analysis of the spectrum of certain operators in the bi-scalar CFT , dominated by ``fishnet" graphs in planar limit, emerging from the the \(\gamma\)-deformed \({\cal\ N}=4\)  in a certain double scaling limit combining  weak coupling and strong imaginary twist.

The QSC formalism already permitted to obtain outstanding new results in \({\cal\ N}=4\) SYM theory (see references in introduction  and in the review\cite{Gromov:2017blm}). The QSC construction is also known for the ABJM model\cite{Cavaglia:2014exa,Anselmetti2015,Lee2017} and even for the Hubbard model~\cite{Cavaglia2015}. The QSC formalism is designed first of all for the spectral problem for local operators. However, with appropriate modifications it was also  used for computing the dimensions of non-local quantities, such as the cusped Wilson loop, the quark-anti-quark potential or the BFKL  limit for twist-2 operators~in planar \({\cal\ N}=4\) SYM \cite{Gromov:2015dfa,Gromov:2016rrp,Alfimov:2014bwa,Gromov:2015wca}. So far, the generalizations of QSC  to more complicated OPE data, such as the structure constants, is not known, though the underlying physical quantities -- the three-point correlation functions -- obey remarkable integrability properties\cite{Basso:2015zoa,Fleury:2016ykk}. Another desirable generalization of QSC would be the \(1/N_c^2\) corrections where the integrability seems to be also helpful\cite{Bargheer:2017nne}. The QSC is also already constructed for the other deformations of SYM related to quantum groups\cite{Klabbers:2017vtw}.

The bi-scalar CFT \eqref{bi-scalarL} and its generalizations \eqref{fullDS}~\cite{Gurdogan:2015csr,Caetano:TBP}, obtained from  \(\gamma\)-deformed \({\cal\ N}=4\) SYM theory in special double scaling (DS) limit,   apart from being new interesting examples of {\it integrable planar four-dimensional CFTs},  also play an important conceptual role. Namely, they are dominated by very particular sets of {\it integrable} planar Feynman graphs, such as  ``fishnet" graphs (of the shape of regular square lattice in the bulk of graph) of bi-scalar model, or the ``brick wall" graphs of the case \(\xi_3=0,\,\xi_1=\xi_2\ne 0\) of the model \eqref{fullDS} formed by regular hexagonal lattice of Yukawa-type vertices\cite{Caetano:2016ydc,Chicherin:2017frs}. Thus these CFTs explicitly demonstrate, for the first time, the all-loop integrability of the original  \({\cal\ N}=4\) SYM theory, at least in this specific DS limit and it might be the key of understanding of the origins of the full AdS\(_5\)/CFT\(_4\) integrability and of the gauge-string duality. Similar considerations are applicable to the three-dimensional regular triangular planar graphs emerging from the  ABJM theory\cite{Caetano:TBP}. A similar chiral CFT dominated by hexagonal graphs can be constructed in six dimensions\cite{Mamroud:2017uyz} but, curiously, its 6D ``mother" SYM theory is unknown. It is also worth asking whether a similar chiral CFT dominated by regular planar graphs could be found in two dimensions and whether it could be an analogous DS limit of the twisted AdS\(_3\)/CFT\(_2\) duality.

The chiral CFT's emerging in the  double scaling are dominated by very few graphs at each order (sometimes only a single one, such as the ``wheel"(Fig.\ref{fig:globe}) and ``spiral"(Fig.\ref{fig:spiral}) graphs for the vacuum and one magnon operators of bi-scalar CFT). This opens an opportunity to compute these graphs exactly, at any number of loops, as we demonstrated it here for the wheel graphs. Another interesting case is the fishnet amplitudes defined and studied in  \cite{Chicherin:2017cns,Chicherin:2017frs}: each of them is dominated by a single fishnet diagram with specific boundary -- a disc cut out from a piece of regular square lattice. Some particular  graphs of this kind have been recently computed\cite{Basso:2017jwq,Bourjaily:2017bsb} and it would be interesting to understand and generalize these results from the point of view of the Yangian symmetry discovered in\cite{Chicherin:2017cns,Chicherin:2017frs}.

The question of existence of a string dual for the double scaling limit considered here remains open. Naively, the classical string picture is gone since in the weak coupling limit the AdS radius goes to zero. On the other hand, we deal with multi-loop Feynman graphs which might provide, at high loop orders, at least for  long operators, a new dual string description. The results of\cite{Gromov:2017cja} in the strong coupling limit, already with respect to the DS coupling \(\xi\), are encouraging in this sense since they show that the bi-scalar model can be described by a classical model of a few non-compact spins. If we increase the R-charge of operator, and thus the number of spins, we could reach a classical string picture. The study of wheel graphs of higher lengths, certainly possible by integrability, would be an important step in this direction.

Many physical quantities which seem prohibitively difficult to compute in the full  \({\cal\ N}=4\) SYM theory appear to be accessible in the DS limit. In particular, the exact dimension \(\Delta_2(\xi,S)\) of \(L=2\) wheel operator and of similar operators with conformal spin \(S\), of the form \(\tr(\phi_1\partial^S_\pm\phi_1)\),  was computed in\cite{Grabner:2017pgm} in explicit form and is given in very simple explicit form, as solutions of
\begin{align}  
\label{L2Delta}
\frac{1}{ 16}{}& (\Delta +S-2) (\Delta +S)
(\Delta -S-2) (\Delta -S-4)=\xi^4.
\end{align}
{Moreover, an exact four-point function for specific scalar operators, the \(L=2\) case of the correlator \eqref{Kkernel}, given by the cylindric graphs of the type drawn on fig.\ref{fig:graphBuild}, was explicitly computed in\cite{Grabner:2017pgm} in all loops -- the only example, to our knowledge, of\hfilneg}

\pagebreak

\noindent
explicit all-loop calculation of a non-trivial four-point function in a 4D CFT.\footnote{A similar computation was done in the D-dimensional analogue of the bi-scalar fishnet model, as proposed in\cite{Kazakov:2018qbr}. Its 2D version looks particularly interesting since it is directly related to the \(SU(2,C)\) spin chain describing a 2D  CFT for BFKL physics\cite{Lipatov:1993qn,Faddeev:1994zg,Korchemsky:1997fy}. Its 1D version describes, in the case of a similar 4-point correlation function, the summation of ladder graphs appearing in the scalar sSYK model~ \cite{Gross:2017aos}.   } The formula \eqref{L2Delta} was obtained  in\cite{Grabner:2017pgm} by analysing the divergencies of this four-point correlation function, as well as from the QSC formalism. This four-point correlation function appears to have a nice OPE expansion generating infinitely many exact structure constants involving the above operators of \(L=2\).

The computation of more complicated correlators, involving longer operators and multi-point correlators,  is a complicated but very promising enterprise. To do it efficiently, we have to learn how to efficiently taylor such quantities from the integrable conformal spin chains, in the spirit of the one-loop procedure of\cite{Escobedo:2011xw}.

\vspace*{-5pt}
\section*{Acknowledgments}
\label{sec:acknowledgments}
\vspace*{-1pt}
We thank V.~Bazhanov, B.~Basso, N.~Gromov, J.~Caetano, S.~Derkachov, G.~Korchemsky, I.~Kostov, S.~Leurent, F.~Levkovich-Maslyuk,
E.~Olivucci, D.~Serban, Z.~Tsuboi, D.~Volin, D.~Zhong for reading the manuscript and providing very useful comments.  The work was supported by
the European Research Council (Programme ``Ideas'' ERC-2012-AdG 320769
``AdS-CFT-solvable").

\vspace*{-5pt}
\bibliographystyle{ws-rv-van} 
\bibliography{biblio_Faddeev_Volume_arxiv} 



\end{document}